\documentclass[10pt,journal]{IEEEtran}
\IEEEoverridecommandlockouts
\usepackage{cite}
\usepackage{color}
\usepackage{amsmath,amssymb,amsfonts}
\usepackage{algorithmic}
\usepackage{graphicx}
\usepackage{textcomp}
\def\BibTeX{{\rm B\kern-.05em{\sc i\kern-.025em b}\kern-.08em
    T\kern-.1667em\lower.7ex\hbox{E}\kern-.125emX}}

\usepackage[ruled]{algorithm2e}

\usepackage{url}

\def\BibTeX{{\rm B\kern-.05em{\sc i\kern-.025em b}\kern-.08em T\kern-.1667em\lower.7ex\hbox{E}\kern-.125emX}}

\newcommand {\mymarginpar}[1]{\marginpar{#1}}
\renewcommand {\marginpar}[1]{}

\def\_{\rule{.3em}{.15ex}}      

\newcommand{\ls}[1]
   {\dimen0=\fontdimen6\the\font
    \lineskip=#1\dimen0
    \advance\lineskip.5\fontdimen5\the\font
    \advance\lineskip-\dimen0
    \lineskiplimit=.9\lineskip
    \baselineskip=\lineskip
    \advance\baselineskip\dimen0
    \normallineskip\lineskip
    \normallineskiplimit\lineskiplimit
    \normalbaselineskip\baselineskip
    \ignorespaces
   }

\newcommand {\bearn}{\begin{eqnarray*}}
\newcommand {\eearn}{\end{eqnarray*}}
\newcommand {\barr}{\begin{array}}
\newcommand {\earr}{\end{array}}

\newcommand {\N}{{\cal N}}

\newtheorem{definition}{Definition}
\newtheorem{property}[definition]{Property}
\newtheorem{proposition}[definition]{Proposition}
\newtheorem{lemma}[definition]{Lemma}
\newtheorem{theorem}[definition]{Theorem}
\newtheorem{corollary}[definition]{Corollary}
\newtheorem{example}{Example}
\newtheorem{remark}[definition]{Remark}

\newcommand {\benum} {\begin{enumerate}}
\newcommand {\eenum} {\end{enumerate}}

\newcommand {\bdesc} {\begin{description}}
\newcommand {\edesc} {\end{description}}

\newcommand {\bfig}[2] {\begin{figure}
  \centering
  \includegraphics[width=#2]{#1}}
\newcommand {\brotatefig}[2] {\begin{figure}[htbp]
                        \centerline {
                         \epsfig{figure={#1},clip=,angle=-90,width={#2}}}}
\newcommand {\bfigfirst}[2] {\begin{figure}[h]
                        \centerline {
                        \setlength{\epsfxsize}{#2}
                        \epsffile{#1}}}
\newcommand {\efig}[2]{ \caption{#2}
                        \label{fig:#1}
                        \end{figure}
                        \mymarginpar{fig:#1}}
\newcommand {\erotatefig}[2]{ \caption{#2}
                        \label{fig:#1}
                        \end{figure}
                        \mymarginpar{fig:#1}}
\newcommand {\rfig}[1]{Figure \ref{fig:#1}}

\newcommand {\btab}[1]{
                       \begin{table}
                       \centering
                       \begin{tabular}{#1}}
\newcommand {\etab}[3] {
                       \end{tabular}
                       \caption[#3]{#2}
                       \label{tab:#1}
                       \end{table}
                       \mymarginpar{tab:#1}
                       \vspace{.1in}}

\newcommand {\btabular}[1]{\begin{center}
                       \begin{tabular}{#1}}
\newcommand {\etabular}{\end{tabular}
                       \end{center}}

\newcommand {\bdefin}[1]{\begin{definition}
                      \mymarginpar{def:#1}
                      \label{def:#1} }
\newcommand {\edefin}       {\end{definition}}

\newcommand {\bpro}[1]{\begin{property}
                      \mymarginpar{pro:#1}
                      \label{pro:#1} }
\newcommand {\epro}   {\end{property}}

\newcommand {\bprop}[1]{\begin{proposition}
                      \mymarginpar{prop:#1}
                      \label{prop:#1} }
\newcommand {\eprop}       {\end{proposition}}

\newcommand {\blem}[1]{\begin{lemma}
                      \mymarginpar{lem:#1}
                      \label{lem:#1} }
\newcommand {\elem}   {\end{lemma}}

\newcommand {\bthe}[1]{\begin{theorem}
                      \mymarginpar{the:#1}
                      \label{the:#1} }
\newcommand {\ethe}   {\end{theorem}}

\newcommand {\bcor}[1]{\begin{corollary}
                      \mymarginpar{cor:#1}
                      \label{cor:#1} }
\newcommand {\ecor}   {\end{corollary}}

\newcommand {\bax}[1]{\begin{axiom}
                      \mymarginpar{ax:#1}
                      \label{ax:#1} }
\newcommand {\eax}       {\vspace{-.1in} \end{axiom}}

\newcommand {\bex}[2]{\vspace{.1in}
                      \begin{example}
                      \mymarginpar{ex:#1}
                       {\bf #2}
                      \label{ex:#1} }
\newcommand {\eex}       {\end{example} \vspace{.3cm} }
\newcommand {\rex}[1]{Example \ref{ex:#1}}

\newcommand {\brem}[1]{\begin{remark}
                      \mymarginpar{rem:#1}
                      \label{rem:#1} \em }
\newcommand {\erem}   {\end{remark}}

\newcommand {\beq}[1]{\mymarginpar{eq:#1}
                      \begin{equation}
                      \label{eq:#1} }

\newcommand {\beqno}[1]{\mymarginpar{eq:#1}
                      \begin{eqnarray}
                      \nonumber}

\newcommand {\eeq}       {\end{equation}}
\newcommand {\eeqno}       { && \end{eqnarray}}
\newcommand {\req}[1]{(\ref{eq:#1})}

\newcommand {\bear}[1]{\mymarginpar{eq:#1}
                       \begin{eqnarray}
                       \label{eq:#1} }

\newcommand {\bearno}[1]{\mymarginpar{eq:#1}
                       \begin{eqnarray}
                       \nonumber}

\newcommand {\eear}{\end{eqnarray}}
\newcommand {\eearno}{\end{eqnarray}}
\newcommand {\bsel}{\left \{ \begin{array}{cl}}
\newcommand {\esel}{\end{array} \right.}

\newcommand {\bmat}[1]{\left [ \begin{array}{#1}}
\newcommand {\emat}{\end{array} \right ]}
\newcommand {\bsec}[2]{\mymarginpar{sec:#2}
                       \section{#1}
                       \label{sec:#2} }

\newcommand {\rsec}[1]{Section \ref{sec:#1}}

\newcommand {\bsubsec}[2]{\mymarginpar{sec:#2}
                       \subsection{#1}
                       \label{sec:#2} }

\def\R{I\kern-0.30em R}
\def\N{I\kern-0.30em N}
\def\P{I\kern-0.30em P}

\def\bfc{{\bf c}}

\newcommand{\rhog}{\rho}

\begin{document}

\title{Poisson Receivers: a Probabilistic Framework for Analyzing Coded Random Access}

\author{Che-Hao Yu, Lin Huang, Cheng-Shang~Chang,~\IEEEmembership{Fellow,~IEEE,} and Duan-Shin Lee,~\IEEEmembership{Senior Member,~IEEE}
                \thanks{C.-H. Yu, L. Huang, C.-S. Chang, and D.-S. Lee are with the Institute of Communications Engineering, National Tsing Hua University, Hsinchu 30013, Taiwan, R.O.C. Email: chehaoyu@gapp.nthu.edu.tw;   hl13236597660@outlook.com; cschang@ee.nthu.edu.tw;  lds@cs.nthu.edu.tw.  This work was supported in part by the Ministry of
Science and Technology, Taiwan, under Grant 106-2221-E-007-023-MY3,
and in part by Qualcomm Technologies under Grant SOW NAT-414899. A 14-page version of this paper was submitted to IEEE/ACM Transactions on Networking on Nov. 7, 2019, for possible publication.}
}

\maketitle
\begin{abstract}
In this paper, we develop a probabilistic  framework for analyzing coded random access. Our framework is based on a new
abstract receiver (decoder), called a Poisson receiver, that is characterized by a  success probability function of a tagged packet subject to a Poisson offered load.  { We show that various coded slotted ALOHA (CSA) systems are Poisson receivers. Moreover, Poisson receivers have two elegant closure properties: (i) Poisson receivers with {\em packet routing} are still Poisson receivers, and
(ii) Poisson receivers with  {\em packet coding} are still Poisson receivers.
These two closure properties enable us to use smaller Poisson receivers as building blocks for analyzing a larger Poisson receiver.
As such, we can analyze complicated systems that are not possible by the classical tree evaluation method.
In particular, for CSA systems with both spatial diversity and temporal diversity, we can use the framework of Poisson receivers to compute the exact (asymptotic) throughput.}
We demonstrate that
our {\em framework} can be used to provide differentiated services between ultra-reliable low-latency communication (URLLC) traffic and enhanced mobile broadband (eMBB) traffic. By conducting extensive simulations, we also verify that our theoretical results  match extremely well with the simulation results.

\end{abstract}





%

\bsec{Introduction}{introduction}

In the past few years due to the demand for the massive machine-type communications (mMTC), there is a re-surged interest in random access. In the setting of mMTC, there are a large number of {\em uncoordinated} devices contending for a shared medium.
The slotted ALOHA (SA) protocol \cite{ALOHA}, in which active devices transmit their packets at random (over a frame of time slots), appears to fit very well for the setting of mMTC. One of the key issues of SA is the collision problem when multiple active devices transmit at the same time.
Collisions can greatly degrade the system performance of SA, in particular, the (system) throughput. To improve the throughput of SA, various approaches that exploit the diversity gains, including the spatial diversity and the temporal diversity, were proposed in the literature (see, e.g., the surveys \cite{paolini2012random,paolini2015coded,munari2015multi,stefanovic2018coded}).

For spatial diversity, Zorzi \cite{zorzi1995mobile} derived the throughput formula under the Rayleigh fading channel model with {\em two} independent receivers.
Munari et al. \cite{munari2013throughput} considered the on-off fading (erasure) channel with $J$ ($J \ge 2$)  independent receivers.
In their setting, each transmitted packet reaches a receiver independently with probability $1-\epsilon$. They
then derived a closed-form expression for the throughput of SA. On the other hand, for temporal diversity, Casini et al. \cite{casini2007contention} proposed the
 Contention Resolution Diversity Slotted ALOHA (CRDSA) protocol that uses the successive interference cancellation (SIC) technique to increase throughput. In CRDSA, two copies of each packet from an active  user are transmitted randomly in the system. If any one of these two copies of a packet is successfully received by a receiver, then the other copy can be removed (cancelled) from the system to further reduce possible collisions. Such a process can  then be repeatedly carried out to decode the rest of the packets. As shown in \cite{casini2007contention}, such an approach results in a significant  improvement in throughput.
Instead of using a fixed number of repetitive copies, one can also use an irregular (random) number of copies or other coding schemes to further optimize the throughput. These systems are commonly referred to as the {\em coded slotted ALOHA (CSA)} systems.
 In particular, Liva \cite{liva2011graph} developed a framework for analyzing the throughput with a random number of copies (from a distribution) by using the and-or tree evaluation in \cite{luby1998analysis}.
 It was further shown by Narayanan  and Pfister in \cite{narayanan2012iterative} that 100\% throughput is asymptotically feasible (by choosing a proper distribution) { in the CSA system considered in  \cite{liva2011graph}} if the number of active devices is known.
 The problem of estimating the number of active devices was previously addressed in \cite{stefanovic2013joint}.

The idea of using SIC for temporal diversity can also be applied for spatial diversity.
In order to do this, receivers need to exchange information regarding the packets that are successfully decoded.
Receivers with (resp. without) such capability are called {\em cooperative (resp. non-cooperative)} receivers \cite{jakovetic2015cooperative,ogata2017multi,stefanovic2018coded}.
However, exact throughput analysis for CSA with cooperative receivers using the and-or tree evaluation in \cite{luby1998analysis}
appears to be very difficult as it needs to tackle both spatial diversity and temporal diversity at the same time.
In particular, for the geometric graph model with spatial diversity and temporal diversity in \cite{jakovetic2015cooperative,stefanovic2018coded}, no strong results for throughput exist,  and only bounds on throughput are available. On the other hand, the walk graph approach in \cite{ogata2017multi} is computationally difficult as it needs to examine every possible walk graph { in every SIC iteration} that leads to a successfully decoded packet.

{ The and-or tree  evaluation method in \cite{luby1998analysis,liva2011graph,paolini2011graph,paolini2012random} is an iterative decoding  method on a random {\em bipartite graph}, where one side of nodes uses the AND-gate to decode incoming edges and the other side of nodes uses the OR-gate to decode incoming edges. By iteratively tracking the evolution of the probability that an incoming edge can be decoded, the and-or tree evaluation method then derives a limit on the probability that an incoming edge can be decoded. It is not necessary to restrict the decoding method to the AND-gate
and the OR-gate. One can use a more complicated decoding method at each node. In particular, the and-or tree evaluation method was extended in \cite{luby1998analysisb,richardson2001capacity} to derive the capacity of low-density parity-check (LDPC) codes \cite{gallager1962low}.
Such a tree evaluation method tracks the evolution of the densities of the symbols in an LDPC code, and it is known as the density evolution (DE) method.
However, the tree evaluation method in \cite{liva2011graph,paolini2011graph,paolini2012random} only works on a {\em single} (tree-like) bipartite graph. In a CSA system with multiple cooperative receivers,
the decoding process can no longer be modelled by a single bipartite graph. Although they might be modelled by a complicated graph that consists of multiple  bipartite subgraphs (with each receiver being mapped to a bipartite subgraph), using the tree  evaluation method directly to decode incoming edges in that graph is no longer valid
as the probability of having small cycles does not vanish asymptotically with the size of the
graph \cite{stefanovic2018coded}.

To analyze CSA systems with both spatial diversity and temporal diversity, one needs an abstract formulation that can hide the complexity of a complicated graph.
The key innovation of this paper is to propose a new concept of an abstract receiver (decoder), called a {\em Poisson receiver}, that is characterized by a success probability function $P_{\rm suc}(\cdot)$.
If the number of packets  arriving at a Poisson receiver follows a Poisson distribution with mean $\rhog$, a tagged packet is successfully received with probability $P_{\rm suc}(\rhog)$. As such, the throughput of a Poisson receiver subject to a Poisson offered load $\rho$ is $\rhog \cdot P_{\rm suc}(\rhog)$.
 Such a concept can be generalized to the setting with multiple classes of input traffic. With $K$ classes of input traffic, a  tagged class $k$ packet is successfully received with probability $P_{{\rm suc},k}(\rhog)$, for $k=1,2, \ldots, K$ when the receiver is subject to a Poisson offered load $\rho=(\rho_1, \rho_2, \ldots, \rho_K)$. For such a Poisson receiver with multiple classes of input traffic, the throughput for class $k$ traffic is $\rhog_k \cdot P_{{\rm suc},k}(\rhog)$.

 There are several elegant properties of Poisson receivers.
\begin{description}
\item[(i)] Many CSA systems are Poisson receivers, including SA (see \rsec{collision}), SA with multiple {\em non-cooperative} receivers (see \rsec{onoff}), and SA with multiple {\em cooperative} receivers (see \rsec{twoSIC} and \rsec{SAmul}).
\item[(ii)]  Poisson receivers with {\em packet routing} are still Poisson receivers (see \rsec{routing} for detailed descriptions).
\item[(iii)] Poisson receivers with  {\em packet coding} are still Poisson receivers (see \rsec{cprmul} for detailed descriptions).
\end{description}
The first property shows that many CSA systems are Poisson receivers, and
these Poisson receivers serve as basic building blocks (by hiding the complexity of spatial diversity).
 The last two properties are known as {\em closure properties}, and they enable us to use {\em smaller}  Poisson receivers as {\em building blocks} for analyzing a {\em larger} Poisson receiver. As such, we can analyze complicated systems that are not possible by the classical tree evaluation method. In particular, to analyze a CSA system with temporal diversity,
 we model it by
a system of Poisson receivers with  packet coding. In that system, the tree evaluation method is carried out iteratively on a bipartite graph with one side of nodes decoded by (abstract) Poisson receivers. For these Poisson receivers, they
 have  Poisson degree distributions in the bipartite graph. The Poisson degree distribution plays a crucial
role in proving the closure property for Poisson receivers with  packet coding. Two important closure properties of the Poisson degree distribution
are used in the tree  evaluation method: (i) the excess degree distribution (defined as the degree distribution along a randomly selected edge) of a Poisson degree distribution is still a Poisson degree distribution, and (ii)  random thinning (that removes each edge independently with a certain probability) of a Poisson degree distribution is still a Poisson
degree distribution. Without the assumption on the Poisson degree distribution, tracking the decodability probabilities in the tree  evaluation method is extremely complicated.

 Poisson receivers with multiple classes of input traffic enable us to provide differentiated services in CSA systems with both spatial diversity and temporal diversity.  For such a CSA system, we consider a correlated on-off fading channel.
Our correlated on-off fading channel model is
more general than the independent on-off fading (erasure) channel model in \cite{munari2013throughput} as it can model the correlation of the channel states of the receivers.
To the best of our knowledge,  our method seems to be the first analytical method that can compute the exact (asymptotic) throughput of each class in such a CSA system.
In particular, we demonstrate how Poisson receivers can be used to provide differentiated services between ultra-reliable low-latency communication (URLLC) traffic and enhanced mobile broadband (eMBB) traffic.

The rest of the paper is organized as follows.
In \rsec{aloha}, we briefly review the collision channel model in the SA system with a single receiver and then extend it to
the correlated on-off fading channel model in the SA system with multiple receivers. Closed-form expressions for the throughput of two non-cooperative receivers and that of two cooperative receivers are shown in \rsec{onoff} and \rsec{twoSIC}, respectively.
In \rsec{poisson}, we introduce the concept of Poisson receivers
and
 develop the tree evaluation method for coded Poisson receivers. We then extend our probabilistic  framework
 to Poisson receivers with multiple classes of input traffic.
 Various numerical results are shown in \rsec{num} to further verify our theoretical results and provide insights on the effects of the offered load and the channel parameters on the throughput. In particular, we compute our theoretical results and conduct extensive simulations to estimate the throughputs of  various CSA systems in a correlated on-off fading channel with two receivers in \rsec{numerical}.
We also demonstrate how the theory of Poisson receivers can be used for providing differentiated services between URLLC traffic and eMBB traffic in \rsec{usecase}.
The paper is then concluded in \rsec{con}.

In this paper, we use the two phrases ``successfully received'' and ``successfully decoded'' interchangeably. They both mean the receiver has the same information of a packet as the sender.  In Table \ref{table:notations}, we provide a list of notations used in the paper.

{
	\tiny
	\begin{table}[ht]
		\begin{center}
			\caption{List of Notations\label{tab:one}}{%
				\begin{tabular}{||l|l||}
					\hline\hline
					$A(\bf c)$ & The set of receivers reached by this tagged packet in \\
&state $\bf c$\\
					$B_k$ & The set of receivers associated with class $k$ users. \\
					$\bf c$ & The channel state\\
					${\bf c}_i(t)$ & The binary $J$-vector that represents the channel state \\
&of the on-off fading channel for user $i$ at time $t$ \\
					$F(m)$ & The configuration graph for the vector $m$ \\
					$G$ & The normalized offered load \\
					$J$ & The number of receivers \\
					$K$ & The number of classes of input traffic \\
					$L$ & The number of copies of a packet \\
					$M$ & The number of packets transmitted in a particular \\
&time slot \\
					$M_A$ & The number of packets that reach at least one of the\\
&receivers in the set $A$ \\
					$z_k$ & The number of class $k$ packets transmitted in a \\
&particular  time slot \\
					$N$ & The number of active users \\
					$m$ & The vector $(m_1, \ldots, m_K)$ \\
					$m_k$ & $\min[z_k,2]$ \\
					$P_{\bf c}$ & The probability of channel state ${\bf c}$ \\
					$P_{\rm suc}$ & The success probability of a tagged packet. \\
					$P(A)$ & The probability that a packet reaches at least one of \\
&the receivers in the set $A$ in that time slot \\
					$P_{\rm suc}^{\rm all}(A)$ & The probability that a tagged packet is successfully\\
& received by all the receivers in the set $A$\\
					$P_{\rm suc}(A)$ & The probability that a tagged packet is successfully\\
& received by at least one of the receivers in the set $A$\\
					$p(n)$ & The probability of the configuration $n$ subject to the \\
&(independent) Poisson offered load $\rho$ \\
					$p_k^{(i)}$ & The probability that the receiver end of a randomly \\
&selected class $k$ edge has not been successfully\\
& received after the $i^{th}$ SIC iteration \\
					$q_k^{(i)}$ & The probability that the user end of a randomly \\
& selected class $k$ edge has not been successfully \\
&received after the $i^{th}$ SIC iteration\\
					$R$ & The routing probability matrix \\
					$r_{k_1,k_2}$ & The probability that a class $k_1$ external packet \\
&transmitted to the Poisson receiver becomes  \\
					& a class $k_2$ packet at the Poisson receiver \\
					$S$ & The throughput \\
					$T$ & The number of Poisson receivers/cooperative receivers\\
					$w_k(n)$ & The number of class $k$ packets that are successfully\\
& received  in the configuration graph $F(n)$\\
					$\epsilon$ & The erased probability \\
					$\rho$ & The Poisson offered load $\rho=(\rho_1, \ldots, \rho_K)$\\
                   $\rho_k$ & The Poisson offered load of class $k$\\
					$\Lambda_{k,\ell}$ & The probability that a class $k$ packet is transmitted\\
& $\ell$ times \\
					$\Lambda_k(x)$ & The generating function of the degree distribution of\\
& a class $k$ user node \\
                   $\Lambda_k^{\langle j \rangle} (x)$ & The $j^{th}$ derivative of $\Lambda_k(x)$  \\
					$\lambda_{k,\ell}$ & The probability that the user end of a randomly \\
&selected class $k$ edge has additional $\ell$ edges\\
&excluding the randomly selected  edge \\
					$\lambda_k(x)$ & The generating function of the excess degree \\
&distribution of a class $k$ user node \\
$\lambda_k^{\langle j \rangle} (x)$ & The $j^{th}$ derivative of $\lambda_k(x)$ \\
					\hline
					\hline
			\end{tabular}}
			\label{table:notations}
		\end{center}
	\end{table}
}

\bsec{Slotted ALOHA}{aloha}

\bsubsec{SA in the collision channel model with a single receiver}{collision}

\begin{figure}[ht]
	\centering
	\includegraphics[width=0.45\textwidth]{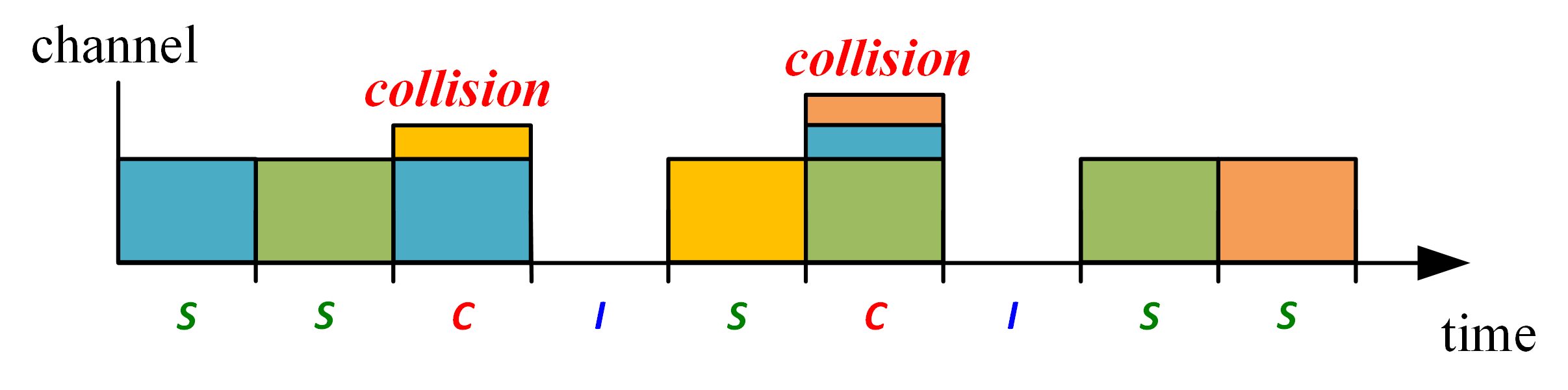}
	\caption{Slotted ALOHA.}
	\label{fig:saloha}
\end{figure}

In this section, we briefly review the
SA system \cite{ALOHA} (see \rfig{saloha} for an illustration) in the classical collision channel model (with a single receiver). In such a system, time is partitioned into fixed-length time slots. In each time slot, an active user (who has a packet to send) transmits its packet with a certain probability. When there is more than one packet transmitted in a time slot, we say there is a collision (see the time slots marked with the letter ``c'' in \rfig{saloha}). Collided packets are assumed to be lost. On the other hand, if there is exactly one packet transmitted in a time slot, then that packet is assumed to be successfully received.

When the number of active users is large, the number of packets transmitted in a time slot can be well approximated by a Poisson random variable with mean $\rhog$ \cite{ALOHA}.
As such, the probability that there is exactly one packet transmitted in a time slot is $\rhog e^{-\rhog}$ and this leads to the following well-known throughput-offered load formula:
\beq{aloha1111}
S=\rhog e^{-\rhog}.
\eeq
Moreover, a randomly selected packet, called a {\em tagged} packet, is successfully
received only if there are no other transmissions in the same time slot. Thus, a tagged packet is successfully received with the probability
\beq{aloha1122}
P_{\rm suc}=e^{-\rhog}.
\eeq

\bsubsec{SA in the correlated on-off fading channel model with multiple non-cooperative receivers}{onoff}

In this section, we consider  SA with {\em multiple receivers}. Such a system has been previously analyzed in the literature (see, e.g., \cite{munari2015multi} for a survey and \cite{ogata2017multi} for more references) by using the on-off fading channel \cite{munari2013throughput} that assumes each packet reaches a receiver independently with probability $1-\epsilon$ and is erased with probability $\epsilon$.
Instead of using the independent on-off fading channel model,
we consider a more general on-off fading (erasure) channel model that can model the correlation of the channel states of the receivers.
Suppose that there are $J$ receivers and $N$ active users. The $J$ receivers are {\em non-cooperative} and they are not able to exchange any information among them. For the $i^{th}$ active user,
let $\bfc_i (t)=(c_{i,1}(t), c_{i,2}(t), \ldots, c_{i,J}(t)) \in \{0,1\}^J$ be the binary $J$-vector that represents the channel state of the on-off fading
channel model at time $t$.
In such an on-off fading channel model, a packet transmitted by the $i^{th}$ active user at time $t$ {\em reaches} {(resp. is {\em erased} at)} the $j^{th}$ receiver if $c_{i,j}(t)=1$ (resp. $c_{i,j}(t)=0$).
The channel states are assumed to be independent and identically distributed ({\em i.i.d.}) with respect to time and it is in the state $\bfc=(c_1, c_2, \ldots, c_J)$
with probability $P_{\bfc}$ (that is assumed to be identical for all the $N$ active users).

As described in the SA system with a single receiver in the previous section, { a packet that reaches a receiver in a time slot may not be successfully received by that receiver.}
 A packet is said to be successfully received by a receiver in a time slot if that packet is the only packet that reaches the receiver in that time slot.
 For a multi-receiver SA system in  such an on-off fading channel, a packet is said to be successfully received if it is successfully received by at least one of the $J$ receivers.

To analyze such a system,
let $M$ be the number of packets transmitted in a particular time slot. As discussed in the previous section, $M$
can be well approximated by  a Poisson random variable  with mean $\rhog$, where $\rhog$ is the offered load.
Also, let $M_A$ be the number of packets that reach at least one of the receivers in a set $A$ in that time slot, and
$P(A)$ be the probability that a packet reaches at least one of the receivers in $A$ in that time slot.
Clearly,
\beq{erase1111}
P(A)=\sum_{\{\bfc: c_j=1, \;\mbox{for}\;\mbox{some}\;j \in A\}}P_\bfc .
\eeq

 { Since $M$ is a Poisson random variable  with mean $\rhog$, and a packet in $M$ is randomly erased with probability $1-P(A)$, $M_A$ is a Poisson random variable with mean
$\rhog \cdot P(A)$. Thus, the probability that no packet reaches any of the receivers in a set $A$ in a time slot is
\beq{erase1125}
e^{-\rhog \cdot P(A)}.
\eeq
Let $P_{\rm suc}^{\rm all}(A)$ be the probability that a tagged packet is
successfully received by  {\em all} the receivers in $A$.
For a SA system, this happens if no other packets reach any of the receivers in $A$ in that time slot.
Thus, we have from \req{erase1125} that
\beq{erase1122}
P_{\rm suc}^{\rm all}(A)=e^{-\rhog \cdot P(A)}.
\eeq
Let $P_{\rm suc}(A)$ be the probability that a tagged packet is
successfully received  by {\em at least one} of the receivers in $A$.
Suppose that
$A=\{{j_1}, {j_2}, \ldots, j_{|A|}\}$.
Using the inclusion-exclusion principle,
we have
\bear{erase1127}
&&P_{\rm suc}(A)=P_{\rm suc}(\cup_{\ell=1}^{|A|}\{j_\ell\})\nonumber\\
&&=\sum_{\ell=1}^{|A|}P_{\rm suc}^{\rm all}(\{j_\ell\})-\sum_{\ell_1 <\ell_2}P_{\rm suc}^{\rm all}(\{j_{\ell_1},j_{\ell_2}\})+ \ldots
\nonumber\\
&&\quad \quad+(-1)^{|A|-1}P_{\rm suc}^{\rm all}(A).
\eear
Thus, $P_{\rm suc}(A)$ can be easily computed by using \req{erase1122} and \req{erase1127}.

Let $A({\bfc})=\{j: c_j=1\}$ be the set of receivers reached by a tagged packet when the channel (seen by this tagged packet) is in state $\bfc$.
Since the channel is in state $\bfc$ with probability $P_\bfc$, the probability that a tagged packet is successfully received is
\beq{erase1155}
P_{\rm suc}=\sum_{\bfc} P_{\bfc} \cdot P_{\rm suc}(A({\bfc})).
\eeq
}


For the case with two receivers, i.e.,
$J=2$, we can further derive a closed-form expression for the success probability $P_{\rm suc}$.
For this, we denote by
$P_{11}$ the probability that the tagged packet reaches both receivers,
$P_{10}$ the probability that the tagged packet reaches receiver 1,
$P_{01}$ the probability that the tagged packet reaches receiver 2, and
$P_{00}$ the probability that the tagged packet does not reach any one of the two receivers.

From \req{erase1155}, we have
\bear{erase3344}
&&P_{\rm suc}
=(P_{11}+P_{10}) e^{-\rhog(P_{11}+P_{10})}\nonumber \\
&&\quad+(P_{11}+P_{01}) e^{-\rhog(P_{11}+P_{01})} -P_{11} e^{-\rhog  (1-P_{00})}.
\eear
As in \req{aloha1111} and \req{aloha1122}, the throughput for such a system subject to the offered load $\rho$ is then
\bear{erase3355}
S=\rhog\cdot P_{\rm suc},
\eear
with $P_{\rm suc}$ in \req{erase3344}.
In particular, for the independent on-off fading channel with the erase probability $\epsilon$, we have $P_{10}=P_{01}=(1-\epsilon)\epsilon$
and $P_{11}=(1-\epsilon)^2$. Using these in \req{erase3355}
recovers the throughput formula for $J=2$ in (6) of \cite{munari2013throughput}.

\bsubsec{SA in the correlated on-off fading channel model with two cooperative receivers with spatial SIC}{twoSIC}

In our analysis in the previous section, we assume that the two receivers cannot exchange information.
In this section, we show how one can further improve the throughput in the SA system  by allowing the two receivers to exchange information. Our approach is to use the well-known SIC technique \cite{casini2007contention,liva2011graph}.
To illustrate the SIC technique, let us consider the scenario shown in  \rfig{spatialk2}. Suppose that at some time $t$, there are two active users, user 1 and user 2. The packet transmitted by user 1 reaches both receivers and the packet transmitted by user 2 only reaches receiver 2.
In such a scenario, the packet transmitted by user 1 is successfully received by receiver 1. Then receiver 1 can send that packet to receiver 2
so that receiver 2 can remove that packet from its received ``signal.'' By doing so, there is only one packet left at receiver 2 and
the packet transmitted by user 2 is thus successfully received by receiver 2. In this scenario, both packets can be successfully received by using SIC, and that improves the system throughput. As SIC is done in the receiver domain, we will call it {\em spatial SIC}.

\begin{figure}[ht]
	\centering
	\includegraphics[width=0.3\textwidth]{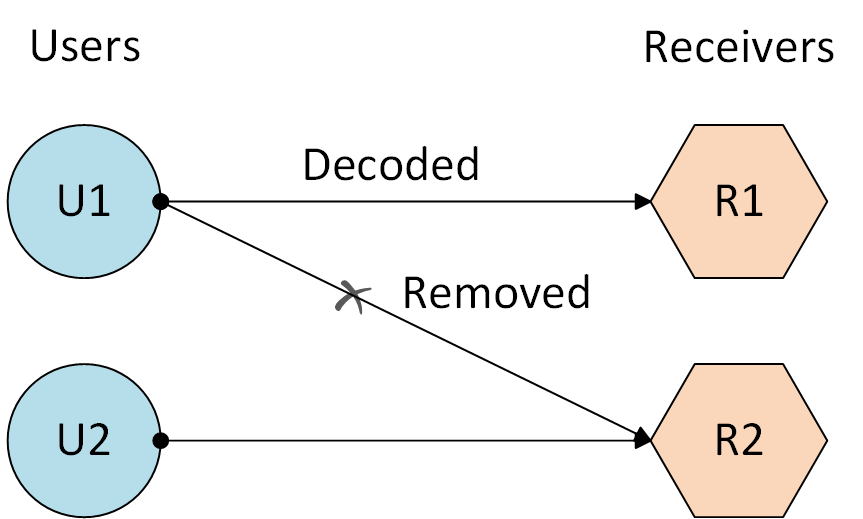}
	\caption{An illustrative example of spatial SIC.}
	\label{fig:spatialk2}
\end{figure}

Now we show the  probability that a packet is
successfully received by at least one of the two receivers in the setting with spatial SIC.
Note that the scenario shown in \rfig{spatialk2} is exactly one of the two scenarios that the success probability can be increased in the setting with spatial SIC.
The other scenario is the exact opposite when the packet transmitted by user 2 only reaches receiver 1 and the packet transmitted by user 1 reaches both receivers.
Consider the scenario in \rfig{spatialk2} and call the packet transmitted by user 2 the tagged packet.  The probability that the tagged packet only reaches receiver 2 is $P_{01}$ and the probability that the tagged packet ``sees'' another packet that is successfully received by the two receivers is
$\rho P_{11} e^{-\rho(1-P_{00})}$. Thus, the success probability of a packet is increased by
$ P_{01}\rho P_{11} e^{-\rho(1-P_{00})}$. Similarly, for the other scenario,
 the success probability of a packet is increased by $ P_{10}\rho P_{11}\ e^{-\rho(1-P_{00})}$.
In conjunction with \req{erase3344},
 the probability that a packet is successfully received in two cooperative receivers with spatial SIC is
\bear{erase3377}
&&P_{\rm suc}
=(P_{11}+P_{10}) e^{-\rhog(P_{11}+P_{10})}\nonumber \\
&&\;+(P_{11}+P_{01}) e^{-\rhog(P_{11}+P_{01})} -P_{11} e^{-\rho (1-P_{00})} \nonumber \\
&&\;+ P_{01} \rho P_{11} e^{-\rho(1-P_{00})}  +P_{10}\rho P_{11} e^{-\rho(1-P_{00})}.
\eear
Once again, the throughput $S$ can be computed by $\rho P_{\rm suc}$ with
$P_{\rm suc}$ in \req{erase3377}.

\bsec{Poisson receivers}{poisson}

\bsubsec{Definitions and examples of Poisson receivers}{pr}

In addition to the spatial diversity gain from multiple receivers, another approach is to exploit the temporal diversity gain as in the CSA systems (see, e.g., \cite{casini2007contention,liva2011graph,narayanan2012iterative,paolini2012random,jakovetic2015cooperative,stefanovic2018coded}).
To analyze CSA systems, a common approach is to use the and-or tree evaluation method in \cite{luby1998analysis,liva2011graph}.
Such an approach works well for a CSA system with SIC in a single receiver.
However, as pointed out in \cite{stefanovic2018coded}, such an approach cannot be directly applied for CSA systems with multiple cooperative receivers as we now have both spatial and temporal SIC.
Our approach to tackling such a problem is to hide the complexity of spatial SIC by treating it as a receiver (decoder) with a certain success probability that is a function of the {\em average} incoming degree of the decoder (the {\em average} number of incoming packets).
For this, we propose a new concept of an abstract receiver, called a {\em Poisson receiver}.

\bdefin{Poisson}{(\bf Poisson receiver)}
An abstract receiver (see \rfig{poissonrec}) is called a {\em $P_{\rm suc}(\rhog)$-Poisson receiver} { if
the number of packets arriving at the receiver follows a Poisson distribution with mean $\rho$ (Poisson offered load $\rho$), a tagged (randomly selected) packet is successfully received with probability $P_{\rm suc}(\rhog)$.}
Moreover, a Poisson receiver is called {\em normal} if the success probability function $P_{\rm suc}(\rhog)$ is decreasing in $\rho$.
\edefin

\begin{figure}
	\centering
	\includegraphics[width=0.35\textwidth]{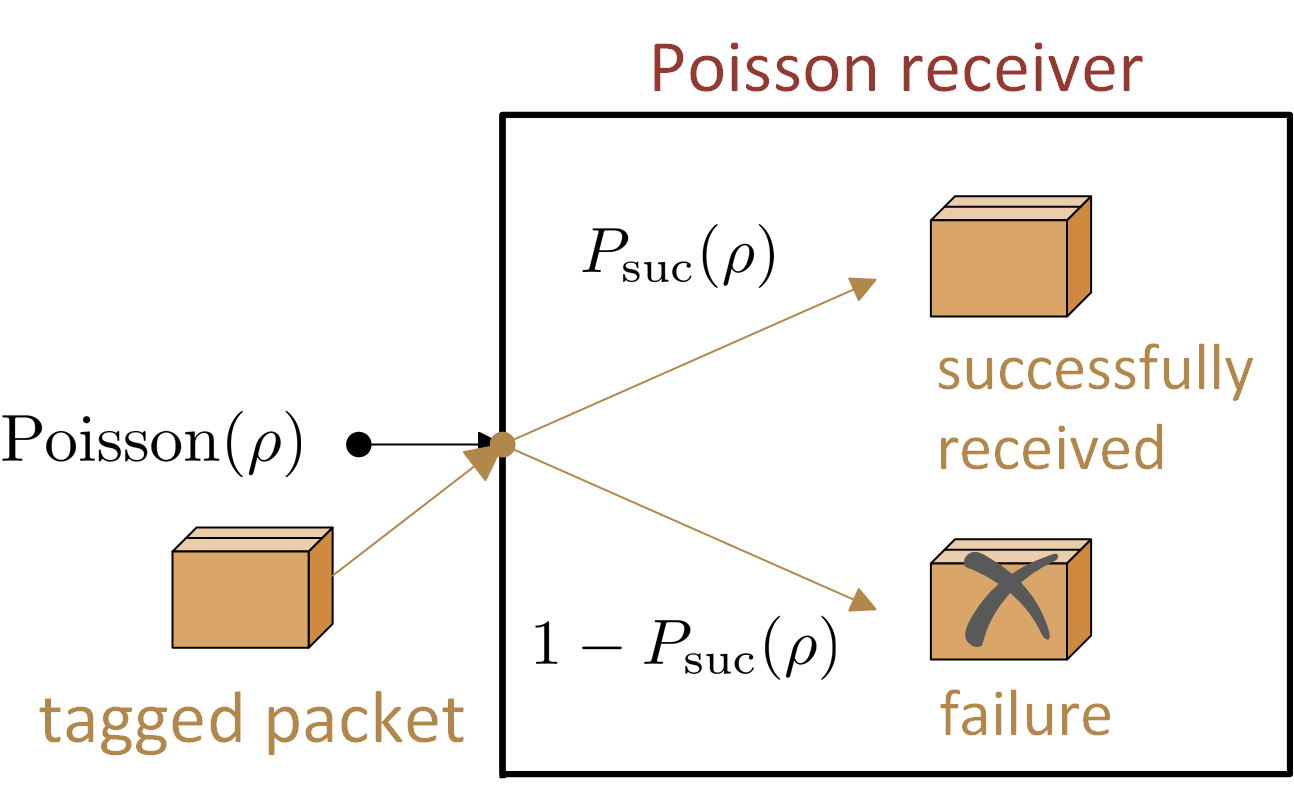}
	\caption{A Poisson receiver.}
	\label{fig:poissonrec}
\end{figure}

The throughput (defined  as the expected number of packets that are successfully received) for a {\em $P_{\rm suc}(\rhog)$-Poisson receiver} subject to a Poisson offered load $\rho$ is thus
\beq{Poithr}
S=\rho \cdot P_{\rm suc}(\rhog).
\eeq

Clearly, as shown in \req{aloha1122}, the SA system with a single receiver is a $P_{\rm suc}(\rhog)$-Poisson receiver with $P_{\rm suc}(\rho)=e^{-\rhog}$.
Moreover,  the SA system
with two non-cooperative receivers  is a $P_{\rm suc}(\rhog)$-Poisson receiver with
$P_{\rm suc}(\rhog)$ in \req{erase3344}.
Similarly,  the SA system
with two cooperative receivers  is a $P_{\rm suc}(\rhog)$-Poisson receiver with
$P_{\rm suc}(\rhog)$ in \req{erase3377}.

\bex{TfoldALOHA}{(T-fold ALOHA)}
$T$-fold ALOHA proposed in \cite{ordentlich2017low} is a generalization of the SA system.
If there are less than or equal to  $T$ packets transmitted in a time slot, then all these packets
can be successfully decoded. On the other hand, if there are more than $T$ packets
transmitted in a time slot, then all these packets are lost. Clearly, the SA system corresponds to
the case that $T=1$. The throughput of the $T$-fold ALOHA subject to a Poisson offered load $\rho$ is
$$S=\sum_{t=0}^T t \cdot \frac{e^{-\rho} \rho^t}{t!}=\rho \sum_{t=0}^{T-1} \frac{e^{-\rho} \rho^{t}}{t!}.$$
As such, $T$-fold ALOHA is a Poisson receiver with
\beq{tfold1111}
P_{\rm suc}(\rhog)=\sum_{t=0}^{T-1} \frac{e^{-\rho} \rho^{t}}{t!}.
\eeq
\eex

There are other well-known channel models in the literature that can also be modelled as a Poisson receiver, including
the Rayleigh block fading channel with capture \cite{clazzer2017irregular}.


\bsubsec{Coded Poisson receivers}{cpr}

{ In this section, we show how to use Poisson receivers with packet coding to construct another Poisson receiver.
Our approach is based on the tree evaluation method in \cite{luby1998analysis,luby1998analysisb,richardson2001capacity,liva2011graph}.
Such a construction is feasible due to two important closure properties of Poisson random variables: (i) the excess degree distribution (defined as the degree distribution along a randomly selected edge) of a Poisson degree distribution is still a Poisson degree distribution, and (ii)  random thinning (that removes each edge independently with a certain probability) of a Poisson degree distribution is still a Poisson
degree distribution.}

Analogous to IRSA in \cite{liva2011graph},
let us consider a system with $N$ active users and $T$ (independent) Poisson receivers with the success probability function $P_{\rm suc}(\rho)$. Each user transmits its packet  for a random  number of times (copies).
Let $L$ be the random variable that represents the number of copies of a packet. Each of the $L$ copies is transmitted to one of the $T$ Poisson receivers that is chosen {\em uniformly} and {\em independently}. As in CRDSA, if any one of these $L$ copies of a packet is successfully received by a Poisson receiver, then the other copies can be removed (cancelled) from the system to further reduce the system load. Such a process can  then be repeatedly carried out to decode the rest of the packets.
We call such a system a system of coded Poisson receivers (CPR). Similar to the throughput analysis for IRSA in \cite{liva2011graph},
{ our analysis is based on the tree evaluation method in \cite{luby1998analysis,luby1998analysisb,richardson2001capacity}.}
A realization of a CPR can be represented by
 a bipartite graph with the $N$ active users on one side (user nodes) and the $T$ Poisson receivers on the other side (receiver nodes).
  A link between a user node and a receiver node in the bipartite graph represents a packet transmission from that user node to that receiver node.

For our throughput analysis, we
let $\Lambda_\ell$ be the probability that a packet is transmitted $\ell$ times, i.e.,
\begin{equation}
P(L=\ell)=\Lambda_{\ell}, \;\ell=1,2\dots
\end{equation}
The sequence $\{\Lambda_\ell, \ell \ge 1\}$ is called the {\em degree distribution} of a user node.
Define the generating function
\beq{mean0000}
\Lambda(x)=\sum_{\ell=0}^\infty \Lambda_\ell \cdot x^\ell
\eeq
 of the
degree distribution of a user node.
Clearly,  the
mean  degree of a user node can be represented as follows:
\beq{mean1111}
\Lambda^{\prime}(1)=\sum_{\ell=0}^\infty \ell\cdot \Lambda_{\ell}.
\end{equation}
Let
\beq{mean2222}
\lambda_\ell=\frac{\Lambda_{\ell+1}\cdot (\ell+1)}{\sum_{\ell=0}^\infty\Lambda_{\ell+1}\cdot (\ell+1)}
\eeq
be the probability that the user end of a randomly selected edge has additional $\ell$ edges excluding the randomly selected edge.
Such a probability is called the {\em excess degree distribution} of a user node in the literature (see, e.g., the book \cite{Newman2010}).
Also, let
\beq{mean3333}
\lambda(x)=\sum_{\ell=0}^\infty \lambda_\ell \cdot x^\ell
\eeq
be the corresponding generating function. It is easy to see that these two generating functions are related as follows:
\beq{mean3344}
\lambda(x)=\frac{\Lambda^\prime(x)}{\Lambda^{\prime}(1)}.
\eeq
The
offered load to a Poisson receiver, defined as the expected number of packets transmitted to that receiver, is
\beq{mean4444}
\rho=\frac{N}{T}\cdot\Lambda^{\prime}(1)=G \Lambda^\prime(1),
\eeq
where
\beq{load1111}
G=\frac{N}{T}
\eeq
 is called the {\em normalized} offered load.
When $N$ is large, the number of packets transmitted to a receiver can be assumed to be
a Poisson random variable with mean $\rho$ (as a sum of $N$ independent Bernoulli random variables with mean $\Lambda^\prime(1)/T$ approaches to a Poisson random variable with mean $\rho$).  As such, we can assume  the degree distribution of a receiver node is a Poisson distribution with  mean $\rho$.
It is well known (see, e.g.,  \cite{Newman2010}) that the excess degree distribution of a Poisson degree distribution is also Poisson with the same mean.
Thus, the probability that the receiver end of a randomly selected edge has additional $\ell$ edges excluding the randomly selected edge
is
\beq{mean3377}
\frac{e^{-\rho} \rho^\ell}{\ell!}.
\eeq

Consider a randomly selected edge from the bipartite graph.
Analogous to the tree evaluation in \cite{luby1998analysis,luby1998analysisb,richardson2001capacity,liva2011graph},
let $p_i$ (resp $q_i$) be the probability that  the receiver (resp. user)  end of a randomly selected edge has not been successfully received after the $i^{th}$ SIC iteration.
Since the receiver end of an edge corresponds to a transmission of a tagged packet from a user to a $P_{\rm suc}(\rho)$-Poisson receiver, we have
\beq{tag1111}
p_1=1-P_{\rm suc}(\rho).
\eeq
Recall that a packet sent from a user (the user end of the bipartite graph) can be successfully received if at least one of its copies is successfully received at the {\em receiver} end.
Since the probability that the user end of a randomly selected edge  has additional $\ell$ edges is $\lambda_\ell$, 
the probability that the {\em user} end of a randomly selected edge  cannot be successfully received after the first iteration is
thus
\beq{tag2222}
q_1=1-\sum_{\ell=0}^\infty \lambda_\ell \cdot \Big (1-p_1^{\ell} \Big).
\eeq
This then leads to
\beq{tag3333}
q_1=\lambda(p_1)=\lambda(1-P_{\rm suc}(\rho)).
\eeq
To compute $p_2$, note that the excess degree distribution of the receiver end is Poisson with mean $\rho$  (under the tree assumption in \cite{luby1998analysis,liva2011graph}). With probability
$1-q_1$, an (excess) edge at the receiver end of a randomly selected edge is removed by SIC  after the first iteration.
Since random thinning of a Poisson random variable is still a Poisson random variable, the number of the remaining (excess)  edges at the receiver end of a randomly selected edge at the second iteration is Poisson with mean $q_1 \rho$.
Thus, the offered load at the second iteration is effectively reduced from $\rho$ to $q_1 \rho$.
This leads to
\beq{tag4444}
p_2=1-P_{\rm suc}(q_1\rho),
\eeq
and
\beq{tag5555}
q_2=\lambda(p_2)=\lambda(1-P_{\rm suc}(q_1\rho)).
\eeq
In general, we have the following recursive equations:
\bear{tag6666c}
p_{i+1}&=&1-P_{\rm suc}(q_i\rho), \label{eq:tag6666a}\\
q_{i+1}&=&\lambda(1-P_{\rm suc}(q_i\rho)).\label{eq:tag6666}
\eear
Moreover, if $P_{\rm suc}(\rho)$ is decreasing in $\rho$, then  $q_{i+1} \le q_i$ and there is a limit $0 \le q \le 1$ if we start from $q_0=1$.

To illustrate our tree analysis, let us consider the bipartite graph in  \rfig{bipart} with user nodes on the left and slot nodes on the right. In this bipartite graph, each user transmits exactly twice, i.e., $\Lambda(x)=x^2$ and $\lambda(x)=x$. Suppose we randomly select an edge in the bipartite graph, say edge $e_1$. We are interested in finding out the probability $q_2$ that the user end  of that edge, i.e., user node 3 (U3), has not been successfully received after the second SIC iteration. For this, we enumerate the set of nodes and edges from user node 3 in two iterations and that results in the tree shown in \rfig{tree}. As each user transmits exactly twice, the probability $q_2$ is the same as the probability that the receiver end of edge $e_4$, i.e., receiver node 4 (T4), has not been successfully ``decoded'' after the second iteration, which is exactly $p_2$ (from the symmetry of users). The excess degree of receiver node 4 is three. Thus, there are three incoming packets with each packet being removed with probability $1-q_1$.
This corresponds to a Poisson receiver where the offered load is reduced from $\rho$ by $q_1\rho$.
Finally, we have $q_1=p_1$ by using the same argument for $q_2=p_2$.

\begin{figure}[ht]
	\centering
\includegraphics[width=0.4\textwidth]{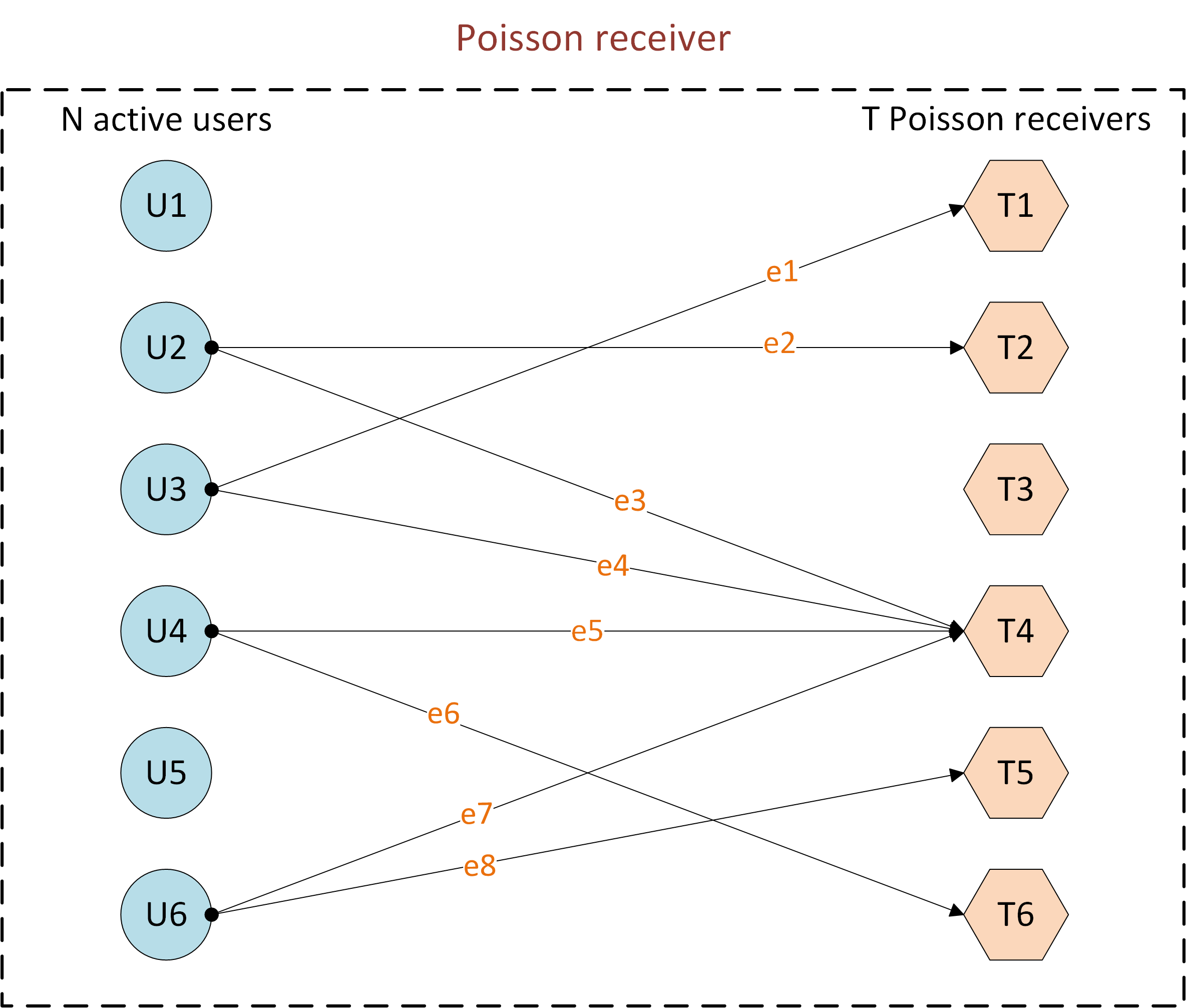}
		\caption{The bipartite graph representation for a system of CPRs.}
		\label{fig:bipart}
\end{figure}

\begin{figure}[ht]
		\centering
\includegraphics[width=0.35\textwidth]{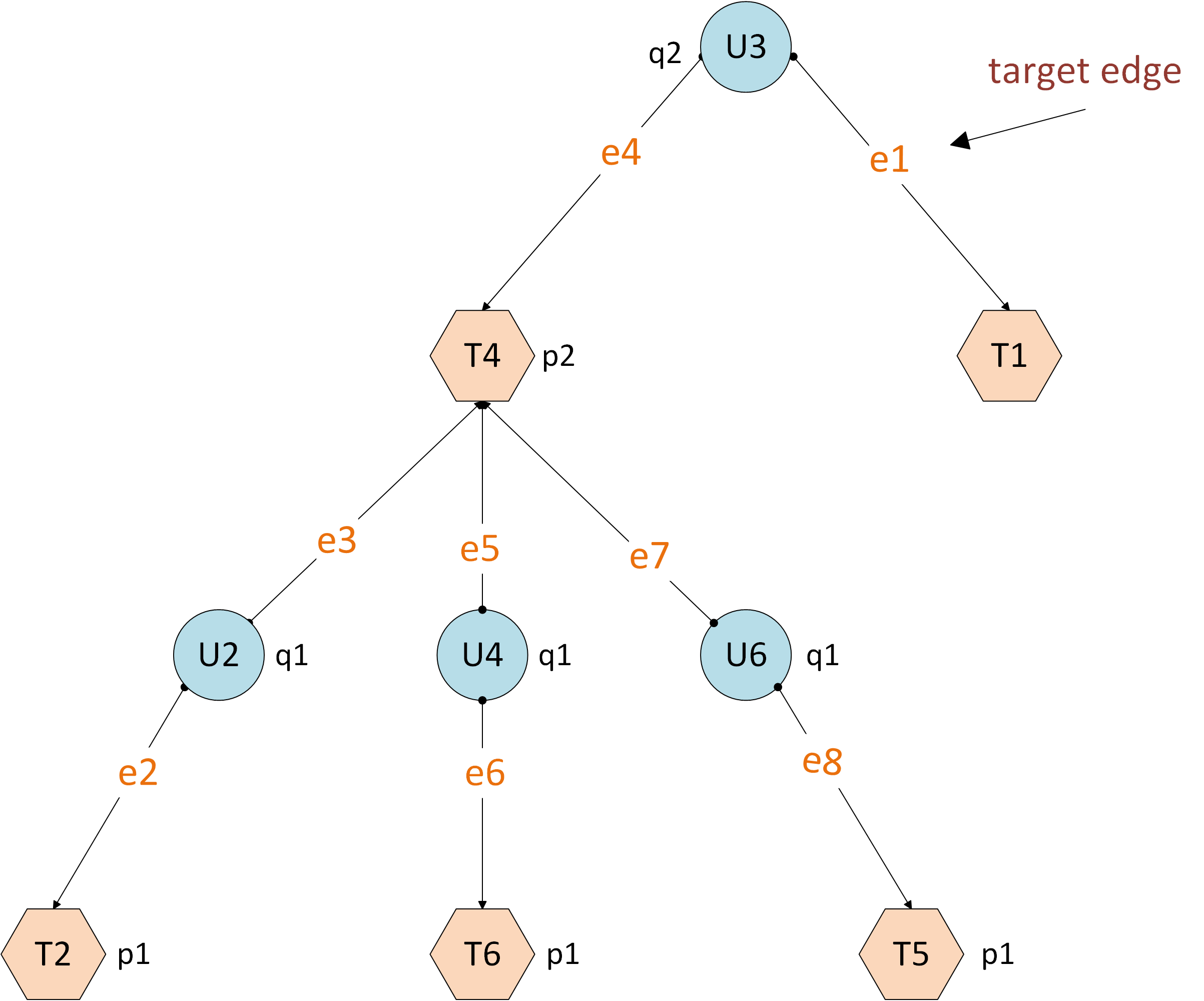}
		\caption{Enumeration of the tree from a specific user node.}
		\label{fig:tree}
	\end{figure}


Now we derive the success probability of a user.
{ Let $\tilde P_{\rm suc}^{(i)}(G)$
be  the success probability for the CPR system after the $i^{th}$ SIC iteration when the system is subject to a (normalized) Poisson offered load $G$.}
 Once again, note that a packet sent from a user can be successfully received if at least one of its copies is successfully received at the {\em receiver} end. Since the probability that a randomly selected {\em user} has  $\ell$ edges is $\Lambda_\ell$, the probability that a packet sent from a randomly selected {\em user} can be successfully received after the $i^{th}$ iteration is
\bear{mean5555d}
&&\sum_{\ell=0}^\infty \Lambda_\ell \cdot \Big (1-p_i^{\ell} \Big) =\sum_{\ell=0}^\infty \Lambda_\ell \cdot \Big (1-(1-P_{\rm suc}(q_{i-1}\rho))^{\ell} \Big) \nonumber \\
&&=1-\Lambda\Big (1-P_{\rm suc}(q_{i-1}\rho)\Big).
\eear
Then it follows from \req{mean5555d} and \req{mean4444} that
\beq{mean8888thu}
 \tilde P_{\rm suc}^{(i)}(G)=1-\Lambda\Big (1-P_{\rm suc}(q_{i-1} G\Lambda^\prime(1))\Big).
 \eeq
 Clearly, if $P_{\rm suc}(\rho)$ is decreasing in $\rho$, then $\tilde P_{\rm suc}^{(i)}(G)$ is also decreasing in $G$.
 One interesting interpretation of the CPR system constructed from $T$ independent normal Poisson receivers is that it is also a normal Poisson receiver with another success probability $\tilde P_{\rm suc}^{(i)}(G)$ (by viewing the number of users $N$ as a Poisson random variable  with mean $GT$).
Such an interpretation shows that CRDSA \cite{casini2007contention}, IRSA \cite{liva2011graph} and other CSA systems are in fact Poisson receivers with certain success probability functions.

For instance,
in the CSA system with a single receiver, we have from \req{aloha1122} that $P_{\rm suc}(q_i\rho)=e^{-q_i\rho}$.
Then \req{tag6666} implies that
\beq{mean8899}
q_{i+1}=\lambda(1-e^{-q_i \rho}),
\eeq
which is exactly the recursive equation for CSA in (15) of \cite{stefanovic2018coded}.
By using Poisson receivers as building blocks, one can also extend our analysis to convolutional (or hierarchical) CPR
like the convolutional CSA in \cite{Liva2012spatially}.

\bsec{Poisson receivers with multiple classes of input traffic}{poissonmul}


In the following,
we extend Poisson receivers to the setting with multiple classes of input traffic.
One of the main advantages for this is that we can provide differentiated services (Diffserv) for different classes of traffic.
{
We say a system with $K$ classes of input traffic is subject to a Poisson offered load  $\rho=(\rho_1, \rho_2, \ldots, \rho_K)$ if these $K$ classes of input traffic are {\em independent}, and the number of class $k$ packets arriving at the system
follows a Poisson distribution with mean $\rho_k$, for $k=1,2, \ldots, K$.}

\bdefin{Poissonmul}{(\bf Poisson receiver with multiple classes of input traffic)}
An abstract receiver (see \rfig{Poissonmul}) is called a {\em $(P_{{\rm suc},1}(\rhog), P_{{\rm suc},2}(\rhog), \ldots, P_{{\rm suc},K}(\rhog))$-Poisson receiver} with $K$ classes of input traffic { if the receiver is subject to a Poisson offered load  $\rho=(\rho_1, \rho_2, \ldots, \rho_K)$, a tagged (randomly selected) class $k$  packet
is successfully received with probability $P_{{\rm suc},k}(\rhog)$, for $k=1,2, \ldots, K$.}
Moreover, a Poisson receiver is called {\em normal} if the success probability function of class $k$ packets $P_{{\rm suc},k}(\rhog)$ is decreasing in $\rho$ for all $k$.
\edefin


\begin{figure}[ht]
	\centering
	\includegraphics[width=0.47\textwidth]{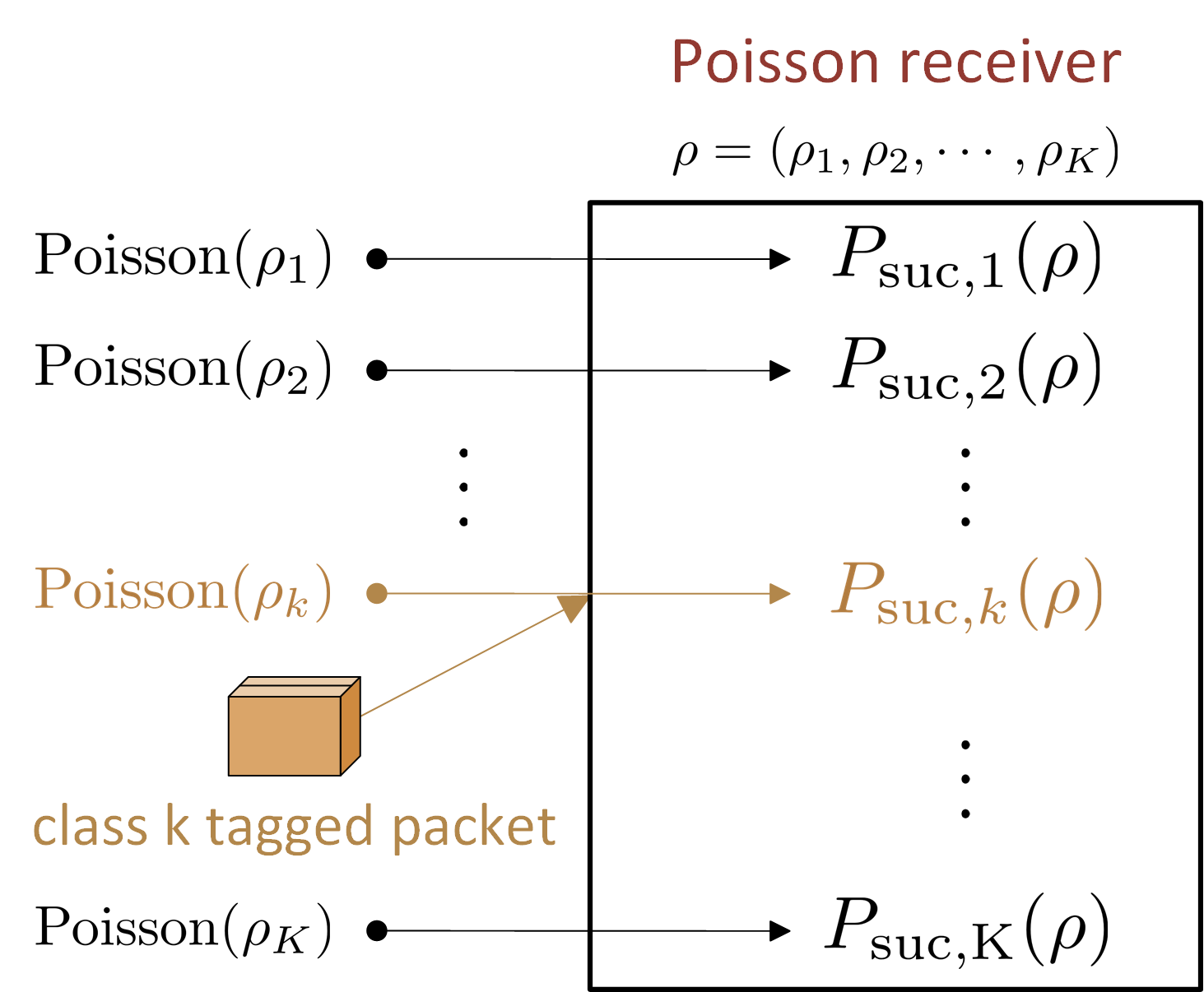}
	\caption{A Poisson receiver with multiple classes of input traffic.}
	\label{fig:Poissonmul}
\end{figure}

The throughput of class $k$ packets  (defined  as the expected number of class $k$ packets that are successfully received) for a {\em $(P_{{\rm suc},1}(\rhog), P_{{\rm suc},2}(\rhog), \ldots, P_{{\rm suc},K}(\rhog))$-Poisson receiver} subject to a Poisson offered load $\rho$ is thus
\beq{Poithrmul}
S_k=\rho_k \cdot P_{{\rm suc},k}(\rhog),
\eeq
$k=1,2, \ldots, K$.

{ One can view a Poisson receiver as a loss system where the loss probability of a randomly selected class $k$ arrival is  $1-P_{{\rm suc},k}(\rhog)$ when the system is subject to a Poisson offered load $\rho$. Such an interpretation from queueing theory allows us to analyze a network of loss systems interconnected by routers in \rsec{routing} and coders in \rsec{cprmul}. The key thing is to maintain {\em independence} among various classes of input traffic, as described in \cite{chang1995sample} for intree networks.}

\bex{tworeceivers}{(Two non-cooperative receivers)}
Consider the SA system
with two non-cooperative receivers in a correlated on-off fading channel in \rsec{onoff}.
For such a system, we show that it can be viewed as a Poisson receiver with three classes of input traffic.
Class 1 packets with a Poisson load $\rho_1=\rho P_{01}$  are sent to receiver 1, class 2 packets with a Poisson load $\rho_2=\rho P_{10}$ are sent to receiver 2, and class 3 packets with a Poisson load $\rho_3=\rho P_{11}$ are sent to both receivers.
In view of \req{erase3344},
one can see that the success probability function for class 1 packets is
$$P_{{\rm suc},1}(\rho_1,\rho_2, \rho_3)=e^{-(\rho_1+\rho_3)}.$$
Similarly,
the success probability function for class 2 packets is
$$P_{{\rm suc},2}(\rho_1,\rho_2, \rho_3)=e^{-(\rho_2+\rho_3)}.$$
Finally, the success probability function for class 3 packets is
$$P_{{\rm suc},3}(\rho_1,\rho_2, \rho_3)=e^{-(\rho_1+\rho_3)}+e^{-(\rho_2+\rho_3)}-e^{-(\rho_1+\rho_2+\rho_3)}.$$
\eex

\bex{tworeceiversb}{(Two cooperative receivers)}
Consider the SA system
with two cooperative receivers in a correlated on-off fading channel in \rsec{twoSIC}.
For such a system, we show that it can also be viewed as a Poisson receiver with three classes of input traffic.
Class 1 packets with a Poisson load $\rho_1=\rho P_{01}$  are sent to receiver 1, class 2 packets with a Poisson load $\rho_2=\rho P_{10}$ are sent to receiver 2, and class 3 packets with a Poisson load $\rho_3=\rho P_{11}$ are sent to both receivers.
In view of \req{erase3377},
one can see that the success probability function for class 1 packets is
\beq{invm1111}
P_{{\rm suc},1}(\rho_1,\rho_2, \rho_3)=e^{-(\rho_1+\rho_3)}+\rho_3e^{-(\rho_1+\rho_2+\rho_3)}.
\eeq
Similarly,
the success probability function for class 2 packets is
\beq{invm2222}
P_{{\rm suc},2}(\rho_1,\rho_2, \rho_3)=e^{-(\rho_2+\rho_3)}+\rho_3e^{-(\rho_1+\rho_2+\rho_3)}.
\eeq
Finally, the success probability function for class 3 packets is
\beq{invm3333}
P_{{\rm suc},3}(\rho_1,\rho_2, \rho_3)=e^{-(\rho_1+\rho_3)}+e^{-(\rho_2+\rho_3)}-e^{-(\rho_1+\rho_2+\rho_3)}.
\eeq
\eex

\bsubsec{SA with multiple cooperative receivers}{SAmul}

In this section, we extend the result for the SA system
with two cooperative receivers in \rex{tworeceiversb} to the SA system
with multiple cooperative receivers. These cooperative receivers are assumed to be capable of performing spatial SIC.
For such a system, we show that it can also be viewed as a Poisson receiver with success probability functions that can be computed from a set of bipartite graphs.

Consider the SA system
with $T$ cooperative receivers and $K$ classes of users. Let $B_k$, $k=1,2, \ldots, K$, be the set of receivers associated with class $k$ users. A packet sent from a class $k$ user reaches the set of receivers $B_k$ (with probability 1). For this, we can construct a $K \times T$ bipartite graph with $K$ user nodes on the left and $T$ receiver nodes on the right. For  user node $k$, we connect an edge to each receiver node in $B_k$. Call such a graph the {\em association graph} as it represents the association between users and receivers. To illustrate the concept of the association graph, we consider
an example with four classes of users ($K=4$) and six receivers ($T=6$) in \rfig{configuration} (a), where $B_1=\{1,6\}$, $B_2=\{1,2,4,6\}$, $B_3=\{3,5\}$ and $B_4=\{2,5,6\}$.

 \begin{figure*}[tb]
    \begin{center}
    \begin{tabular}{p{0.3\textwidth}p{0.3\textwidth}p{0.3\textwidth}}
      \includegraphics[width=0.3\textwidth]{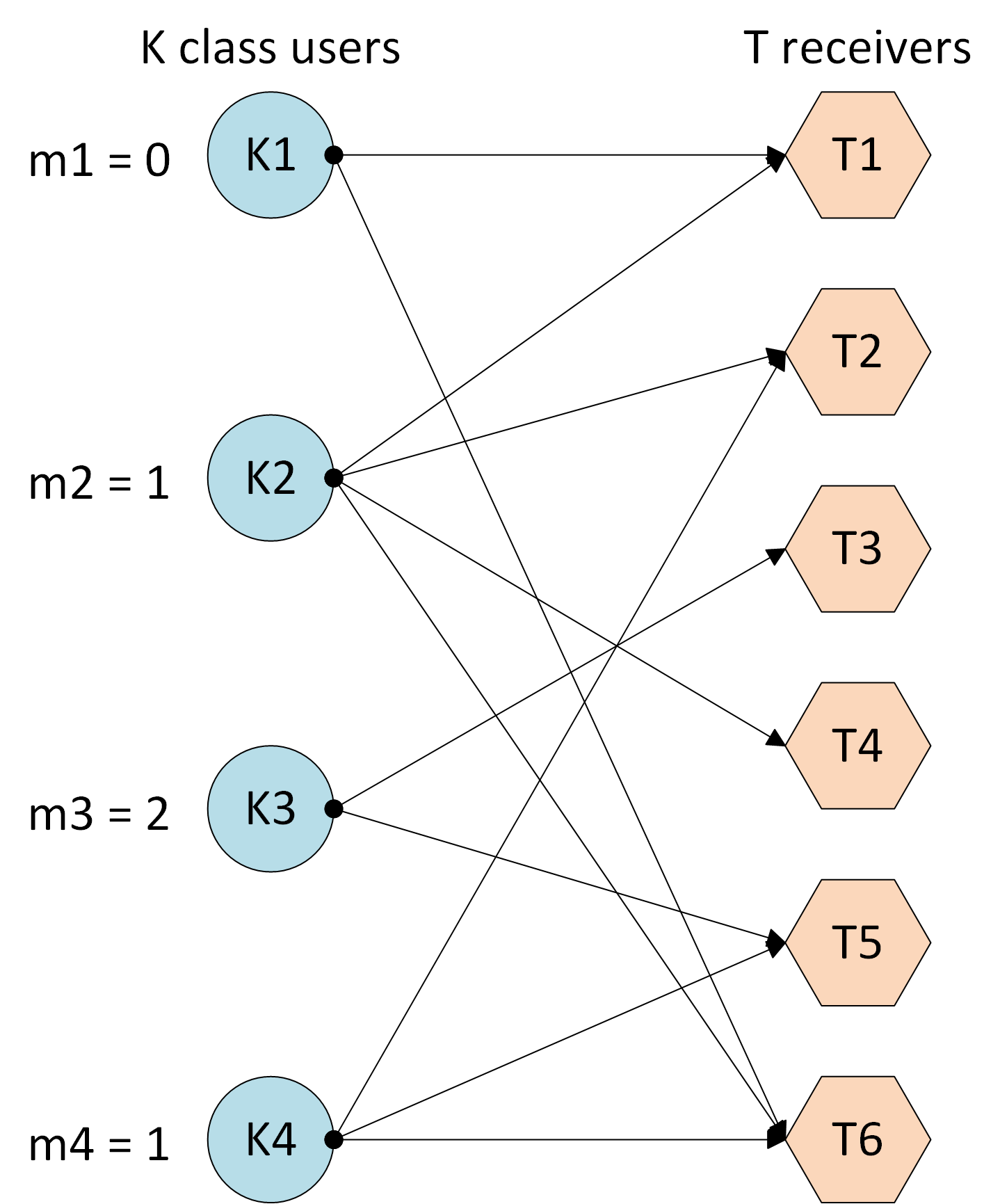} &
      \includegraphics[width=0.3\textwidth]{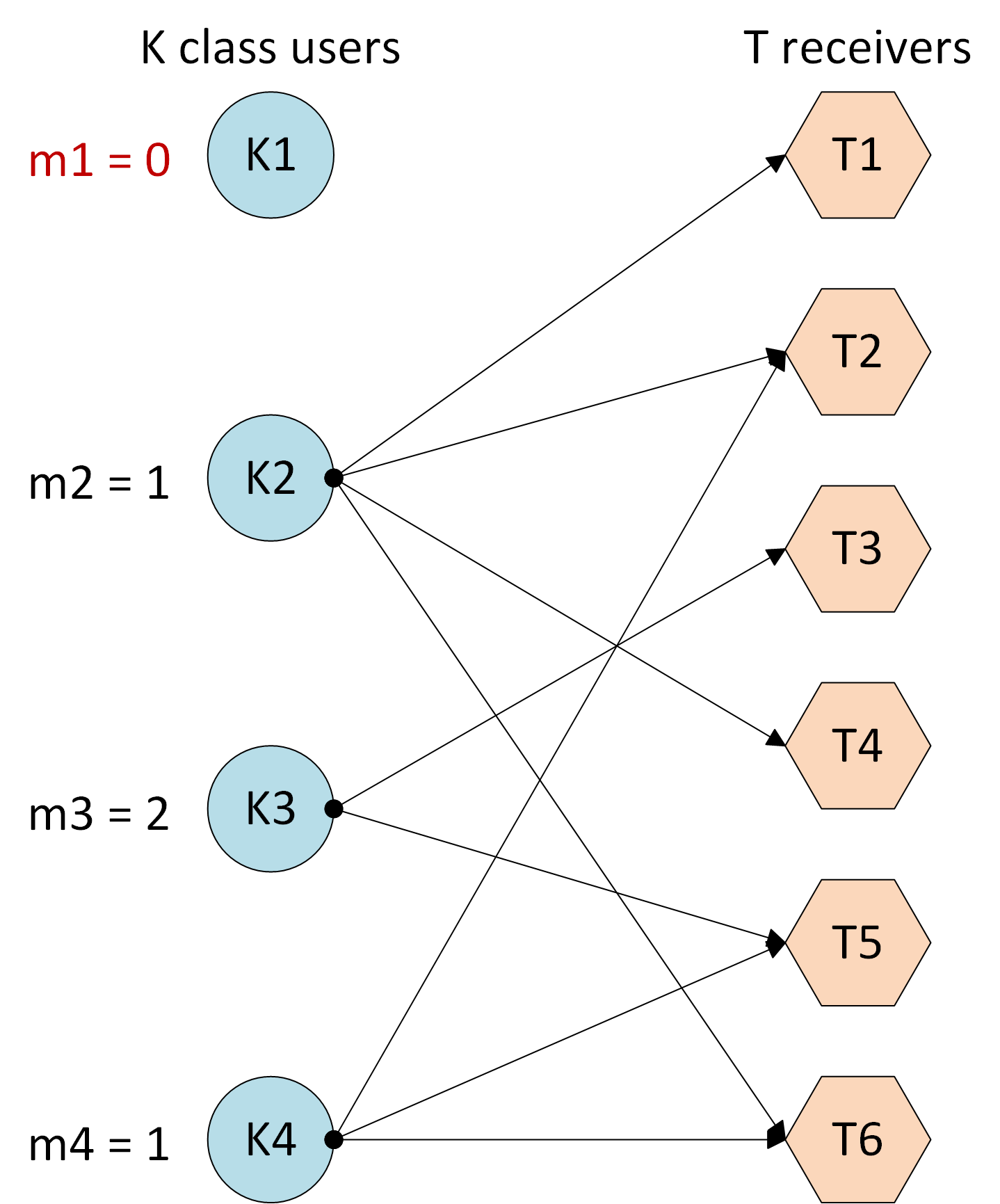} &
      \includegraphics[width=0.3\textwidth]{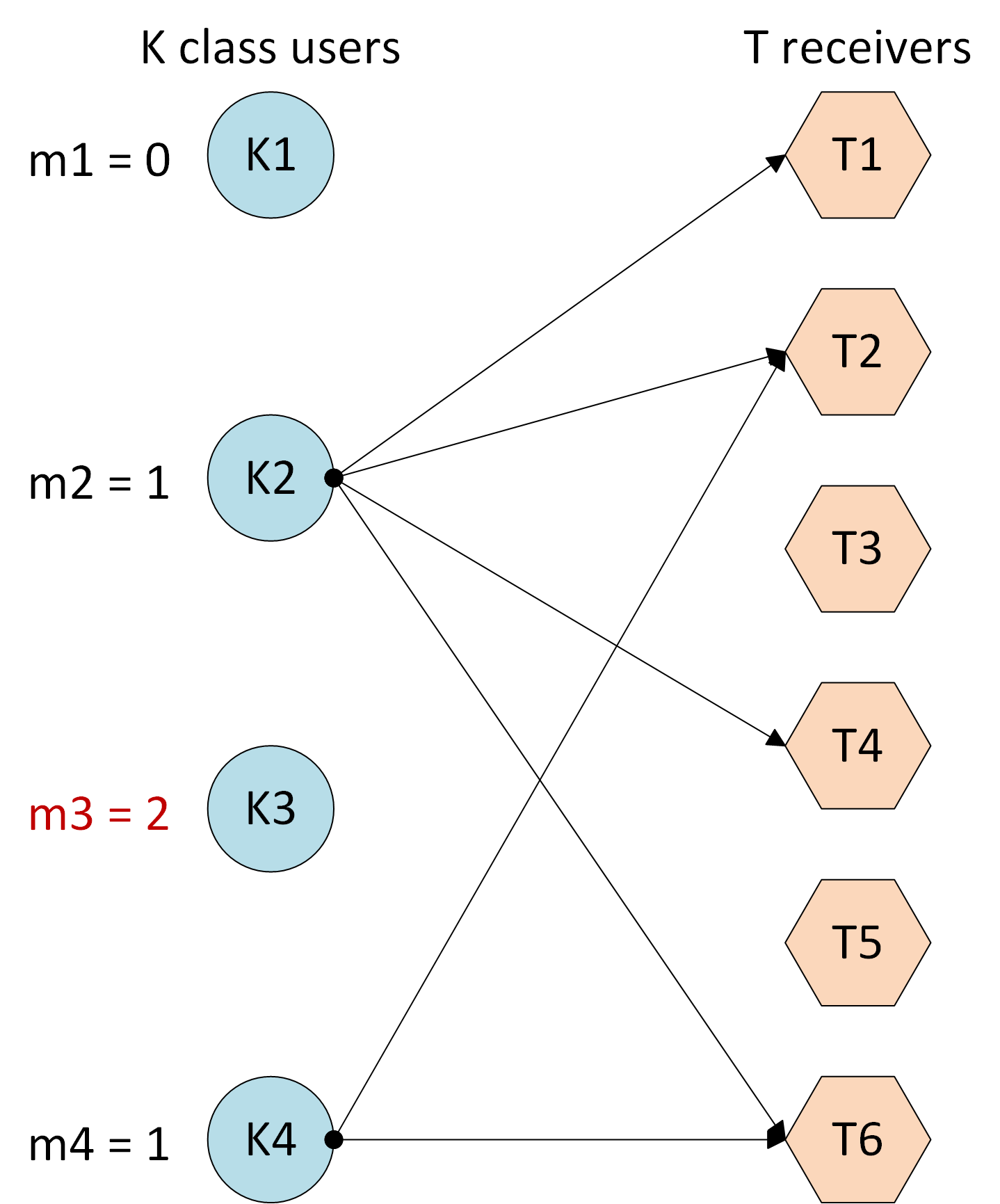}\\
      (a) the association graph with $B_1=\{1,6\}$, $B_2=\{1,2,4,6\}$, $B_3=\{3,5\}$ and $B_4=\{2,5,6\}$ & (b) removing the edges from user node 1 for $m_1=0$ & (c) removing the edges to the receiver nodes in $B_3$  for $m_3=2$ \\
         \end{tabular}
    \caption{Construct the configuration graph from the association graph for the configuration $m_1=0$, $m_2=1$, $m_3=2$ and $m_4=1$. The packet from the class 2 user and the packet from the class 4 user are successfully received by using the iterative decoding method.}
    \label{fig:configuration}
  \end{center}
\end{figure*}

Now we compute the throughput of class $k$ users subject to the (independent) Poisson offered load $\rho=(\rho_1, \rho_2, \ldots, \rho_K)$ for such a SA system.
Consider a particular time slot and construct a ``configuration'' graph according to the number of packets transmitted in that time slot.
Specifically, let $z_k$, $k=1,2, \ldots, K$, be the number of class $k$ packets transmitted in that time slot and
$m_k=\min[z_k,2]$.
For the SA system, it suffices to consider the three cases for each $k$:
$m_k = 0$, $m_k=1$ and $m_k=2$.
If $m_k=0$, then no class $k$ packets are transmitted and we can remove all the edges connected to the user node $k$ (see \rfig{configuration} (b)).
If $m_k = 2$, then there are at least two packets transmitted to the receivers in $B_k$, As such, no packets can be successfully received by the receivers in $B_k$ and we can remove all the edges connected to the receive nodes in $B_k$ in the association graph (see \rfig{configuration} (c)). If $m_k=1$, then there is exactly one packet transmitted from a class $k$ user to the receiver nodes in $B_k$ and we simply leave the edges from user node $k$ to the set of receiver nodes in $B_k$ in the association graph.
Let $m=(m_1, \ldots, m_K)$ and $F(m)$ be the configuration graph for
the configuration $n$.
As $m_k$ takes values 0,1, and 2, there are $3^K$ configuration graphs.

For the configuration graph $F(m)$, we then use the SIC technique \cite{casini2007contention,liva2011graph}
(as in \rsec{twoSIC})
to compute the number of class $k$ packets that are successfully received. Denote by $w_k(m)$ the number of class $k$ packets that are successfully received in the configuration graph $F(m)$. For the configuration graph  in \rfig{configuration} (c), we have $w_1(m)=0$, $w_2(m)=1$, $w_3(m)=0$ and $w_4(m)=1$. Then we can compute the throughput for class $k$ users as follows:
\beq{mulr2222}
S_k=\sum_{m} w_k(m) p(m),
\eeq
where $p(m)$ is the probability of the configuration $m$ subject to the (independent) Poisson offered load $\rho$.
Let
\bearn
h_0(\rho_k)&=&e^{-\rho_k},\\
h_1(\rho_k)&=&\rho_k e^{-\rho_k},\;\mbox{and}\\
h_2(\rho_k)&=&1-e^{-\rho_k}-\rho_k e^{-\rho_k}.
\eearn
For the Poisson distribution with mean $\rho_k$, the probabilities for $m_k=0$, $m_k=1$, and $m_k=2$ are
$h_0(\rho_k)$, $h_1(\rho_k)$, and $h_2(\rho_k)$, respectively.
Since the Poisson offered loads from the $K$ classes are independent, we have
\beq{mulr2255}
p(m)=\prod_{k=1}^K h_{m_k}(\rho_k).
\eeq
Using \req{Poithrmul} yields
\beq{mulr3333}
 P_{{\rm suc},k}(\rhog)=\frac{S_k}{\rho_k}=\frac{1}{\rho_k}\sum_{m} w_k(m) \prod_{k=1}^K h_{m_k}(\rho_k).
 \eeq
 When $K$ is very small, it is possible to obtain closed-form expressions for the success probability functions like those in \rex{tworeceiversb}. For a moderate $K$, one can still compute the success probability functions by using \req{mulr3333}.
 However, it becomes computationally difficult for a large $K$ as there are $3^K$ configuration graphs.
 In that setting, we have to resort to the random graph approach (that will be discussed in \rsec{cprmul}).

\bsubsec{Poisson receivers with packet routing}{routing}

In this section, we show that Poisson receivers with packet routing are still Poisson receivers.
Consider a
Poisson receiver with $K_2$ classes of input traffic and the success probability functions $P_{{\rm suc},1}(\rhog), P_{{\rm suc},2}(\rhog), \ldots, P_{{\rm suc},K_2}(\rhog)$.
There are $K_1$ classes of {\em external} input traffic to the Poisson receiver.
With probability $r_{k_1, k_2}$,
a class $k_1$ external packet transmitted to the Poisson receiver becomes a class $k_2$ packet at the Poisson receiver (see \rfig{Poissonrout}).
Such a probability is called the {\em routing} probability as in the classical queueing analysis.

\begin{figure}[ht]
	\centering
	\includegraphics[width=0.47\textwidth]{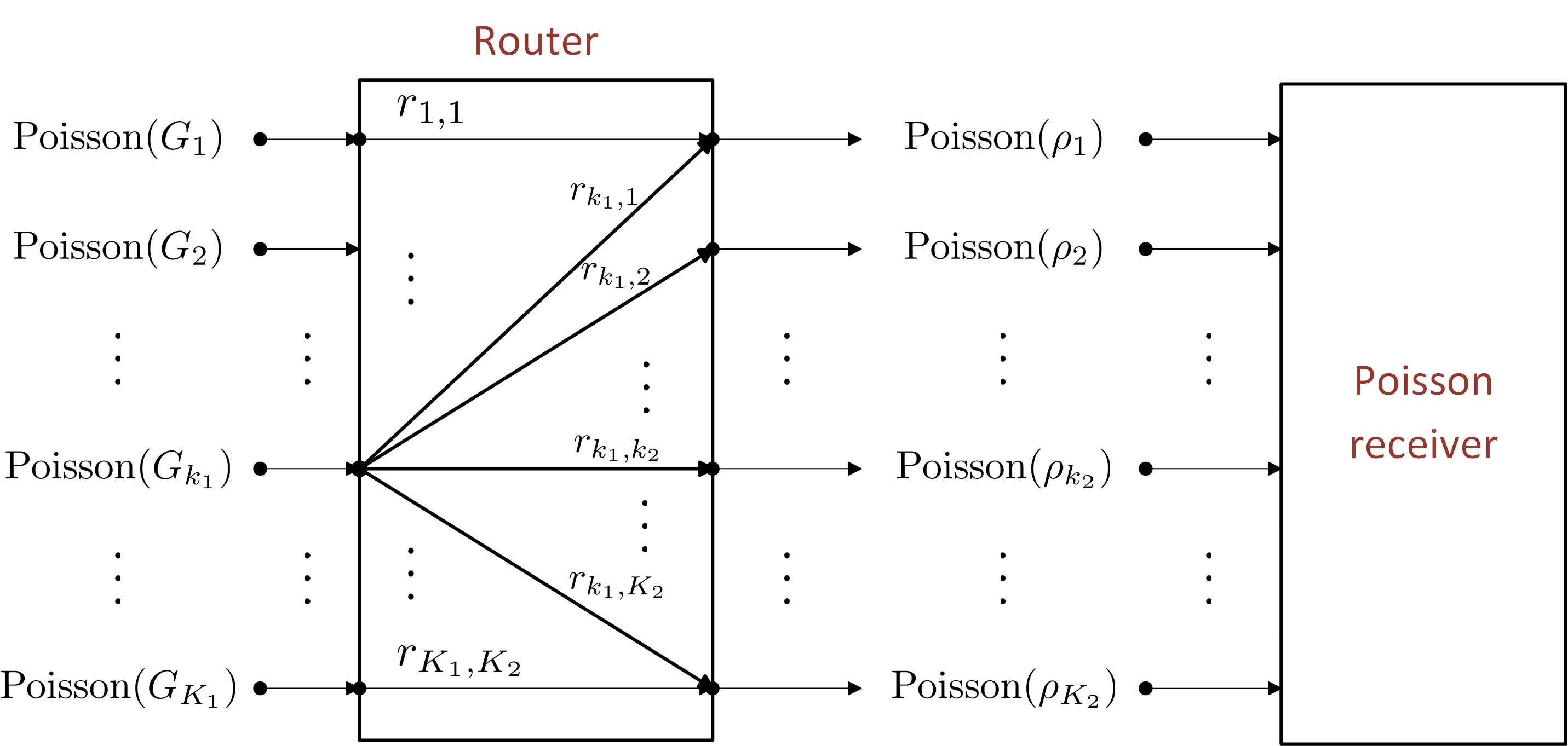}
	\caption{A Poisson receiver with packet routing.}
	\label{fig:Poissonrout}
\end{figure}

Now suppose that the external Poisson offered load for the $K_1$ classes of external input traffic is $G=(G_1, G_2, \ldots, G_{K_1})$.
Since (i) randomly splitting of Poisson random variables yields independent Poisson random variables, and (ii)  superposition of independent Poisson random variables is still a Poisson random variable, we have from the packet routing mechanism that  the offered load for the $K_2$ classes of input traffic at the Poisson receiver is
$\rho=(\rho_1,\rho_2, \ldots, \rho_{K_2})$, where
\beq{mean4444rou}
\rho_{k_2}=\sum_{k_1=1}^{K_1} G_{k_1}  r_{k_1,k_2},
\eeq
$k_2=1,2,\ldots, K_2$.
Since a tagged class $k_1$ external packet becomes a class $k_2$ packet at the Poisson receiver with probability $r_{k_1,k_2}$,
this tagged class $k_1$ packet is successfully received with probability
\beq{routing1111}
\tilde P_{{\rm suc}, k_1}(G)= \sum_{k_2=1}^{K_2} r_{k_1, k_2} P_{{\rm suc}, k_2}(\rho).
\eeq
This shows that the $(P_{{\rm suc},1}(\rhog), P_{{\rm suc},2}(\rhog), \ldots, P_{{\rm suc},K_2}(\rhog))$-Poisson receiver with the $K_1 \times K_2$ packet routing probability matrix $R=(r_{k_1, k_2})$ is a $(\tilde P_{{\rm suc},1}(G), \tilde P_{{\rm suc},2}(G), \ldots, \tilde P_{{\rm suc},K_1}(G))$-Poisson receiver.

\bex{tworeceiversc}{(Inverse multiplexer)}
Consider the Poisson receiver with three classes of input traffic in \rex{tworeceiversb}.
In addition to the three classes of input traffic, there is another external traffic, called class 4 traffic, that has a Poisson offered load $\rho_4$.
Suppose that we would like to operate such a system as an inverse multiplexer by
splitting class 4 traffic into class 1 traffic and class 2 traffic.
Let $p$ be the splitting probability of the class 4 traffic into class 1 traffic.
From the result of Poisson receivers with packet routing, we know that the inverse multiplexing system is a
Poisson receiver with two classes of input traffic, class 3 and class 4.
Moreover,
\bear{invm4444}
&&P_{{\rm suc},4}(\rho_3,\rhog_4)=p P_{{\rm suc},1}(p\rhog_4,(1-p)\rhog_4,\rhog_3 )\nonumber\\
&&\quad+ (1-p)P_{{\rm suc},2}(p\rhog_4,(1-p)\rhog_4,\rhog_3),
\eear
where  the success probability functions $P_{{\rm suc},1}(\rhog)$ and $P_{{\rm suc},2}(\rhog)$ are in
\req{invm1111} and \req{invm2222}, respectively.


The throughput for class 4 traffic is
$\rho_4 P_{{\rm suc},4}(\rho_3,\rhog_4)$. In particular,
if $\rho_3=0$, the throughput is
$$p\rho_4 e^{-p\rho_4}+(1-p)\rho_4 e^{-(1-p)\rho_4}.$$
In \rfig{rho4}, we plot the throughput of class 4 traffic for various splitting probabilities $p=0.1, 0.2, 0.3, 0.4$, and 0.5 (when $\rho_3=0$).
Intuitively, one might expect that perfect load balancing, i.e., $p=0.5$, is the optimal strategy to maximize the throughput.
This is indeed the case when $\rho_4 \le 2$. However, if $\rho_4$ is very large, this is no longer the case.
A better strategy is to overload one class of traffic, say class 1,  and control the load of the other traffic, say class 2.
As shown in \rfig{rho4}, when $\rho_4=8$, the throughput with $p=0.1$ is higher than the other four.
\eex

\begin{figure}[ht]
	\centering
	\includegraphics[width=0.30\textwidth]{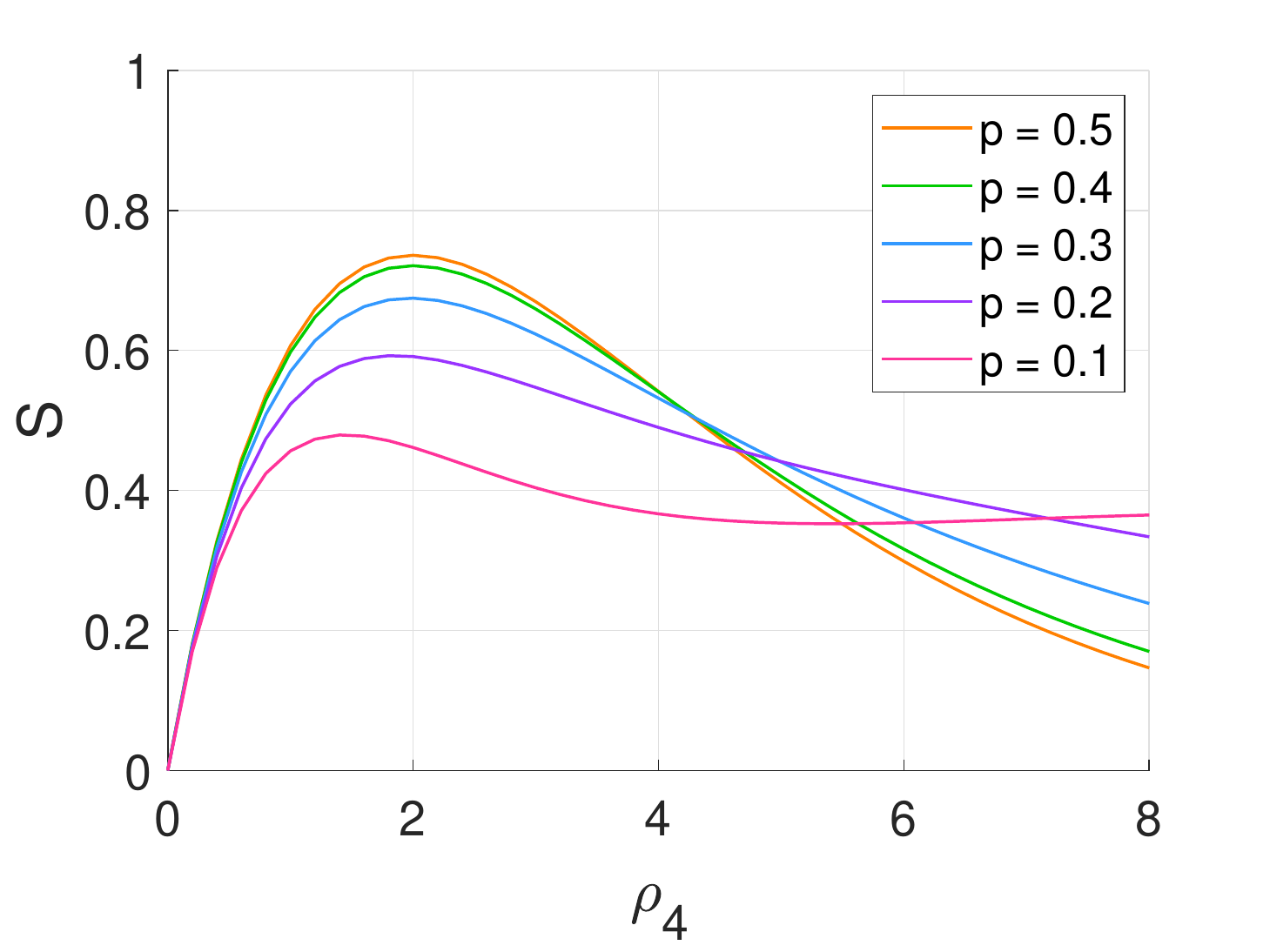}
	\caption{The throughput of class 4 traffic with respect to various splitting probabilities $p=0.1, 0.2, 0.3, 0.4$, and 0.5 (when $\rho_3=0$).}
	\label{fig:rho4}
\end{figure}

\bsubsec{Poisson receivers with packet coding}{cprmul}

{ In this section, we show how to use Poisson receivers with packet coding to construct another Poisson receiver.
Our approach is based on the tree evaluation method in \cite{luby1998analysis,luby1998analysisb,richardson2001capacity,liva2011graph}.
Such a construction is feasible due to two important closure properties of Poisson r.v.'s: (i) the excess degree distribution (defined as the degree distribution along a randomly selected edge) of a Poisson degree distribution is still a Poisson degree distribution, and (ii)  random thinning (that removes each edge independently with a certain probability) of a Poisson degree distribution is still a Poisson
degree distribution.}

\subsubsection{Repetition codes}
\label{sec:repeat}

Analogous to IRSA in \cite{liva2011graph} (for a single class of input traffic),
let us consider a system with $G_{k} T$ class $k$ active users, $k=1,2, \ldots, K$, and $T$ (independent)
Poisson receivers with $K$ classes of input traffic and the success probability functions $P_{{\rm suc},1}(\rhog), P_{{\rm suc},2}(\rhog), \ldots, P_{{\rm suc},K}(\rhog)$. Each class $k$ user transmits its packet  for a random  number of times (copies).
Let $L_{k}$ be the random variable that represents the number of copies of a packet transmitted by a class $k$ user. Each of the $L_{k}$ copies is transmitted to one of the $T$ Poisson receivers that is chosen {\em uniformly} and {\em independently}.
If any one of these $L_k$ copies of a packet is successfully received by a Poisson receiver, then the other copy can be removed (cancelled) from the system to further reduce the system load. Such a process can  then be repeatedly carried out to decode the rest of the packets.
We call the above system a system of coded Poisson receivers (CPR) with multiple classes of input traffic.
We note that in order to remove the other copies of a successfully received packet in a system of coded Poisson receivers, all the (physical) receivers in the system need to exchange the packet information, and thus they need to be cooperative receivers.
Similar to the throughput analysis for IRSA in \cite{liva2011graph},
{ our analysis is based on the tree evaluation method in \cite{luby1998analysis,luby1998analysisb,richardson2001capacity}.}
A realization of a CPR system can also be represented by
 a bipartite graph with the $\sum_{k=1}^{K} G_{k} T$ active users on one side (user nodes) and the $T$ Poisson receivers on the other side (receiver nodes).
  An edge between a user node and a receiver node in the bipartite graph represents a packet transmission from that user node to that receiver node. In particular, an edge is called a class $k$ edge if the corresponding packet transmission is from a class $k$ user, i.e., its user end is connected to a class $k$ user.

For our throughput analysis, we
let $\Lambda_{k,\ell}$ be the probability that a class $k$ packet is transmitted $\ell$ times, i.e.,
\begin{equation}
P(L_{k}=\ell)=\Lambda_{k,\ell}, \;\ell=1,2,\dots
\end{equation}
The sequence $\{\Lambda_{k,\ell}, \ell \ge 1\}$ is called the {\em degree distribution} of a class $k$ user node.
Define the generating function
\beq{mean0000mul}
\Lambda_{k}(x)=\sum_{\ell=0}^\infty \Lambda_{k,\ell} \cdot x^\ell
\eeq
 of the
degree distribution of a class $k$ user node.
Clearly,  the
mean  degree of a user node can be represented as follows:
\beq{mean1111mul}
\Lambda_{k}^{\prime}(1)=\sum_{\ell=0}^\infty \ell\cdot \Lambda_{k,\ell}.
\end{equation}
Let
\beq{mean2222mul}
\lambda_{k,\ell}=\frac{\Lambda_{k,\ell+1}\cdot (\ell+1)}{\sum_{\ell=0}^\infty\Lambda_{k,\ell+1}\cdot (\ell+1)}
\eeq
be the probability that the  user end of a randomly selected class $k$ edge has additional $\ell$ edges excluding the randomly selected class $k$  edge.
Such a probability is called the {\em excess degree distribution} of a class $k$ user node.
Also, let
\beq{mean3333mul}
\lambda_{k}(x)=\sum_{\ell=0}^\infty \lambda_{k,\ell} \cdot x^\ell
\eeq
be the corresponding generating function. Clearly, these two generating functions are related as follows:
\beq{mean3344mul}
\lambda_{k}(x)=\frac{\Lambda_{k}^\prime(x)}{\Lambda_{k}^{\prime}(1)}.
\eeq
The
offered load of class $k$ packets to a Poisson receiver, defined as the expected number of class $k$ packets transmitted to that receiver, is
\beq{mean4444mul}
\rho_{k}= G_{k} \Lambda_{k}^\prime(1) .
\eeq
When $T$ is large, the number of class $k$ packets at a receiver can be assumed to be
a Poisson random variable with mean $\rho_k$.  As such, we can assume  the degree distribution of class $k$ packets at a receiver node is a Poisson distribution with  mean $\rho_k$.
Since the excess degree distribution of a Poisson degree distribution is also Poisson with the same mean (see, e.g.,  \cite{Newman2010}),
the probability that the receiver end of a randomly selected class $k$ edge has additional $\ell$ edges excluding the randomly selected class $k$ edge
is
\beq{mean3377mul}
\frac{e^{-\rho_k} \rho_k^\ell}{\ell!}.
\eeq

Let
\beq{rho0000}
\rho=(\rho_{1}, \rho_2, \ldots, \rho_{K}).
\eeq
Denote by $\circ$ the element-wise multiplication of two vectors, i.e., for two vectors
$(x_1, x_2, \ldots, x_K)$ and $(y_1, y_2, \ldots, y_K)$,
\beq{element1111}
x \circ y =(x_1 y_1, x_2 y_2, \ldots, x_K y_K).
\eeq
Consider a randomly selected class $k$ edge from the bipartite graph.
{ Analogous to the tree evaluation method in \cite{luby1998analysis,luby1998analysisb,richardson2001capacity}}\cite{liva2011graph},
let $p_{k}^{(i)}$ (resp $q_{k}^{(i)}$) be the probability that  the receiver (resp. user)  end of a randomly selected class $k$ edge has not been successfully received after the $i^{th}$ SIC iteration.
Since the receiver end of an edge corresponds to a transmission of a tagged packet from a user to a $(P_{{\rm suc},1}(\rhog), P_{{\rm suc},2}(\rhog), \ldots, P_{{\rm suc},K}(\rhog))$-Poisson receiver, we have
\beq{tag1111mul}
p_{k}^{(1)}=1- P_{{\rm suc},k}(\rho),
\eeq
where $P_{{\rm suc},k}(\rho)$ is the probability that a randomly selected transmission from a class $k$ user is successfully received at the {\em receiver} end in the system  subject to the offered load $\rho$.
Recall that a packet sent from a user (the user end of the bipartite graph) can be successfully received if at least one of its copies is successfully received at the {\em receiver} end.
Since the probability that the user end of a randomly selected class $k$ edge  has additional $\ell$ edges is $\lambda_{k,\ell}$, the probability that the {\em user} end of a randomly selected class $k$ edge  cannot be successfully received after the first iteration is
thus
\beq{tag2222mul}
q_{k}^{(1)}=1-\sum_{\ell=0}^\infty \lambda_{k,\ell} \cdot \Big (1-(p_{k}^{(1)})^{\ell} \Big).
\eeq
This then leads to
\beq{tag3333mul}
q_{k}^{(1)}=\lambda_{k}(p_{k}^{(1)})=\lambda_{k}(1- P_{{\rm suc},k}(\rho)).
\eeq

To compute $p_{k}^{(2)}$, note that with probability
$1-q_{k}^{(1)}$, an (excess) edge at the receiver end of a randomly selected class $k$ edge is removed by SIC  after the first iteration.
Since random thinning of a Poisson random variable is still a Poisson random variable,
the number of the remaining (excess) class $k$ edges at the receiver end of a randomly selected class $k$ edge at the second iteration is Poisson with mean $q_{k}^{(1)} \rho_k$, $k=1,2, \ldots, K$.
Thus,
 the offered load at the second iteration is effectively reduced from $\rho$ to $q^{(1)} \circ \rho$,
 where
 \beq{rho1111}
 q^{(1)}=(q_{1}^{(1)}, q_{2}^{(1)}, \ldots, q_{K}^{(1)}).
\eeq
{ To the best of our knowledge, the step of using reduced Poisson offered load appears to be new in the tree evaluation method, and it plays a crucial role in the framework of Poisson receivers.}
This leads to
\beq{tag4444mul}
p_{k}^{(2)}=1- P_{{\rm suc},k}(q^{(1)} \circ \rho),
\eeq
and
\beq{tag5555mul}
q_{k}^{(2)}=\lambda_{k}(p_{k}^{(2)})=\lambda_{k}(1- P_{{\rm suc},k}(q^{(1)} \circ \rho)).
\eeq
In general, we have the following recursive equations:
\bear{tag6666dmul}
p_{k}^{(i+1)}&=&1- P_{{\rm suc},k}(q^{(i)} \circ \rho), \label{eq:tag6666amul}\\
q_{k}^{(i+1)}&=&\lambda_{k}(1- P_{{\rm suc},k}(q^{(i)} \circ \rho)),\label{eq:tag6666bmul}
\eear
where
\beq{rhoiiii}
q^{(i)}=(q_{1}^{(i)}, q_{2}^{(i)}, \ldots, q_{K}^{(i)}).
\eeq
Moreover, if $P_{{\rm suc},k}(\rho)$ is decreasing in $\rho$ for all $k$, then  $q_{k}^{(i+1)} \le q_{k}^{(i)}$ and there is a limit $0 \le q_k \le 1$ if we start from $q_k^{(0)}=1$, $k=1,2, \ldots, K$.


Now we derive the success probability of a tagged class $k$ user.
 Once again, note that a packet sent from a user can be successfully received if at least one of its copies is successfully received at the {\em receiver} end. Since the probability that a randomly selected {\em class $k$ user} has  $\ell$ edges is $\Lambda_{k,\ell}$, the probability that a packet sent from a randomly selected {\em class $k$ user} can be successfully received after the $i^{th}$ iteration is
\bear{mean5555dmul}
&&\sum_{\ell=0}^\infty \Lambda_{k,\ell} \cdot \Big (1-(p_{k}^{(i)})^{\ell} \Big)\nonumber\\
 &&=\sum_{\ell=0}^\infty \Lambda_{k,\ell} \cdot \Big (1-(1- P_{{\rm suc},k}(q^{(i-1)} \circ \rho))^{\ell} \Big) \nonumber \\
&&=1-\Lambda_k\Big (1- P_{{\rm suc},k}(q^{(i-1)} \circ \rho)\Big).
\eear
Let $G=(G_1, G_2, \ldots, G_{K})$ and
$\Lambda^\prime (1)=(\Lambda^\prime_1 (1), \Lambda^\prime_2 (1), \ldots, \Lambda^\prime_K (1))$.
Denote by
$\tilde P_{{\rm suc},{k}}^{(i)}(G)$
the success probability of a tagged class $k$ packet for the CPR system after the $i^{th}$ SIC iteration when the system is subject to a (normalized) Poisson offered load $G$.
Then it follows from \req{mean5555dmul} and \req{mean4444mul} that
\beq{mean8888thumula}
 \tilde P_{{\rm suc},k}^{(i)}(G)=1-\Lambda_k \Big  (1- P_{{\rm suc},k}(q^{(i-1)} \circ G \circ \Lambda^\prime(1))\Big).
\eeq
 Clearly, if $P_{{\rm suc},k}(\rho)$ is decreasing in $\rho$ for all $k=1, \ldots, K$, then $\tilde P_{{\rm suc},k}^{(i)}(G)$ is also decreasing in $G$ for all $k=1,2, \ldots, K$.

For the limiting case with an infinite number of SIC iterations, let $q=(q_1, q_2, \ldots, q_K)$ with
\beq{mean7777thumul}
q_k=\lim_{i \to \infty}q_k^{(i)},
\eeq
where $q^{(i)}_k$'s are in \req{tag6666bmul} and $q^{(0)}_k=1$, $k=1,2, \ldots, K$.
Then the success probability of a tagged class $k$ packet for the CPR system after an infinite number of SIC iterations when the system is subject to a (normalized) Poisson offered load $G$ is
\bear{mean8888thumul}
 \tilde P_{{\rm suc},k}(G)&=&\lim_{i \to \infty}\tilde P_{{\rm suc},k}^{(i)}(G)\nonumber\\
 &=&1-\Lambda_k \Big  (1- P_{{\rm suc},k}(q \circ G \circ \Lambda^\prime(1))\Big).
 \eear

 One interesting interpretation of the CPR system constructed from $T$ independent normal Poisson receivers is that it is also a normal Poisson receiver with another success probability function.
 Such an interpretation shows that CRDSA \cite{casini2007contention}, IRSA \cite{liva2011graph} and other CSA systems are in fact Poisson receivers with certain success probability functions.

{
We note a rigorous proof for the above probabilistic argument requires the so-called ``concentration theorem'' (see, e.g.,  Theorem 2 of \cite{richardson2001capacity}) that states that
the average fraction of nodes that have not been decoded is roughly the same as the probability that a randomly selected node has not been decoded.
Also, for a fixed number of SIC iterations, the tree assumption can be shown to be true with high probability when the number of independent Poisson receivers $T$ goes to infinity (see, e.g., Appendix A of \cite{richardson2001capacity}).}

{
\subsubsection{Ideal forward error correction codes}
\label{sec:FEC}

In \rsec{repeat}, we considered CPR systems that use simple repetition codes. In that setting, a packet is successfully decoded when one of its copies is successfully received by a Poisson receiver.
In this section, we further extend CPR systems to an ideal $(n,n_0)$-forward error correction (FEC) code. For an ideal $(n,n_0)$-FEC code, a packet is divided into $n_0$ {\em data} blocks. By encoding with additional $n-n_0$ {\em redundant} blocks, we have a code with $n$ blocks for a packet. A packet can  be
successfully decoded as long as  $n_0$ out of the $n$ blocks are successfully received \cite{lee2005throughput}.

Instead of using a repetition code, we assume that each class $k$ user encodes its packet with an ideal $(n_k,n_{k,0})$-FEC code.
Following the same notations and the analysis in \rsec{repeat}, one can show
the following recursive equations:
\bear{tag6666ckFEC}
p_k^{(i+1)}&=&1-P_{{\rm suc},k}(q^{(i)} \circ \rho), \label{eq:tag6666nkaFEC}\\
q_k^{({i+1})}&=&\sum_{j=0}^{n_{k,0}-1}\Big(\lambda_k^{\langle j \rangle}(p_k^{(i+1)})\cdot\dfrac{(1-p_k^{(i+1)})^j}{j!}\Big) \nonumber\\
& = & \sum_{j=0}^{n_{k,0}-1}\Big(\lambda_k^{\langle j \rangle}(1-P_{{\rm suc},k}(q^{(i)} \circ \rho))\nonumber\\
&&\quad\quad\cdot\dfrac{(P_{{\rm suc},k}(q^{(i)} \circ \rho))^j}{j!}\Big), \label{eq:tag6666nkbFEC}
\eear
where $\lambda_k^{\langle j \rangle}(x)$ is the $j^{th}$ derivative of $\lambda_k(x)$.
Moreover,
\bear{mean5555nkFEC}
&&\tilde P_{\rm suc}^{(i)}(G)\\
&& = 1-\sum_{j=0}^{n_{k,0}-1}\Big(\Lambda_k^{\langle j \rangle}(1-P_{\rm suc}(q^{(i-1)}\circ G \circ \Lambda^\prime(1)))\nonumber \\
&&\quad\quad\quad\quad\quad\quad \cdot\dfrac{(P_{\rm suc}(q^{(i-1)}\circ G \circ \Lambda^\prime(1)))^j}{j!}\Big).\nonumber\\
\eear
where $\Lambda_k^{\langle j \rangle}(x)$ is the $j^{th}$ derivative of $\Lambda_k(x)$.
The detailed analysis of
coded Poisson receivers with ideal FEC codes is given in Appendix B of the supplemental material.
}

\bsec{Numerical results}{num}

\bsubsec{A single class of input traffic in SA systems with two receivers}{numerical}

In this section, we compute our theoretical results and conduct extensive simulations to estimate the throughputs of three SA systems
in a correlated on-off fading channel with two receivers. These three SA systems are (i) SA without SIC (non-coop) that can be viewed as a Poisson receiver with $P_{\rm suc}(\rho)$ in \req{erase3344}, (ii) SA with spatial SIC (spatial) that can be viewed as a Poisson receiver
with $P_{\rm suc}(\rho)$ in \req{erase3377}, and (iii) SA with both spatial SIC and temporal SIC (spatial-temporal) that can be viewed as a Poisson receiver with
$\tilde P_{\rm suc}(G)$ in \req{mean8888thumul} and $P_{\rm suc}(\rho)$ in \req{erase3377}. The numerical results are computed for the setting with two receivers ($J=2$), 1000 time slots ($T=1000$), and two copies of each packet ($\Lambda(x)=x^2$).
In all our simulations, the number of iterations for temporal SIC is 100 ($i=100$). Each data point for the estimated throughput is obtained by averaging over 100 independent runs.

\subsubsection{The effect of the offered load}
\label{sec:offeredload}

\begin{figure*}[tb]
	\begin{center}
		\begin{tabular}{p{0.32\textwidth}p{0.32\textwidth}p{0.32\textwidth}}
			\includegraphics[width=0.32\textwidth]{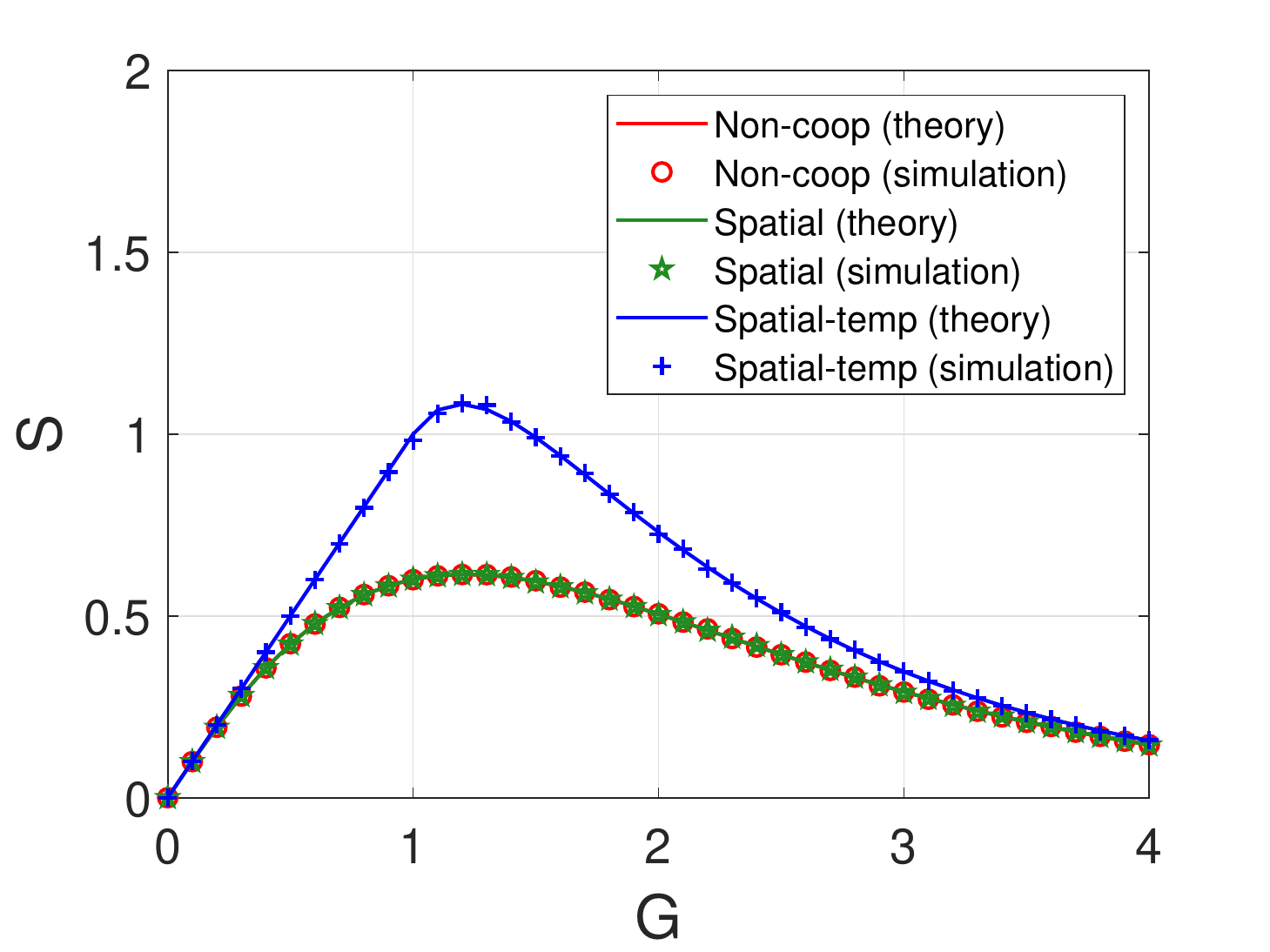} &
			\includegraphics[width=0.32\textwidth]{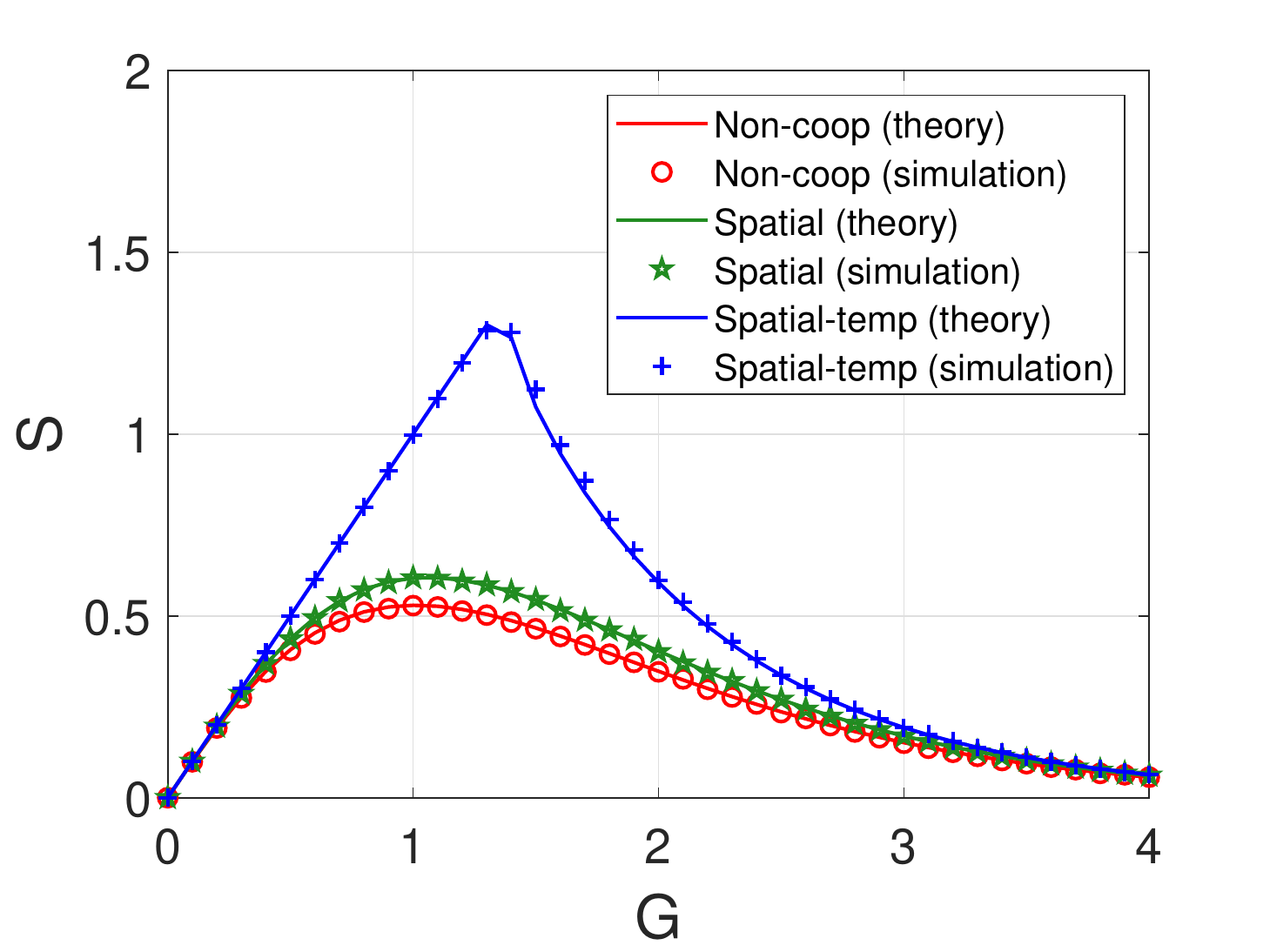} &
			\includegraphics[width=0.32\textwidth]{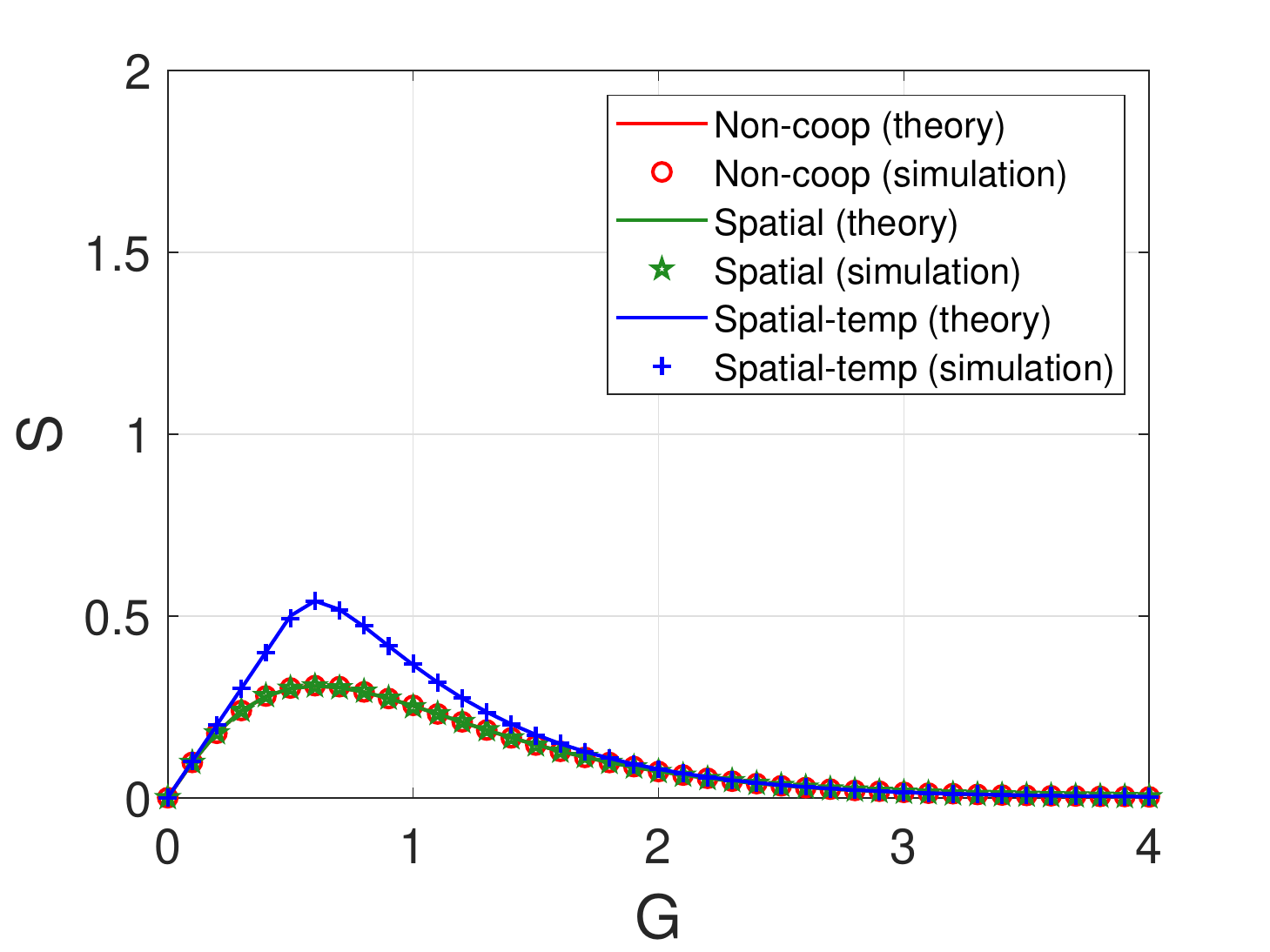}\\
        	(a) $P_{11}=0, P_{10}=P_{01}=0.5$. & (b) $P_{11}=0.3, P_{10}=P_{01}=0.35$. & (c) $P_{11}=1, P_{10}=P_{01}=0$.
		\end{tabular}
		\caption{The effect of the offered load on the throughput.}
	   \label{fig:siccompare}
	\end{center}
\end{figure*}

In this section, we show the effect of the offered load on the throughput.
As the number of time slots $T$ is fixed to be $1000$, the offered load $G=N/T$ is proportional to the number of active users $N$. As such,
we simply increase the number of active users $N$ in our simulations to obtain estimates for the throughputs.
In \rfig{siccompare}, we show both the (asymptotically) theoretical results and the simulation results for the throughputs with the following three different sets of parameters for the correlated on-off fading channel model with two receivers,
  (a) $P_{11}=0, P_{10}=P_{01}=0.5$, (b) $P_{11}=0.3, P_{10}=P_{01}=0.35$,  and (c) $P_{11}=1, P_{10}=P_{01}=0$.
As shown in \rfig{siccompare}, our theoretical results match extremely well with the simulation results.
It seems that  $T=1000$ is large enough for the tree assumption to hold in our tree analysis.
As expected, SA with spatial SIC and temporal SIC has the highest throughput.
In addition, we can observe that when $P_{11}=0$ or $P_{11}=1$, spatial SIC has no improvement for throughput (when compared with those without spatial SIC).
This is because the setting with $P_{11}=1$ reduces to the setting with a single receiver and thus there is no spatial diversity gain.
On the other hand, if $P_{11}=0$, then every packet only reaches exactly one receiver and it is impossible to perform spatial SIC to improve the throughput. However, when $P_{11}=0$, the SA system subject to the offered load $\rho$ is equivalent to two ``separate'' receivers with each receiver subject to the offered load $\rho/2$. As such, there is a significant spatial diversity gain.
In particular,
the maximum throughput in SA with spatial SIC and temporal SIC exceeds 1 in \rfig{siccompare} (a) and (b).

\subsubsection{The effect of the correlation coefficient of the two receivers}
\label{sec:correlation}

In this section, we show the effect of the correlation coefficient of the two receivers on the throughput.
For this, we set $P_{10}=P_{01}=\frac{(1-P_{11})}{2}$.
Let $X_1$ (resp. $X_2$) be the indicator r.v. that has value 1 if a tagged packet reaches receiver 1 (resp. receiver 2) and value 0 otherwise. Then
the correlation coefficient of the two receivers, denoted by $\omega$, can be related to $P_{11}$ as follows:
\bear{correlation2223}
\omega
&=&\frac{E(X_1 X_2)-E(X_1)E(X_2)}{\sqrt{E(X_1^2)-E^2(X_1)}\sqrt{E(X_2^2)-E^2(X_2)}}\nonumber\\
&=&\frac{P_{11}-(P_{11}+P_{10})(P_{11}+P_{01})}{\sqrt{(P_{11}+P_{10})-(P_{11}+P_{10})^2}}  \nonumber\\
&&\quad\quad \frac{1}{\sqrt{(P_{11}+P_{01})-(P_{11}+P_{01})^2}}\nonumber\\
&=&-\frac{1-P_{11}}{1+P_{11}}.
\eear
Clearly, the correlation coefficient is decreasing in $P_{11}$ from 0 to $-1$.
Moreover, the two receivers are  uncorrelated when $P_{11}=1$ as the two indicator random variables are 1.
On the other hand, when $P_{11}=0$, we have the smallest correlation coefficient $\omega=-1$.
In \rfig{influp11}, we  plot the (theoretical) throughput curves as a function of $P_{11}$ for the three SA systems
with $G=1.2$.  One interesting finding is that the throughput for the SA system with both the spatial diversity and the temporal diversity (the blue curve) is not a monotone function of $P_{11}$. As such, it is also not a monotone function of the correlation coefficient.
This is because when $P_{11}=1$, the system reduces to the SA with a single receiver.  On the other hand, when $P_{11}=0$, it reduces to the SA with two separate receivers. With spatial SIC between these two cooperative receivers, it is possible to have some performance gain for a small nonzero $P_{11}$.
But when $P_{11}$ is further increased to 1, we lose spatial diversity and that results in performance degradation.
We note that the effect of the correlation coefficient was discussed in the recent paper \cite{formaggio2020receiver}. However, that paper only considered two non-cooperative receivers with spatial diversity.

\begin{figure}[ht]
	\centering
	\includegraphics[width=0.30\textwidth]{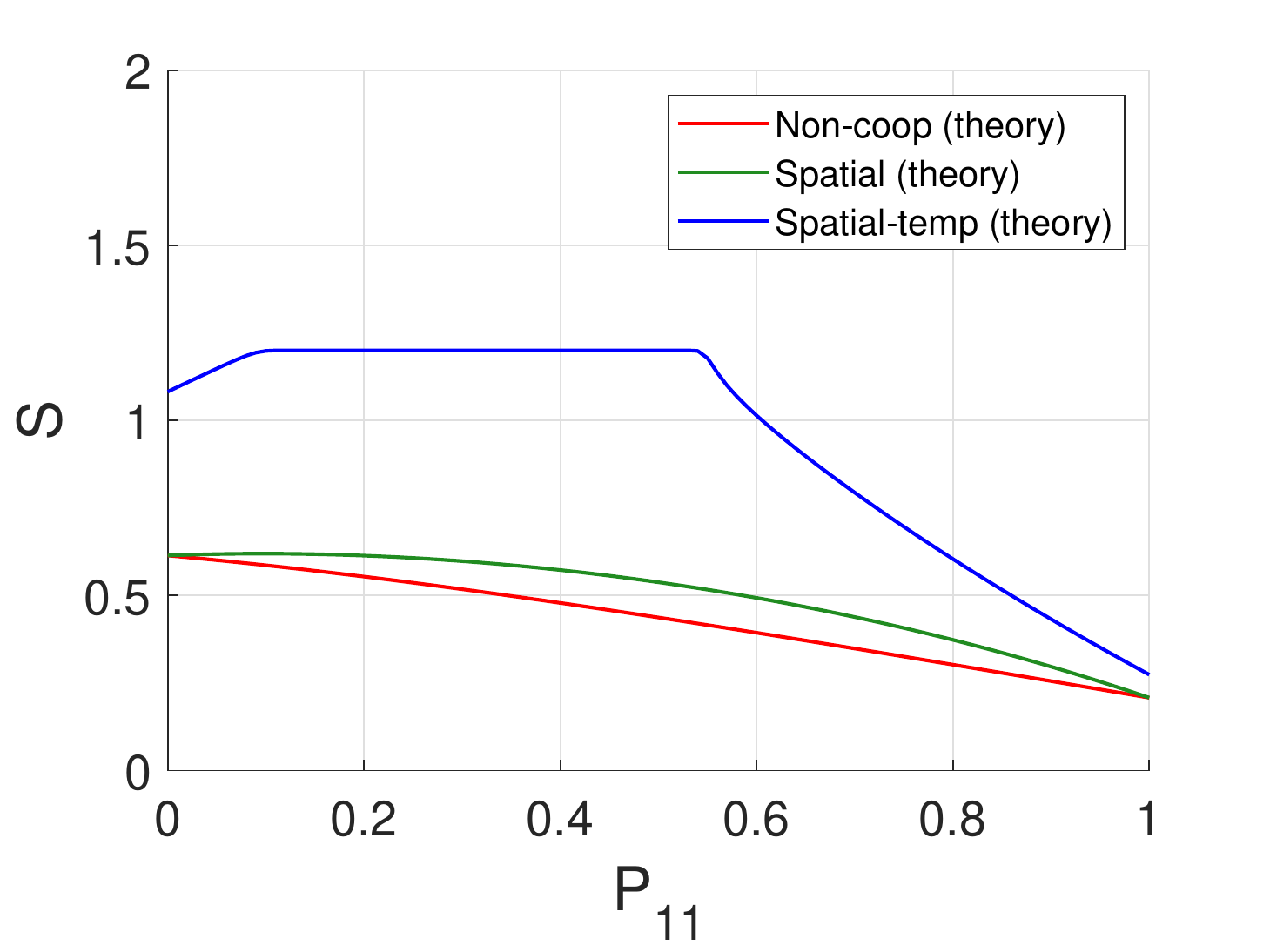}
	\caption{The effect of $P_{11}$ on the throughput.}
	\label{fig:influp11}
\end{figure}

\subsubsection{The effect of the FEC code}
\label{sec:numericalcode}

In this section, we show the effect of the FEC code on the throughput.
For this, we consider four CSA systems with ideal FEC codes in \rsec{FEC}.
 These four FEC codes are (i) $(n,n_0)=(2,1)$, (ii) $(n,n_0)=(4,2)$, (iii) $(n,n_0)=(3,1)$,
 and (iv) $(n,n_0)=(6,2)$. Note that (i) and (iii) are simple repetition codes.
The parameters for the correlated on-off fading channel model with two receivers are $P_{11}=0.3$, and $P_{10}=P_{01}=0.35$.
  For the system with an $(n,n_0)$-FEC code,
 the generating function of the degree distribution  is $\Lambda(x)=x^n$
 and
 the number of time slots is $n_0 T$.
 By doing so, for a fixed $N$ and a fixed $T$, CSA systems with the same code rate $n_0/n$ are subject to the same offered load $$\rho=\frac{N}{n_0T}\Lambda^\prime(1)=\frac{N}{T} \frac{n}{n_0} ,$$
 and thus the comparison for the throughputs of CSA systems with the same code rate is fair.
Specifically, for a fixed $N$ and a fixed $T$,
systems (i) and (ii) have the same offered load, and  systems (iii) and (iv)
have the same offered load. As in the previous section, we set $T=1000$ and
the number of iterations for temporal SIC  100, i.e., $i=100$. Each data point for the estimated throughput is obtained by averaging over 100 independent runs.
In \rfig{code}, we show the throughputs of these four CSA systems. As shown in this figure, the throughput of system (i) is better than that of system (ii) except for a very narrow range of $N/T$. Moreover,
the throughput of system (iii) is always better than that of system (iv). It seems that selecting the simple repetition code is good enough among the family of FEC codes with the same code rate. To see the intuition behind this, one can picture a $(6,2)$-FEC code as two {\em nearly} independent $(3,1)$-FEC codes, where the first three blocks are in the first $T$ time slots and the last three blocks are in the last $T$ time slots. A packet that uses a $(6,2)$-FEC code can be decoded if both $(3,1)$-FEC codes can be decoded. Thus, the success probability of a $(6,2)$-FEC code is lower than that of a $(3,1)$-code. As such, for $n_0 \ge 2$, we do not suggest using $(n,n_0)$-FEC codes with low code rates in CSA systems. Such systems not only increase encoding/decoding complexity but also degrade the throughputs.

\begin{figure}[ht]
	\centering
	\includegraphics[width=0.40\textwidth]{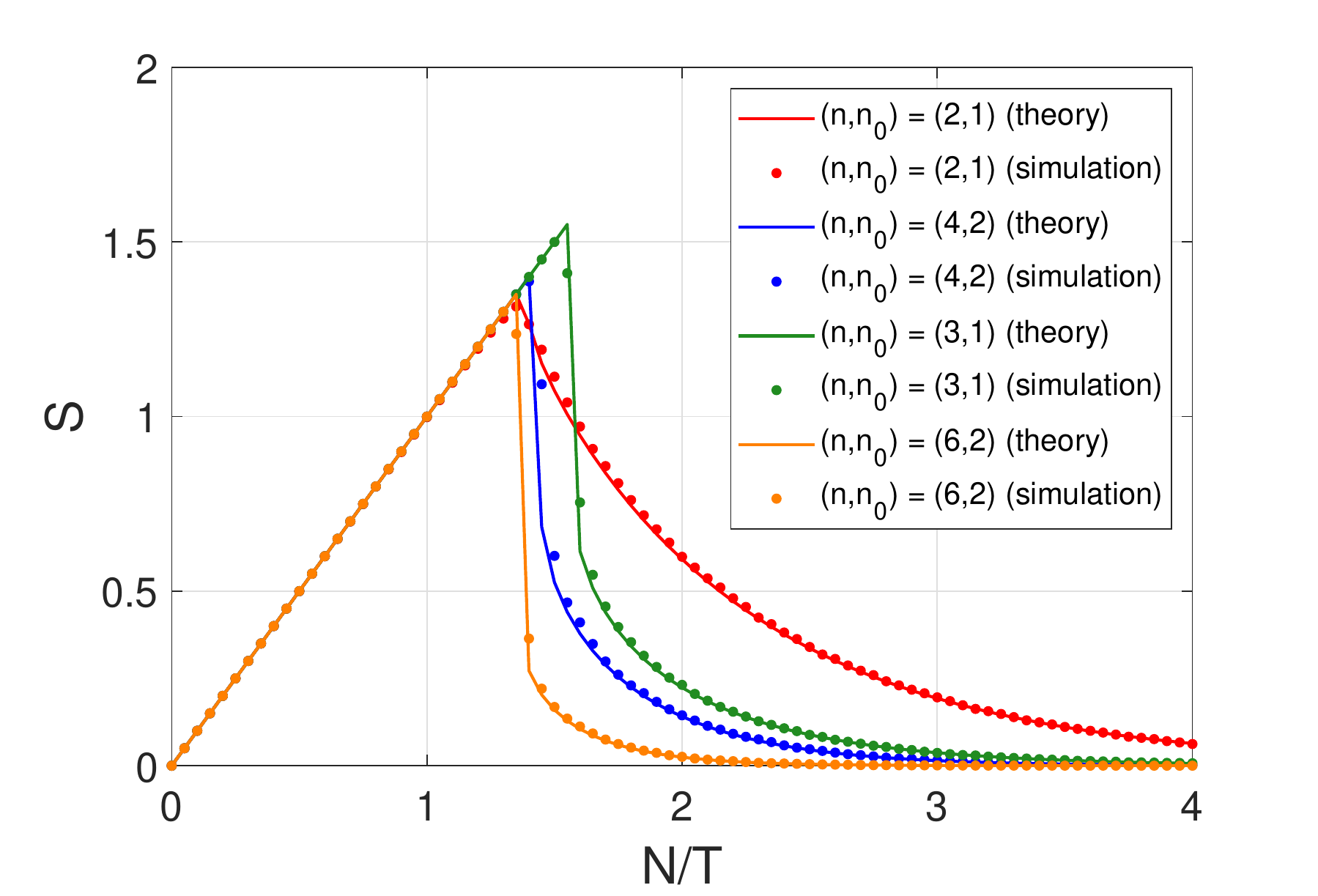}
	\caption{The throughputs of four CSA systems with ideal FEC codes: (i) $(n,n_0)=(2,1)$, (ii) $(n,n_0)=(4,2)$, (iii) $(n,n_0)=(3,1)$,
 and (iv) $(n,n_0)=(6,2)$.}
	\label{fig:code}
\end{figure}

\bsubsec{A use case for URLLC traffic subject to eMBB cross traffic}{usecase}

In this section, we demonstrate how the framework of Poisson receivers can be used for providing differentiated services between URLLC traffic and eMBB traffic.
For grant-free uplink transmissions (see, e.g., \cite{bennis2018ultra,chang2019asynchronous,centenaro2020analysis,liu2020analyzing} and references therein), the URLLC traffic needs to have low latency (e.g., 1ms), and a low error probability (e.g., $10^{-5}$).
Therefore, the service requirement of URLLC traffic is more stringent than that of  eMBB traffic. { Analogous to the current cellular system, we assume that time is partitioned into transmission time intervals (TTI). Each TTI consists of $T$ time slots. Each eMBB user is scheduled to transmit a packet in a (randomly assigned) time slot in each TTI. On the other hand, URLLC users with packets to send in a TTI transmit multiple copies of their packets randomly in that TTI. The packets transmitted by URLLC users are then superposed with the scheduled eMBB traffic (see, e.g., \cite{anand2020joint} for superposition in downlink transmissions), and they are decoded by using the SIC technique in collision channels.
This corresponds to the CPR system with two classes of input traffic, where
the degree distribution of URLLC traffic is $\Lambda_1(x)=x^L$ with some $L >1$ (see \req{mean0000mul}), and the degree distribution of eMBB traffic is $\Lambda_2(x)=x$. In such a system, URLLC transmissions are decoded first by using the SIC technique. eMBB transmissions that overlap with undecodable URLLC transmissions can be treated as {\em erased} or {\em punctured} \cite{popovski20185g} and can be protected by using another layer of error correction codes.}

As suggested by Damanjit Singh,
we focus on a particular use case for supporting precise cooperative robotic motion control defined in use case 1 of mobile robots in \cite{3gpp.22.104}.
The message size for this use case is 40 byte (i.e., 320 bits), and the transfer interval is 1 ms.
Adding overheads (for header information and SIC decoding), the packet size for this use case might be increased to 44 bytes.
On the other hand, from Table 4.1A-2 in \cite{3gpp.36.306}, we know that the maximum number of bits of an uplink shared channel (UL-SCH) transport block transmitted within a transmission time interval (TTI = 1ms) is 105,528 bits (i.e., 13,191 bytes).
As $13,191/44\approx 299$, we can accommodate roughly 299 packet transmissions within 1ms.
To be conservative, we set the number of packet transmissions (minislots) within one TTI  to be 256, i.e., $T=256$ in our analysis in \rsec{repeat}. According to \cite{3gpp.22.104}, the  (maximum) number of URLLC users that can be supported is 100. To stretch to the limit, we set
the number of URLLC users $N_1$ to be 100, and thus $G_1=N_1/T=100/256$. In our simulation, every URLLC user and every eMBB user transmits one packet in a TTI. We are interested in finding out the number of eMBB users $N_2$ that can be admitted to the system while keeping the error probability of URLLC users smaller than $10^{-5}$.

\subsubsection{A single CSA receiver}
\label{sec:singmul}

In this section, we consider the setting with only one CSA receiver.
For such a setting, we have the following success probability function:
  \begin{equation*}
\begin{aligned}
P_{{\rm suc},1}(\rho) & = P_{{\rm suc},2}(\rho) = e^{-(\rho_1+\rho_2)},
\end{aligned}
\end{equation*}
where $\rho_1=G_1\cdot L=(100/256)\cdot 5$ and $\rho_2=G_2\cdot 1=(N_2/256)$ with $N_2$ being the number of eMBB users.
In \rfig{urdegcompare}, we
show the effect of the number of eMBB users on the error probability of URLLC users for  $L=4,5,6,7$.
In our simulations, the number of iterations for successive interference cancellation (SIC) is 100 ($i=100$), and
the error probability is obtained by averaging 200 independent runs of the simulation. As shown in \rfig{urdegcompare}, both the theoretical result and the simulation result are very close to each other for every measurable data point.
Also,  increasing $L$ decreases the error probability of URLLC users. However, increasing $L$ also increases the number of SIC iterations in the decoding process. As our analysis relies on the tree assumption for enumerating a user node, { such a tree assumption, known to hold for a bipartite graph with an infinite number of nodes,  may not hold for a tree with a large depth in  a bipartite graph with a finite number of $T$ receiver nodes.}  As shown in \rfig{urdegcompare}, there are some discrepancies between the theoretical results and simulation results for $L=6$ and $L=7$.
 As such, we choose $L=5$ for all our subsequent experiments.

\begin{figure*}[tb]
	\begin{center}
		\begin{tabular}{p{0.23\textwidth}p{0.23\textwidth}p{0.23\textwidth}p{0.23\textwidth}}
			\includegraphics[width=0.23\textwidth]{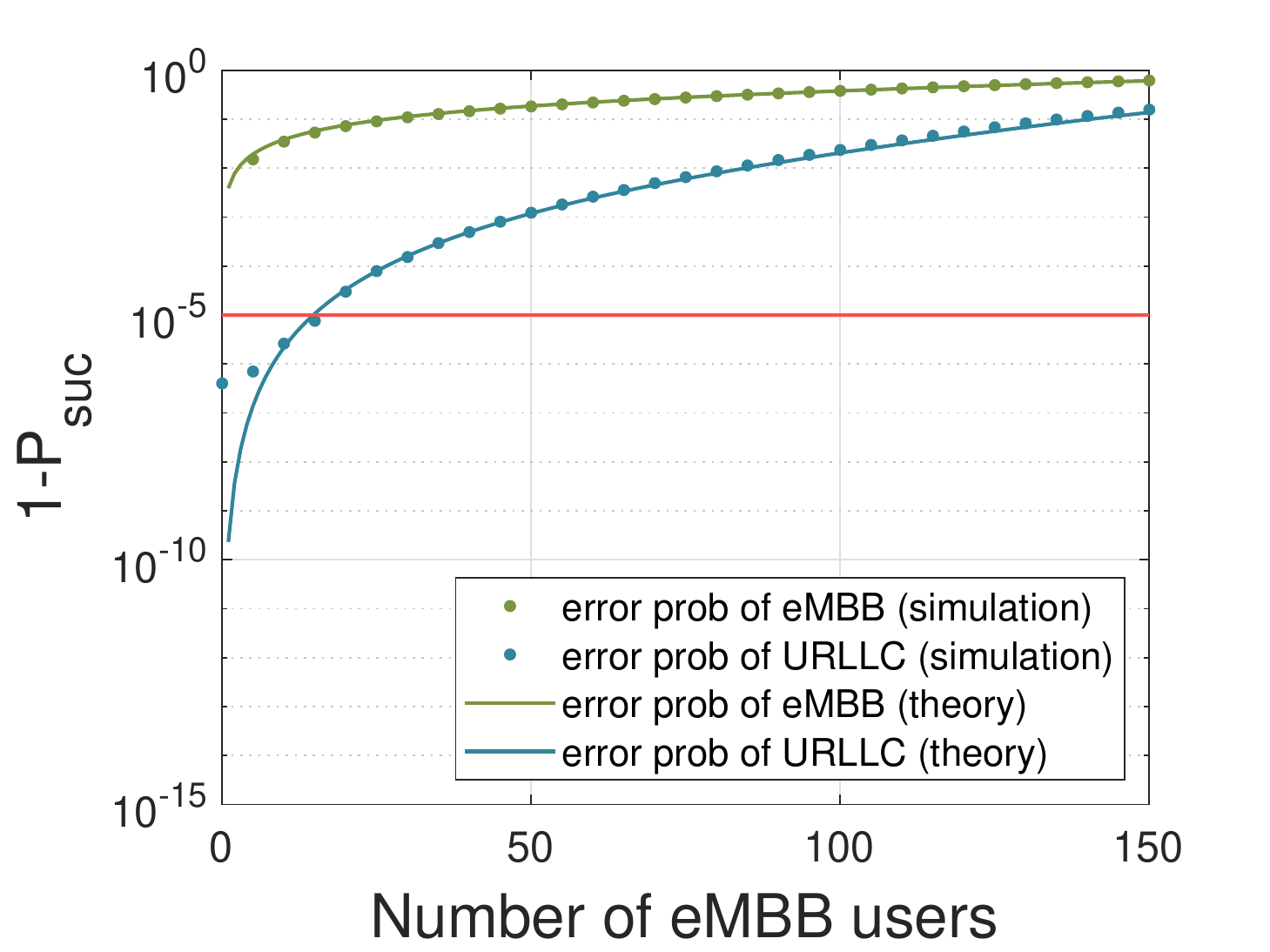} &
			\includegraphics[width=0.23\textwidth]{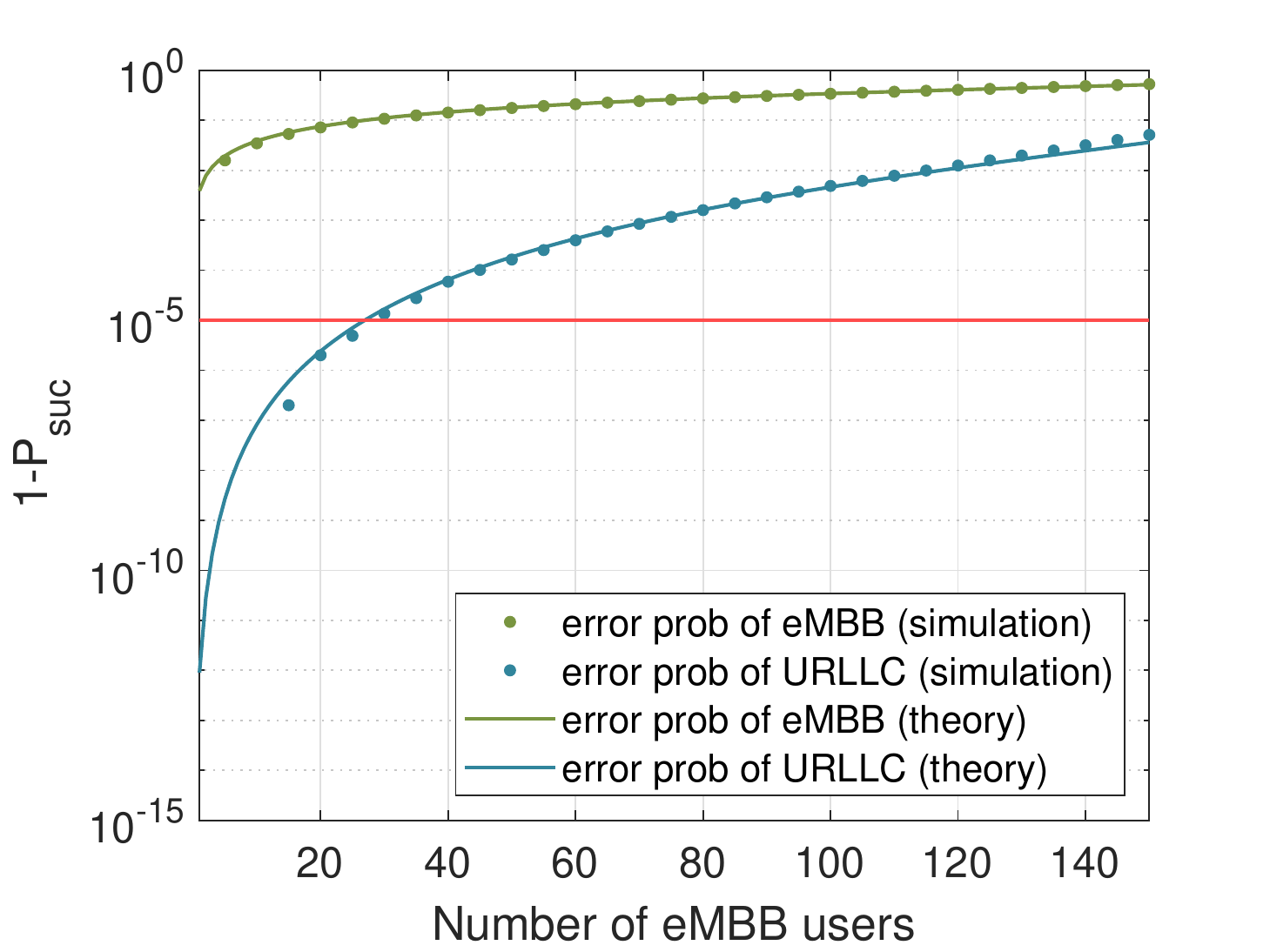} &
			\includegraphics[width=0.23\textwidth]{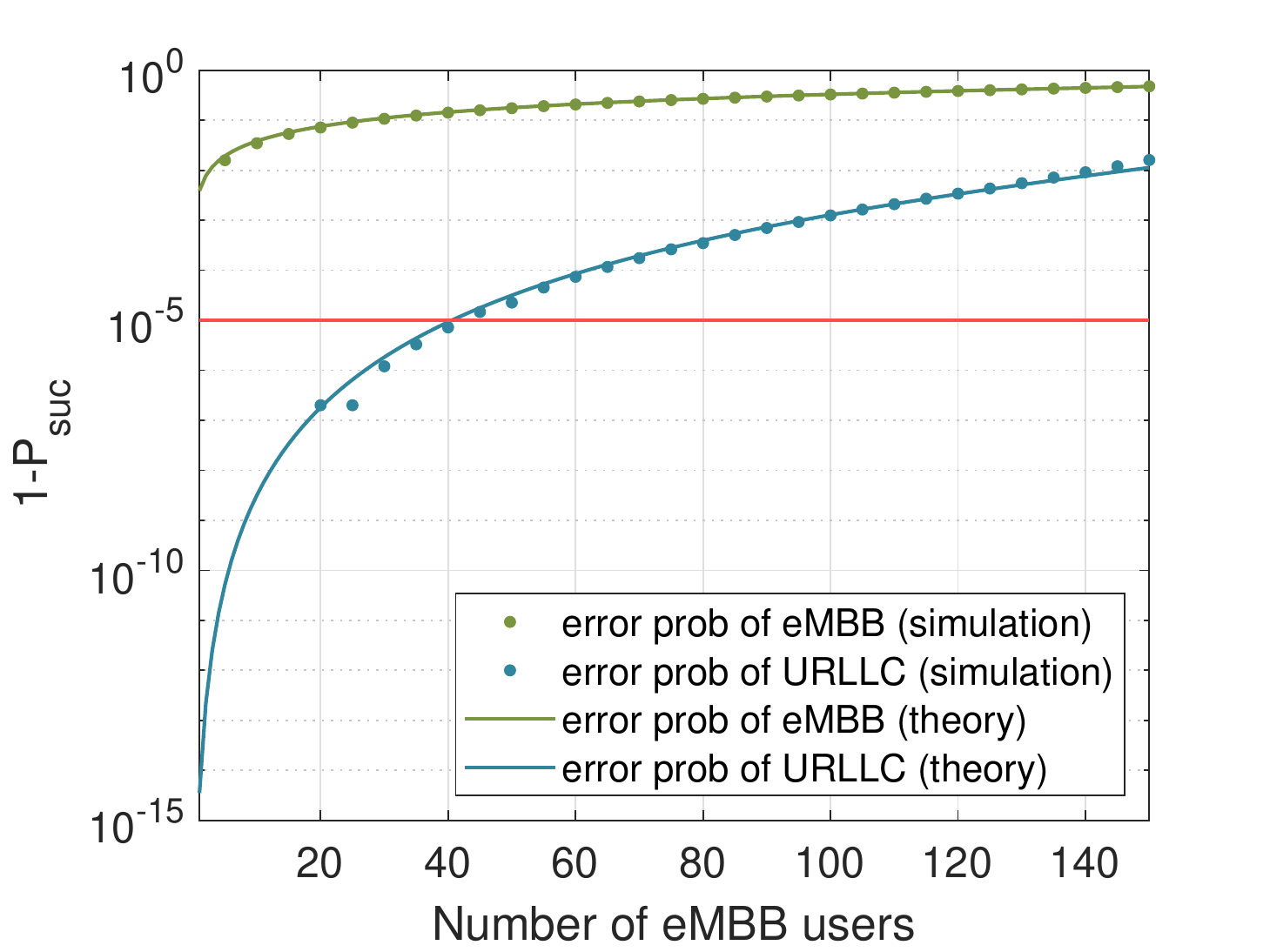} &
			\includegraphics[width=0.23\textwidth]{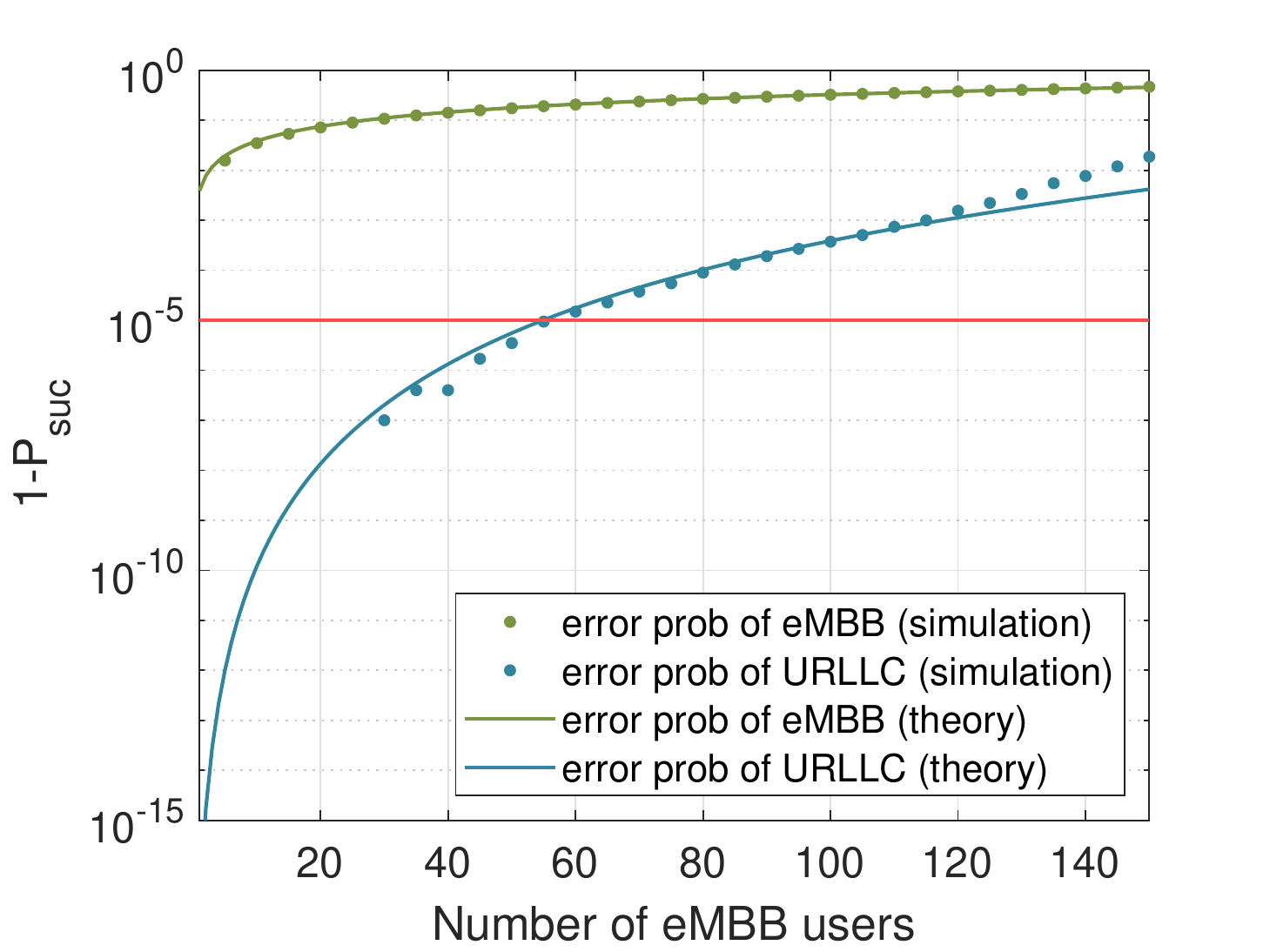}\\
			(a) $\Lambda_1(x) = x^4$. & (b) $\Lambda_1(x) = x^5$. & (c) $\Lambda_1(x) = x^6$. & (d) $\Lambda_1(x) = x^7$.
		\end{tabular}
		\caption{The effect of using different degree distributions of URLLC users.}
		\label{fig:urdegcompare}
	\end{center}
\end{figure*}

From \rfig{urdegcompare}, we also see that increasing the number of eMBB users increases the error probability of URLLC users.
In order for the error probability of URLLC to be smaller than $10^{-5}$, there is a limit on the maximum number of eMBB users that can be admitted to the system. For this, we conduct 100,000 independent runs of the simulation for $L=5$. The results are shown in
 \rfig{different_class_1}.
The theoretical results in \rfig{different_class_1} are calculated by using $\tilde P_{{\rm suc},1}$ and $\tilde P_{{\rm suc},2}$ in \req{mean8888thumul} with $q$ in \req{mean7777thumul}.
For a small number of eMBB users, there are no  errors of URLLC packets in our simulations, and thus there are no data for these points in \rfig{different_class_1}.
As shown in \rfig{different_class_1}, the system of a single receiver can roughly admit (at most) 26 eMBB users while keeping the error probability of URLLC traffic smaller than $10^{-5}$.

\begin{figure}[ht]
	\centering
	\includegraphics[width=0.40\textwidth]{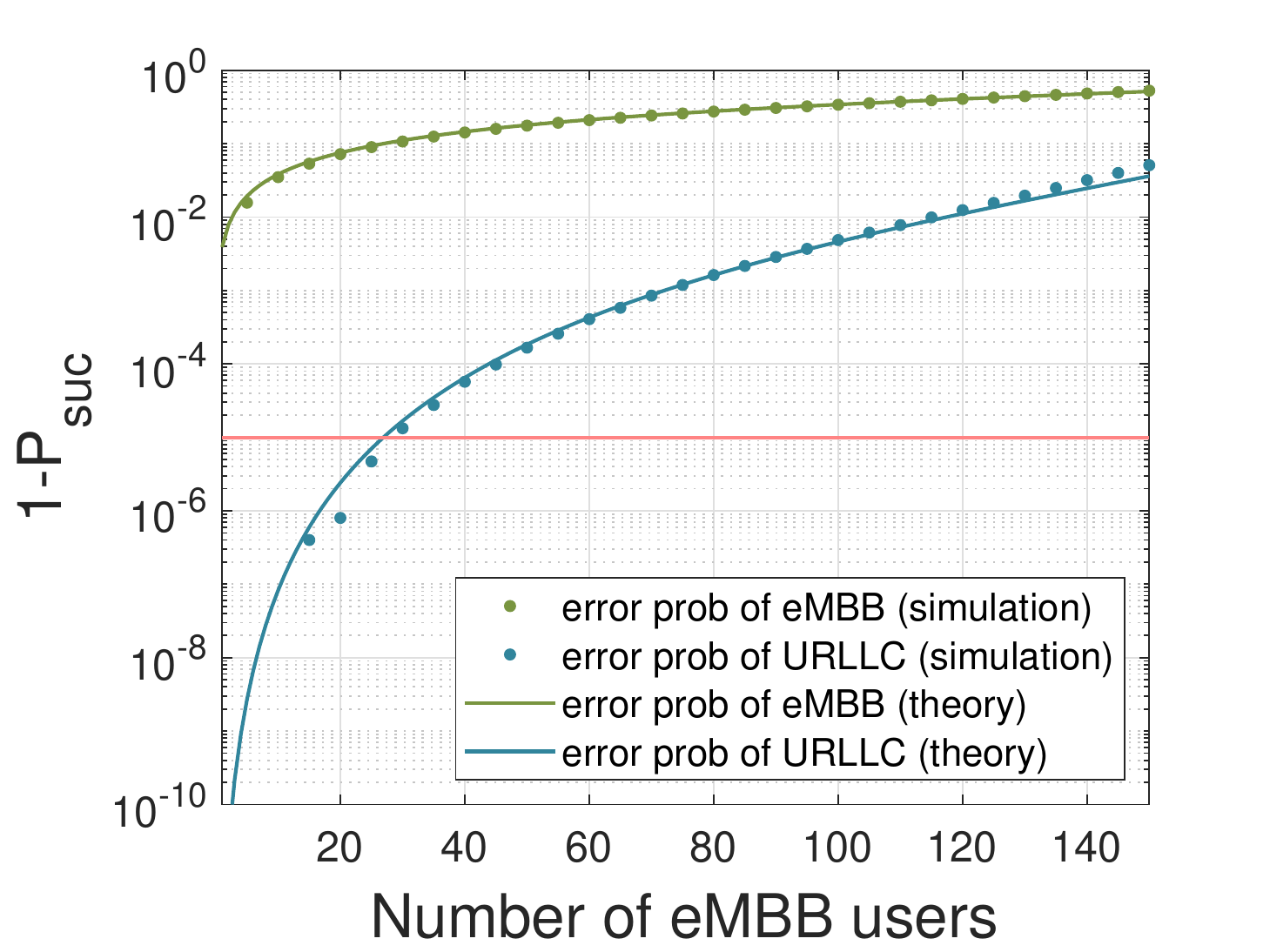}
	\caption{The effect of the number of eMBB users on the error probability of URLLC users for $L=5$ in the setting with a single receiver.}
	\label{fig:different_class_1}
\end{figure}

\subsubsection{Two cooperative receivers with a common fading channel for both eMBB traffic and  URLLC traffic}
\label{sec:twomul}

In this section, we consider the CSA in a correlated on-off fading channel with two receivers in \rsec{numerical}.
We demonstrate that two cooperative receivers can be used to improve the performance of the system.
We assume that both eMBB packets and URLLC packets go through the same fading channel.
For such a setting, we have the following success probability function:
\bearn
&&P_{{\rm suc},1}(\rho)  = P_{{\rm suc},2}(\rho) \\
&& = (P_{11}+P_{10})e^{-(\rho_1+\rho_2)(P_{11}+P_{10})}\\
&&+(P_{11}+P_{01})e^{-(\rho_1+\rho_2)(P_{11}
+P_{01})}-P_{11}e^{-(\rho_1+\rho_2)(1-P_{00})} \\
&& +P_{01}(\rho_1+\rho_2)P_{11}e^{-(\rho_1+\rho_2)(1-P_{00})}\\
&&+P_{10}(\rho_1+\rho_2)P_{11}e^{-(\rho_1+\rho_2)(1-P_{00})}.
\eearn
For our simulations, we use $P_{10} = P_{01} = 0.25$ and $P_{11} = 0.5$.
The other parameters are the same as those used in the single receiver in \rsec{singmul}.
 In \rfig{different_class_2}, we show the  effect of the number of eMBB users on the error probability of URLLC users in the setting with two cooperative receivers.
From \rfig{different_class_2}, the system with two cooperative receivers can roughly admit 65 eMBB users while keeping  the error probability of URLLC traffic smaller than $10^{-5}$. This is more than twice of that for the system with a single receiver (26 eMBB users).
The rationale behind this is the number of Poisson receivers (in terms of time slots in the two receivers) for the system with two cooperative receivers are twice of that for the system with a single receiver, and there are also  packets that can be decoded by spatial SIC.

\begin{figure}[ht]
	\centering
	\includegraphics[width=0.40\textwidth]{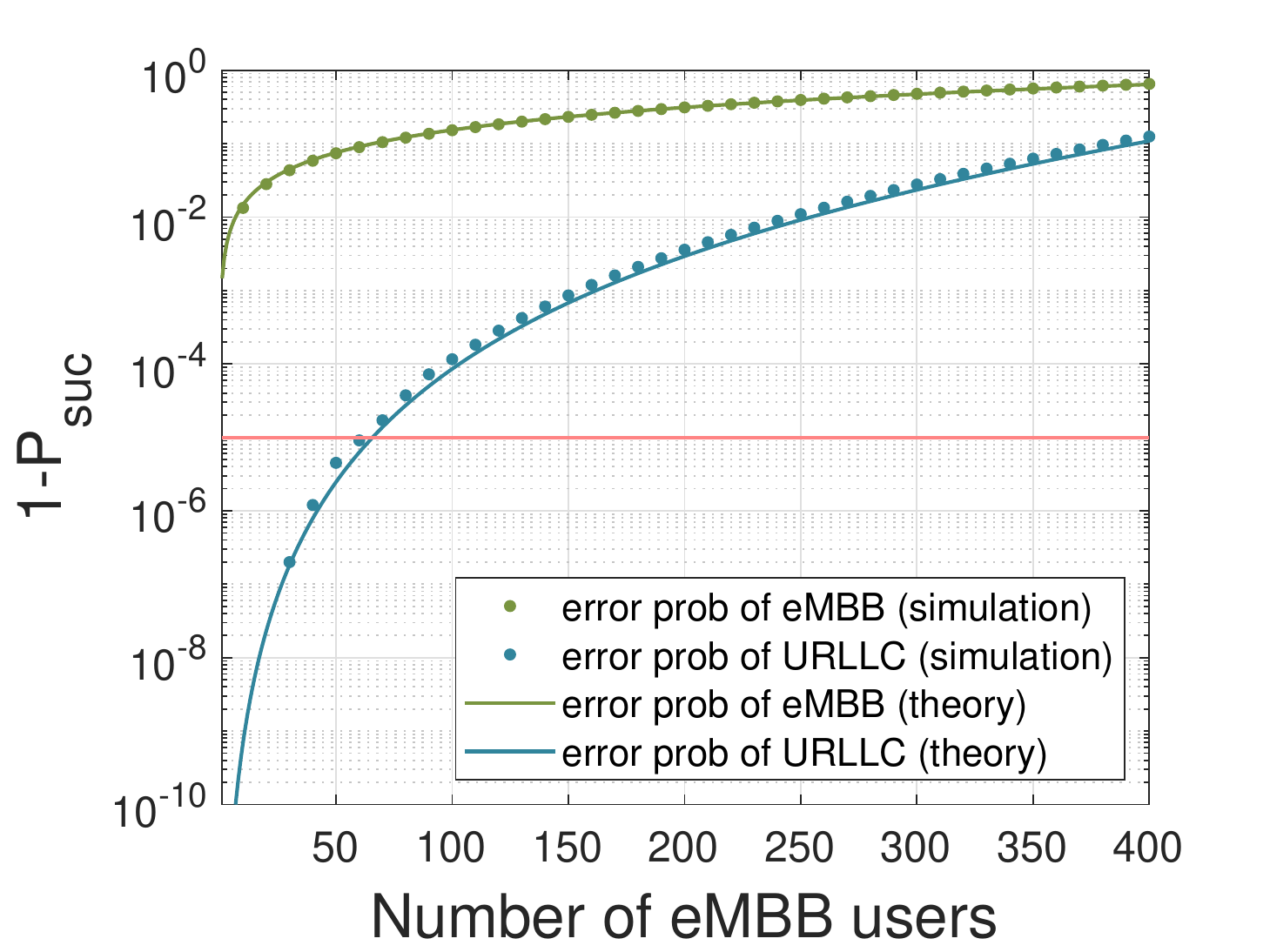}
	\caption{The effect of the number of eMBB users on the error probability of URLLC users in two cooperative receivers with a common fading channel for both eMBB traffic and  URLLC traffic.}
	\label{fig:different_class_2}
\end{figure}

Then we consider the effect of the correlation coefficient between two receivers.
For this, we vary $P_{11}$ by setting $P_{10}=P_{01}=\frac{(1-P_{11})}{2}$ as described in \rsec{correlation}.
For each fixed $P_{11}$, we calculate the maximum number of eMBB users that can be admitted while keeping the error probability of URLLC users (class 1 in this section) smaller than $10^{-5}$.
As a function of $P_{11}$, we plot the (theoretical) result in \rfig{correlation_coefficient_threshold}.

\begin{figure}[ht]
	\centering
	\includegraphics[width=0.22\textwidth]{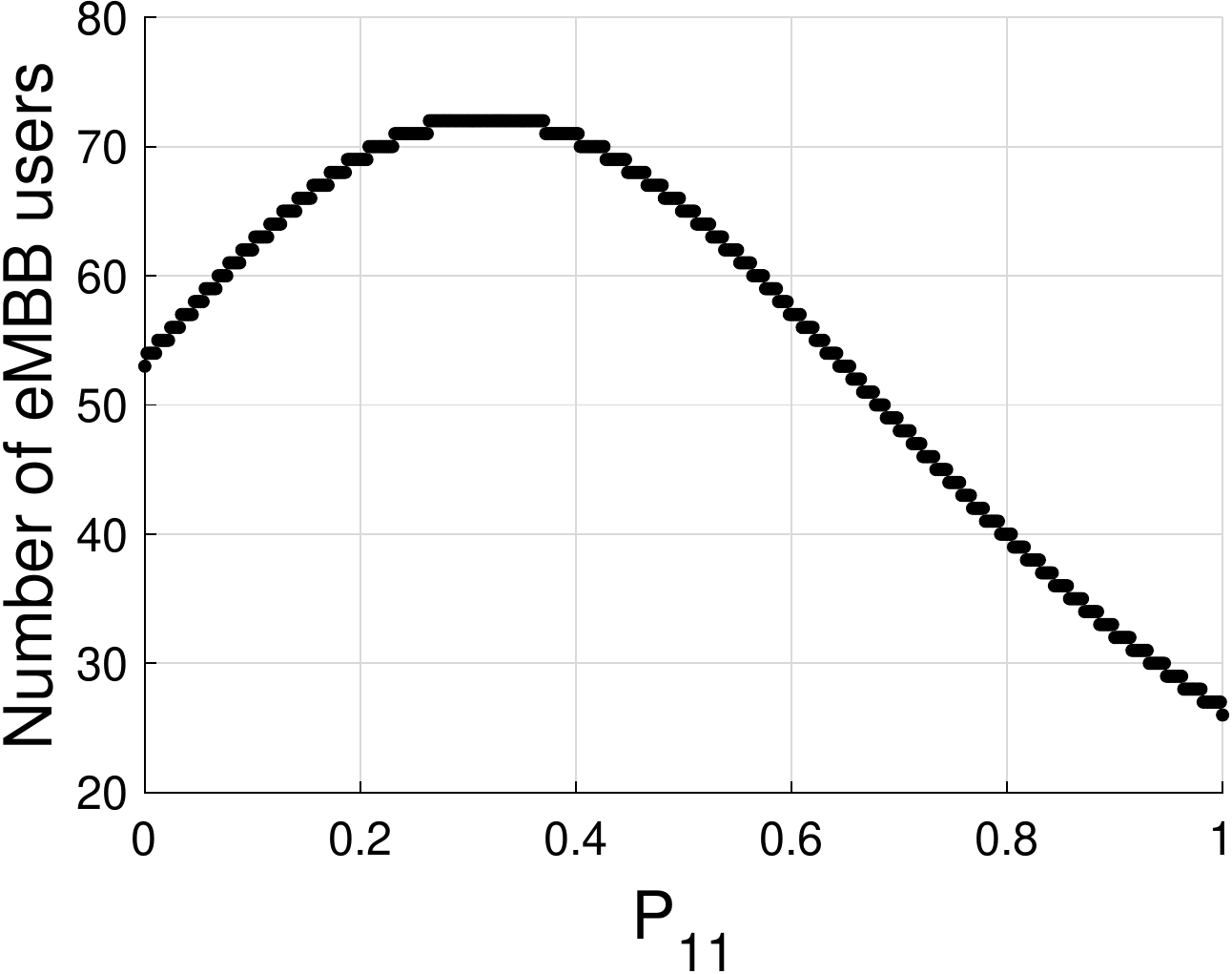}
	\caption{The effect of $P_{11}$ on the number of eMBB users.}
	\label{fig:correlation_coefficient_threshold}
\end{figure}

As shown in \rfig{correlation_coefficient_threshold},
the number of eMBB users that can be admitted to the system is maximized when $P_{11}\approx 0.3$.
The corresponding correlation coefficient is roughly $-0.5$.
As discussed in \rsec{correlation}, when $P_{11}=1$, the system is reduced to  a single receiver. As such, it has the worst performance comparing  to the other values of $P_{11}$.
On the other hand, when $P_{11}=0$, it can be seen as two separate receivers with no spatial diversity gain.
Thus, neither $P_{11}=0$ nor $P_{11}=1$ can lead to the best performance of the system.
}

{

\bsubsec{Two cooperative receivers with different fading channels for eMBB traffic and  URLLC traffic}{mulrec}

\begin{figure}[ht]
	\centering
	\includegraphics[width=0.40\textwidth]{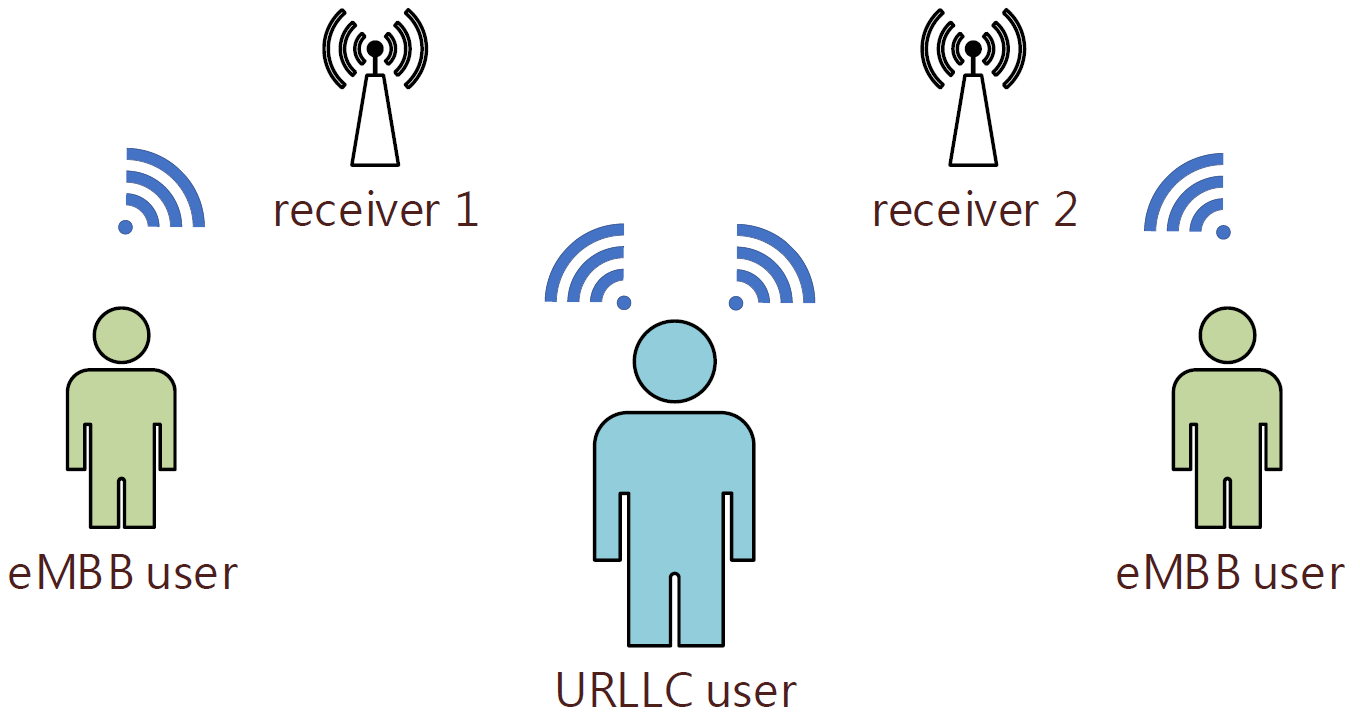}
	\caption{An illustration for two cooperative receivers with different fading channels: URLLC packets are sent to both receivers, while eMBB packets can only be sent to one of the two receivers.}
	\label{fig:difrec}
\end{figure}

In this section, we consider the CSA in a correlated on-off fading channel with two receivers in \rsec{numerical}.
Unlike \rsec{twomul}, we do not assume that both eMBB traffic and URLLC traffic go through the same fading channel.
Instead, we assume that URLLC packets are sent to both receivers, while eMBB packets can only be sent to one of the
two receivers (see \rfig{difrec} for an illustration). This is the setting considered
in \rex{tworeceiversb} and \rex{tworeceiversc}, and it could be feasible by using different power allocations and { directional antennas} for these two types of traffic.
 As described in \rex{tworeceiversb}, there are three classes of input traffic:  class 1 packets that are only sent to receiver 1, class 2 packets that are only sent to receiver 2, and    class 3 packets that are sent to both receivers.
 The first two classes are from
 eMBB traffic and in our simulation each eMBB packet can be either a class 1 packet or a class 2 packet with an equal probability { (see the inverse multiplexer in \rex{tworeceiversc}).}
 On the other hand, class 3 packets are from URLLC traffic.
As  in \rsec{twomul}, only one copy of an eMBB packet is sent, i.e., $\Lambda_1(x)=\Lambda_2(x)=x$, and  five copies of an URLLC packet are sent to both receivers, i.e.,  $\Lambda_3(x)=x^5$.
Each data point for the estimated error probability is obtained by averaging over 100,000 independent runs.
Also, we set the number of iterations for SIC to be 100 ($i=100$), and the number of URLLC users to be 100 ($G_3 = 100/256$).
The result is shown in \rfig{difrec_result}.
Theoretical results are computed by using $\tilde P_{{\rm suc}}$ in \req{mean8888thumul} with $P_{{\rm suc}}$ in \rex{tworeceiversb} and $q$ in \req{mean7777thumul}.

\begin{figure}[ht]
	\centering
	\includegraphics[width=0.40\textwidth]{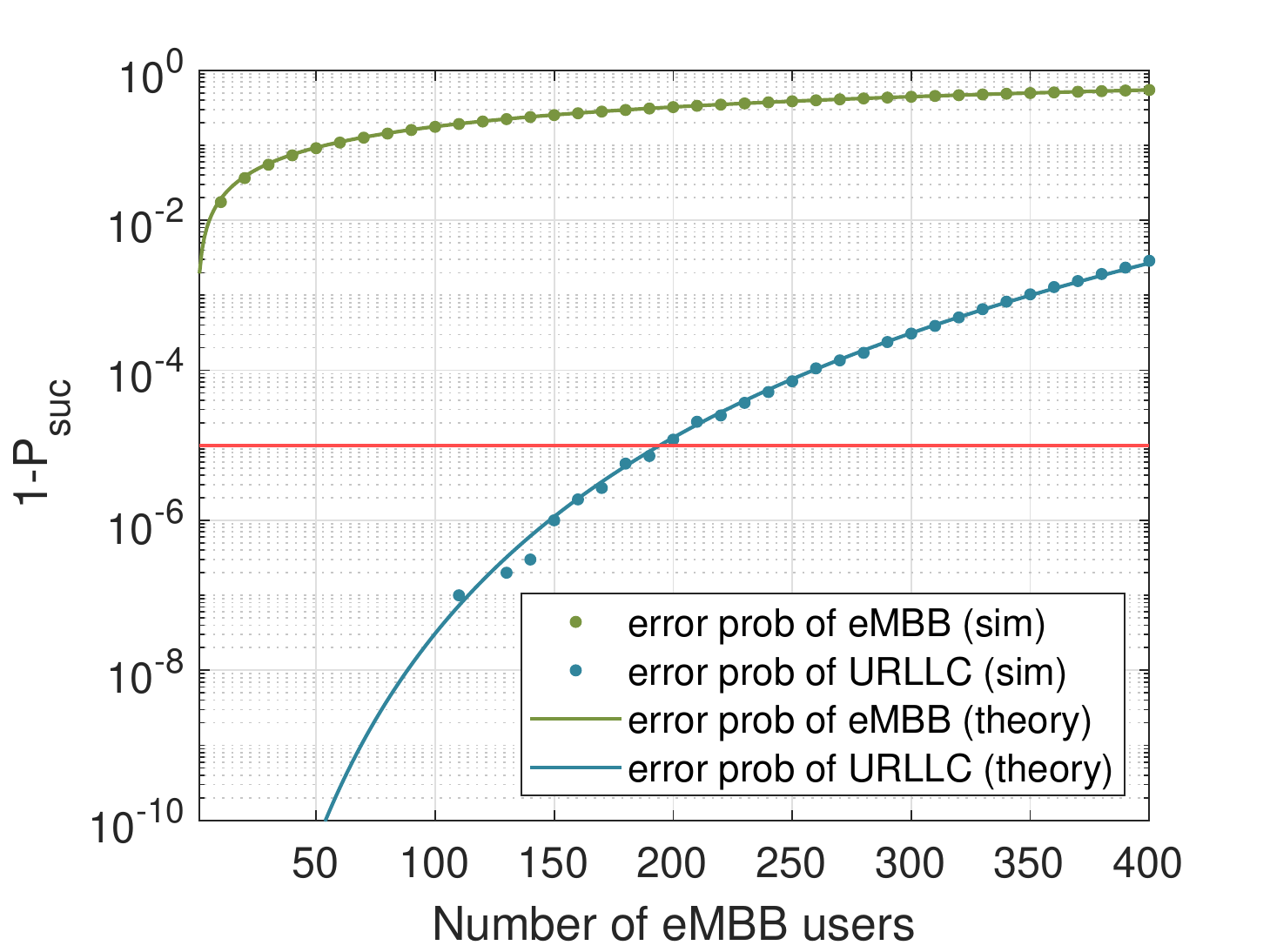}
	\caption{The effect of the number of eMBB users on the error probability of URLLC users in two cooperative receivers with different fading channels for eMBB traffic and  URLLC traffic.}
	\label{fig:difrec_result}
\end{figure}

In \rfig{difrec_result}, the system can admit roughly 194 eMBB users while keeping the error probability of URLLC traffic (class 3 in this section) smaller than $10^{-5}$.
In this section, URLLC users send five copies to both receivers, and thus  ten copies to this system with spatial SIC. On the other hand,
in \rsec{twomul}  URLLC users  send five copies to the two receivers with probability $P_{11}=0.5, P_{10}=P_{01}=0.25$, and that leads to on average 7.5 copies to this system. Similarly, eMBB users in \rsec{twomul} send on average 1.5 copies to this system.
Comparing to the results in \rsec{twomul}, URLLC users in this section send more copies of their packets (10 in this section and 7.5 in \rsec{twomul}), while
the eMBB users in this section  send fewer copies (1 in this section and 1.5 in \rsec{twomul}).
Thus, the error probability of URLLC users is much lower than that in \rsec{twomul}, and one can admit more eMBB users in this system.

}

\bsec{Conclusion}{con}

{
Motivated by the analytical intractability of the CSA system with both spatial diversity and temporal diversity in
a correlated on-off fading channel, we developed a probabilistic  framework for Poisson receivers. We first showed that many CSA systems are Poisson receivers, and then used them as building blocks for analyzing complicated CSA systems.
Such an analysis is feasible due to two important closure properties of Poisson receivers: (i) Poisson receivers with packet routing are still Poisson receivers, and
(ii) Poisson receivers with packet coding are still Poisson receivers.
 We also demonstrated through a use case that our framework could provide differentiated services between URLLC traffic and eMBB traffic.

Poisson receivers unlock the power of analyzing CSA systems by the classical tree evaluation method \cite{luby1998analysis,luby1998analysisb,richardson2001capacity,liva2011graph}. There is an interesting analogy between Poisson receivers and quasi-reversible queues in queueing theory (see, e.g., the two books for detailed descriptions of quasi-reversible queues \cite{kelly2011reversibility,nelson2013probability}).  A Markov process is a very powerful
tool for analyzing a single queue. However, when there is a network of queues interconnected by each other, analysis by using multi-dimensional Markov processes becomes difficult. Quasi-reversibility exploits the closure properties of Poisson processes so that each queue in a quasi-reversible queueing network can be analyzed in isolation \cite{kelly2011reversibility,walrand1983probabilistic}. Moreover, a network of quasi-reversible queues is itself a quasi-reversible queue.
Similarly, the tree evaluation method is a very powerful tool for tracking the density evolution in  a single tree-like bipartite graph.
However, tracking multiple densities in a general graph is difficult. By exploiting the closure properties of Poisson degree distributions,
we are able to show the two closure properties of
Poisson receivers. These two closure properties enable us to treat CSA systems with spatial diversity as Poisson receivers, and thus they can be further used for analyzing CSA systems with both spatial diversity and temporal diversity.

}

\begin{thebibliography}{10}
\providecommand{\url}[1]{#1}
\csname url@samestyle\endcsname
\providecommand{\newblock}{\relax}
\providecommand{\bibinfo}[2]{#2}
\providecommand{\BIBentrySTDinterwordspacing}{\spaceskip=0pt\relax}
\providecommand{\BIBentryALTinterwordstretchfactor}{4}
\providecommand{\BIBentryALTinterwordspacing}{\spaceskip=\fontdimen2\font plus
\BIBentryALTinterwordstretchfactor\fontdimen3\font minus
  \fontdimen4\font\relax}
\providecommand{\BIBforeignlanguage}[2]{{%
\expandafter\ifx\csname l@#1\endcsname\relax
\typeout{** WARNING: IEEEtran.bst: No hyphenation pattern has been}%
\typeout{** loaded for the language `#1'. Using the pattern for}%
\typeout{** the default language instead.}%
\else
\language=\csname l@#1\endcsname
\fi
#2}}
\providecommand{\BIBdecl}{\relax}
\BIBdecl

\bibitem{ALOHA}
N.~Abramson, ``The aloha system: another alternative for computer
  communications,'' in \emph{Proceedings of the November 17-19, 1970, fall
  joint computer conference}.\hskip 1em plus 0.5em minus 0.4em\relax ACM, 1970,
  pp. 281--285.

\bibitem{paolini2012random}
E.~Paolini, G.~Liva, and M.~Chiani, ``Random access on graphs: A survey and new
  results,'' in \emph{Signals, Systems and Computers (ASILOMAR), 2012
  Conference Record of the Forty Sixth Asilomar Conference on}.\hskip 1em plus
  0.5em minus 0.4em\relax IEEE, 2012, pp. 1743--1747.

\bibitem{paolini2015coded}
E.~Paolini, C.~Stefanovi{\'c}, G.~Liva, and P.~Popovski, ``Coded random access:
  applying codes on graphs to design random access protocols,'' \emph{IEEE
  Communications Magazine}, vol.~53, no.~6, pp. 144--150, 2015.

\bibitem{munari2015multi}
A.~Munari, F.~Clazzer, and G.~Liva, ``Multi-receiver aloha systems-a survey and
  new results,'' in \emph{Communication Workshop (ICCW), 2015 IEEE
  International Conference on}.\hskip 1em plus 0.5em minus 0.4em\relax IEEE,
  2015, pp. 2108--2114.

\bibitem{stefanovic2018coded}
{\v{C}}.~Stefanovi{\'c} and D.~Vukobratovi{\'c}, ``Coded random access,'' in
  \emph{Network Coding and Subspace Designs}.\hskip 1em plus 0.5em minus
  0.4em\relax Springer, 2018, pp. 339--359.

\bibitem{zorzi1995mobile}
M.~Zorzi, ``Mobile radio slotted aloha with capture and diversity,''
  \emph{Wireless Networks}, vol.~1, no.~2, pp. 227--239, 1995.

\bibitem{munari2013throughput}
A.~Munari, M.~Heindlmaier, G.~Liva, and M.~Berioli, ``The throughput of slotted
  aloha with diversity,'' in \emph{Communication, Control, and Computing
  (Allerton), 2013 51st Annual Allerton Conference on}.\hskip 1em plus 0.5em
  minus 0.4em\relax IEEE, 2013, pp. 698--706.

\bibitem{casini2007contention}
E.~Casini, R.~De~Gaudenzi, and O.~D.~R. Herrero, ``Contention resolution
  diversity slotted aloha (crdsa): An enhanced random access schemefor
  satellite access packet networks,'' \emph{IEEE Transactions on Wireless
  Communications}, vol.~6, no.~4, 2007.

\bibitem{liva2011graph}
G.~Liva, ``Graph-based analysis and optimization of contention resolution
  diversity slotted aloha,'' \emph{IEEE Transactions on Communications},
  vol.~59, no.~2, pp. 477--487, 2011.

\bibitem{luby1998analysis}
M.~Luby, M.~Mitzenmacher, and M.~A. Shokrollahi, ``Analysis of random processes
  via and-or tree evaluation,'' in \emph{SODA}, vol.~98, 1998, pp. 364--373.

\bibitem{narayanan2012iterative}
K.~R. Narayanan and H.~D. Pfister, ``Iterative collision resolution for slotted
  aloha: An optimal uncoordinated transmission policy,'' in \emph{Turbo Codes
  and Iterative Information Processing (ISTC), 2012 7th International Symposium
  on}.\hskip 1em plus 0.5em minus 0.4em\relax IEEE, 2012, pp. 136--139.

\bibitem{stefanovic2013joint}
{\v{C}}.~Stefanovi{\'c}, K.~F. Trilingsgaard, N.~K. Pratas, and P.~Popovski,
  ``Joint estimation and contention-resolution protocol for wireless random
  access,'' in \emph{2013 IEEE International Conference on Communications
  (ICC)}.\hskip 1em plus 0.5em minus 0.4em\relax IEEE, 2013, pp. 3382--3387.

\bibitem{jakovetic2015cooperative}
D.~Jakoveti{\'c}, D.~Bajovi{\'c}, D.~Vukobratovi{\'c}, and V.~Crnojevi{\'c},
  ``Cooperative slotted aloha for multi-base station systems,'' \emph{IEEE
  Transactions on Communications}, vol.~63, no.~4, pp. 1443--1456, 2015.

\bibitem{ogata2017multi}
S.~Ogata, K.~Ishibashi, and G.~Abreu, ``Multi-access diversity gain via
  multiple base station cooperation in frameless aloha,'' in \emph{Signal
  Processing Advances in Wireless Communications (SPAWC), 2017 IEEE 18th
  International Workshop on}.\hskip 1em plus 0.5em minus 0.4em\relax IEEE,
  2017, pp. 1--5.

\bibitem{paolini2011graph}
E.~Paolini, G.~Liva, and M.~Chiani, ``Graph-based random access for the
  collision channel without feedback: Capacity bound,'' in \emph{2011 IEEE
  Global Telecommunications Conference-GLOBECOM 2011}.\hskip 1em plus 0.5em
  minus 0.4em\relax IEEE, 2011, pp. 1--5.

\bibitem{luby1998analysisb}
M.~Luby, M.~Mitzenmacher, A.~Shokrollah, and D.~Spielman, ``Analysis of low
  density codes and improved designs using irregular graphs,'' in
  \emph{Proceedings of the thirtieth annual ACM symposium on Theory of
  computing}, 1998, pp. 249--258.

\bibitem{richardson2001capacity}
T.~J. Richardson and R.~L. Urbanke, ``The capacity of low-density parity-check
  codes under message-passing decoding,'' \emph{IEEE Transactions on
  Information Theory}, vol.~47, no.~2, pp. 599--618, 2001.

\bibitem{gallager1962low}
R.~Gallager, ``Low-density parity-check codes,'' \emph{IRE Transactions on
  information theory}, vol.~8, no.~1, pp. 21--28, 1962.

\bibitem{ordentlich2017low}
O.~Ordentlich and Y.~Polyanskiy, ``Low complexity schemes for the random access
  gaussian channel,'' in \emph{2017 IEEE International Symposium on Information
  Theory (ISIT)}.\hskip 1em plus 0.5em minus 0.4em\relax IEEE, 2017, pp.
  2528--2532.

\bibitem{clazzer2017irregular}
F.~Clazzer, E.~Paolini, I.~Mambelli, and {\v{C}}.~Stefanovi{\'c}, ``Irregular
  repetition slotted aloha over the rayleigh block fading channel with
  capture,'' in \emph{2017 IEEE International Conference on Communications
  (ICC)}.\hskip 1em plus 0.5em minus 0.4em\relax IEEE, 2017, pp. 1--6.

\bibitem{Newman2010}
M.~Newman, \emph{Networks: an introduction}.\hskip 1em plus 0.5em minus
  0.4em\relax OUP Oxford, 2009.

\bibitem{Liva2012spatially}
G.~Liva, E.~Paolini, M.~Lentmaier, and M.~Chiani, ``Spatially-coupled random
  access on graphs,'' in \emph{2012 IEEE International Symposium on Information
  Theory Proceedings}.\hskip 1em plus 0.5em minus 0.4em\relax IEEE, 2012, pp.
  478--482.

\bibitem{chang1995sample}
C.-S. Chang, ``Sample path large deviations and intree networks,''
  \emph{Queueing Systems}, vol.~20, no. 1-2, pp. 7--36, 1995.

\bibitem{lee2005throughput}
I.-C. Lee, C.-S. Chang, and C.-M. Lien, ``On the throughput of multicasting
  with incremental forward error correction,'' \emph{IEEE transactions on
  information theory}, vol.~51, no.~3, pp. 900--918, 2005.

\bibitem{formaggio2020receiver}
F.~Formaggio, A.~Munari, and F.~Clazzer, ``On receiver diversity for grant-free
  based machine type communications,'' \emph{Ad Hoc Networks}, p. 102245, 2020.

\bibitem{bennis2018ultra}
M.~Bennis, M.~Debbah, and H.~V. Poor, ``Ultra-reliable and low-latency wireless
  communication: Tail, risk and scale,'' \emph{arXiv preprint
  arXiv:1801.01270}, 2018.

\bibitem{chang2019asynchronous}
C.-S. Chang, D.-S. Lee, and C.~Wang, ``Asynchronous grant-free uplink
  transmissions in multichannel wireless networks with heterogeneous qos
  guarantees,'' \emph{IEEE/ACM Transactions on Networking}, vol.~27, no.~4, pp.
  1584--1597, 2019.

\bibitem{centenaro2020analysis}
M.~Centenaro, L.~Vangelista, and S.~Saur, ``Analysis of 5g radio access
  protocols for uplink urllc in a connection-less mode,'' \emph{IEEE
  Transactions on Wireless Communications}, vol.~19, no.~5, pp. 3104--3117,
  2020.

\bibitem{liu2020analyzing}
Y.~Liu, Y.~Deng, M.~Elkashlan, A.~Nallanathan, and G.~K. Karagiannidis,
  ``Analyzing grant-free access for urllc service,'' \emph{arXiv preprint
  arXiv:2002.07842}, 2020.

\bibitem{anand2020joint}
A.~Anand, G.~De~Veciana, and S.~Shakkottai, ``Joint scheduling of urllc and
  embb traffic in 5g wireless networks,'' \emph{IEEE/ACM Transactions on
  Networking}, vol.~28, no.~2, pp. 477--490, 2020.

\bibitem{popovski20185g}
P.~Popovski, K.~F. Trillingsgaard, O.~Simeone, and G.~Durisi, ``5g wireless
  network slicing for embb, urllc, and mmtc: A communication-theoretic view,''
  \emph{IEEE Access}, vol.~6, pp. 55\,765--55\,779, 2018.

\bibitem{3gpp.22.104}
\BIBentryALTinterwordspacing
3GPP, ``Service requirements for cyber-physical control applications in
  vertical domains,'' {3rd Generation Partnership Project (3GPP)}, Technical
  Specification (TS) 22.104, 03 2019, version 16.1.0. [Online]. Available:
  \url{https://portal.3gpp.org/desktopmodules/Specifications/SpecificationDetails.aspx?specificationId=3528}
\BIBentrySTDinterwordspacing

\bibitem{3gpp.36.306}
\BIBentryALTinterwordspacing
------, ``{Evolved Universal Terrestrial Radio Access (E-UTRA); User Equipment
  (UE) radio access capabilities},'' {3rd Generation Partnership Project
  (3GPP)}, Technical Specification (TS) 36.306, 06 2019, version 15.5.0.
  [Online]. Available:
  \url{https://portal.3gpp.org/desktopmodules/Specifications/SpecificationDetails.aspx?specificationId=2434}
\BIBentrySTDinterwordspacing

\bibitem{kelly2011reversibility}
F.~P. Kelly, \emph{Reversibility and stochastic networks}.\hskip 1em plus 0.5em
  minus 0.4em\relax Cambridge University Press, 2011.

\bibitem{nelson2013probability}
R.~Nelson, \emph{Probability, stochastic processes, and queueing theory: the
  mathematics of computer performance modeling}.\hskip 1em plus 0.5em minus
  0.4em\relax Springer Science \& Business Media, 2013.

\bibitem{walrand1983probabilistic}
J.~Walrand, ``A probabilistic look at networks of quasi-reversible queues,''
  \emph{IEEE Transactions on Information Theory}, vol.~29, no.~6, pp. 825--831,
  1983.

\end{thebibliography}


\appendix
\section*{Appendix A}

\setcounter{section}{1}


In \rsec{repeat}, we considered coded Poisson receivers that use simple repetition codes. In that setting, a packet is successfully decoded when one of its copies is successfully received by a Poisson receiver.
In this section, we further extend coded Poisson receivers to an ideal $(n,n_0)$-forward error correction (FEC) code. For an ideal $(n,n_0)$-FEC code, a packet is divided into $n_0$ {\em data} blocks. By encoding with additional $n-n_0$ {\em redundant} blocks, we have a code with $n$ blocks for a packet. A packet can  be
successfully decoded as long as  $n_0$ out of the $n$ blocks are successfully received.

Analogous to the throughput analysis in \rsec{repeat}, we consider a system with $G_k T$ class $k$ active users, $k=1,2,\ldots, K$, and $T$ (independent) Poisson receivers with $K$ classes of input traffic and
the success probability functions $P_{{\rm suc},k}(\rho)$, $k=1,2,\ldots, K$. Each class $k$ user encodes its packet with an ideal $(n_k,n_{k,0})$-FEC code and transmits its packet  for a random  number of blocks.
Let $L_k$ be the random variable that represents the number of blocks transmitted from a class $k$ packet. Each of the $L_k$ blocks is transmitted to one of the $T$ Poisson receivers chosen {\em uniformly} and {\em independently}. If (at least) $n_{k,0}$ blocks of these $L_k$ blocks of a class $k$ packet are successfully received by Poisson receivers, then the other blocks of that packet can be removed (cancelled) from the system. Such a process can  then be repeatedly carried out to decode the rest of the packets.

Let  $\Lambda_{k,\ell}$, $\ell={n_{k,0}},n_{k,0}+1, \ldots,n_k$, be the probability that $\ell$ blocks are transmitted for a class $k$ packet.
Define the generating function $\Lambda_k(x)$ for the degree distribution as in \req{mean0000mul} and the generating function $\lambda_k(x)$
for the excess degree distribution as in \req{mean3333mul}. Also,
$$\rho=(\rho_1, \rho_2, \ldots, \rho_K),$$
where $\rho_k=G_k \Lambda_k^\prime(1)$ is the Poisson offered load for the class $k$ traffic at a receiver.

As in \rsec{repeat}, we model the decoding process by a user-receiver bipartite graph. Let
$p_k^{(i)}$ (resp $q_k^{(i)}$) be the probability that  the receiver (resp. user)  end of a randomly selected class $k$ edge has not been successfully received/decoded  after the $i^{th}$ SIC iteration.
Clearly, we still have
\beq{tag1111nk}
p_k^{(1)}=1-P_{{\rm suc},k}(\rho).
\eeq
In our setting, a class $k$ packet sent from a user is successfully received if at least $n_{k,0}$ blocks are successfully received at the {\em receiver} end.
Thus,

\bear{tag6666cnk}
&&q_k^{(1)}\nonumber\\
&&= 1-\sum_{\ell={n_{k,0}}}^{n_k} \Big(\lambda_{k,\ell}\cdot(1-\sum_{i=0}^{n_{k,0}-1}{\ell \choose i}(p_k^{(1)})^{\ell-i}(1-p_k^{(1)})^{i})\Big)\nonumber \\
&&= \sum_{\ell={n_{k,0}}}^{n_k} \Big(\lambda_{k,\ell}\cdot\sum_{i=0}^{n_{k,0}-1}{\ell \choose i}(p_k^{(1)})^{\ell-i}(1-p_k^{(1)})^{i}\Big) \nonumber \\
&&=\sum_{\ell={n_{k,0}}}^{n_k} \Big(\lambda_{k,\ell}\cdot({\ell \choose 0}(p_k^{(1)})^{\ell}+{\ell \choose 1}(p_k^{(1)})^{\ell-1}(1-p_k^{(1)}) \nonumber\\
&&\quad+ \cdots + {\ell \choose n_{k,0}-1}(p_k^{(1)})^{\ell-(n_{k,0}-1)}(1-p_k^{(1)})^{n_{k,0}-1})\Big)\nonumber \\
&&= \lambda_k(p_k^{(1)}) + \lambda_k^\prime(p_k^{(1)})\cdot(1-p_k^{(1)}) + \cdots\nonumber\\
&&\quad + \lambda_k^{\langle n_{k,0}-1\rangle }(p_k^{(1)})\cdot \frac{(1-p_k^{(1)})^{n_{k,0}-1}}{(n_{k,0}-1)!}\nonumber \\
&&= \sum_{j=0}^{n_{k,0}-1}\Big(\lambda_k^{\langle j \rangle}(p_k^{(1)})\cdot\dfrac{(1-p_k^{(1)})^j}{j!}\Big),
\eear
where $\lambda_k^{\langle j \rangle}(x)$ is the $j^{th}$ derivative of $\lambda_k(x)$.
In general, we have the following recursive equations:
\bear{tag6666ck}
p_k^{(i+1)}&=&1-P_{{\rm suc},k}(q^{(i)} \circ \rho), \label{eq:tag6666nka}\\
q_k^{({i+1})}&=&\sum_{j=0}^{n_{k,0}-1}\Big(\lambda_k^{\langle j \rangle}(p_k^{(i+1)})\cdot\dfrac{(1-p_k^{(i+1)})^j}{j!}\Big) \nonumber\\
& = & \sum_{j=0}^{n_{k,0}-1}\Big(\lambda_k^{\langle j \rangle}(1-P_{{\rm suc},k}(q^{(i)} \circ \rho))\nonumber\\
&&\quad\quad\cdot\dfrac{(P_{{\rm suc},k}(q^{(i)} \circ \rho))^j}{j!}\Big). \label{eq:tag6666nkb}
\eear

Let $G=(G_1, G_2, \ldots, G_{K})$ and
$\Lambda^\prime (1)=(\Lambda^\prime_1 (1), \Lambda^\prime_2 (1), \ldots, \Lambda^\prime_K (1))$.
Also, let $\tilde P_{{\rm suc},k}^{(i)}(G)$
be  the success probability of a tagged class $k$ packet for the CPR system after the $i^{th}$ SIC iteration when the system is subject to a (normalized) Poisson offered load $G$.
Recall that a packet sent from a user can be successfully received if at least $n_{k,0}$ blocks are successfully received at the {\em receiver} end.
Thus, we have from \req{tag6666nka} and \req{tag6666nkb} that
\bear{mean5555nk}
&&\tilde P_{\rm suc}^{(i)}(G)\\
&&=\sum_{\ell={n_{k,0}}}^{n_k} \Big(\Lambda_{k,\ell} \cdot \sum_{j={n_{k,0}}}^\ell {\ell\choose j} (1-p_k^{(i)})^j(p_k^{(i)})^{\ell-j}\Big). \nonumber\\
&& = 1-\sum_{j=0}^{n_{k,0}-1}\Big(\Lambda_k^{\langle j \rangle}(p_k^{(i)})\cdot\dfrac{(1-p_k^{(i)})^j}{j!}\Big) \nonumber \\
&& = 1-\sum_{j=0}^{n_{k,0}-1}\Big(\Lambda_k^{\langle j \rangle}(1-P_{\rm suc}(q^{(i-1)}\circ G \circ \Lambda^\prime(1)))\nonumber \\
&&\quad\quad\quad\quad\quad\quad \cdot\dfrac{(P_{\rm suc}(q^{(i-1)}\circ G \circ \Lambda^\prime(1)))^j}{j!}\Big).\nonumber\\
\eear
where $\Lambda_k^{\langle j \rangle}(x)$ is the $j^{th}$ derivative of $\Lambda_k(x)$.

\begin{IEEEbiography}[{\includegraphics[width=1in,height=1.25in,clip,keepaspectratio]{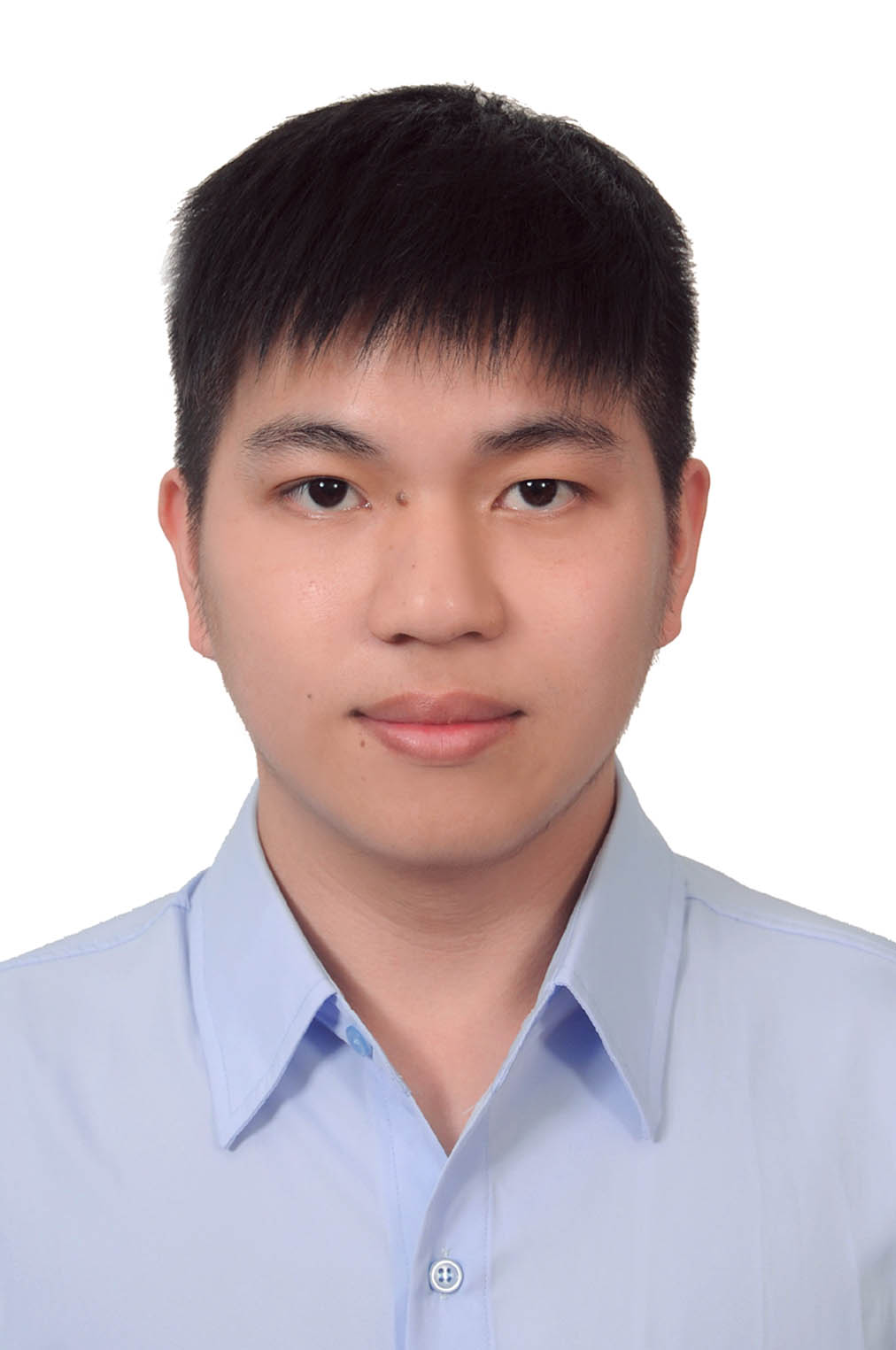}}]
{Che-Hao Yu} received his B.S. degree in mathematics from National Tsing-Hua University, Hsinchu, Taiwan (R.O.C.), in 2018, and the M.S. degree in communications engineering from National Tsing Hua University, Hsinchu, Taiwan (R.O.C.), in 2020. His research interest is in 5G wireless communication.
\end{IEEEbiography}

\begin{IEEEbiography}[{\includegraphics[width=1in,height=1.25in,clip,keepaspectratio]{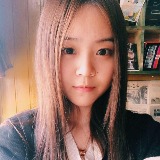}}]
	{Lin Huang} received her B.S. degree in Science and Technology from Nanjing University, Nanjing, China, in 2017, and the M.S. degree in communications engineering from National Tsing Hua University, Hsinchu, Taiwan (R.O.C.), in 2019. Her research interest is in 5G wireless communications. Currently, she is a communication algorithm engineer.
\end{IEEEbiography}

\begin{IEEEbiography}[{\includegraphics[width=1in,height=1.25in,clip,keepaspectratio]{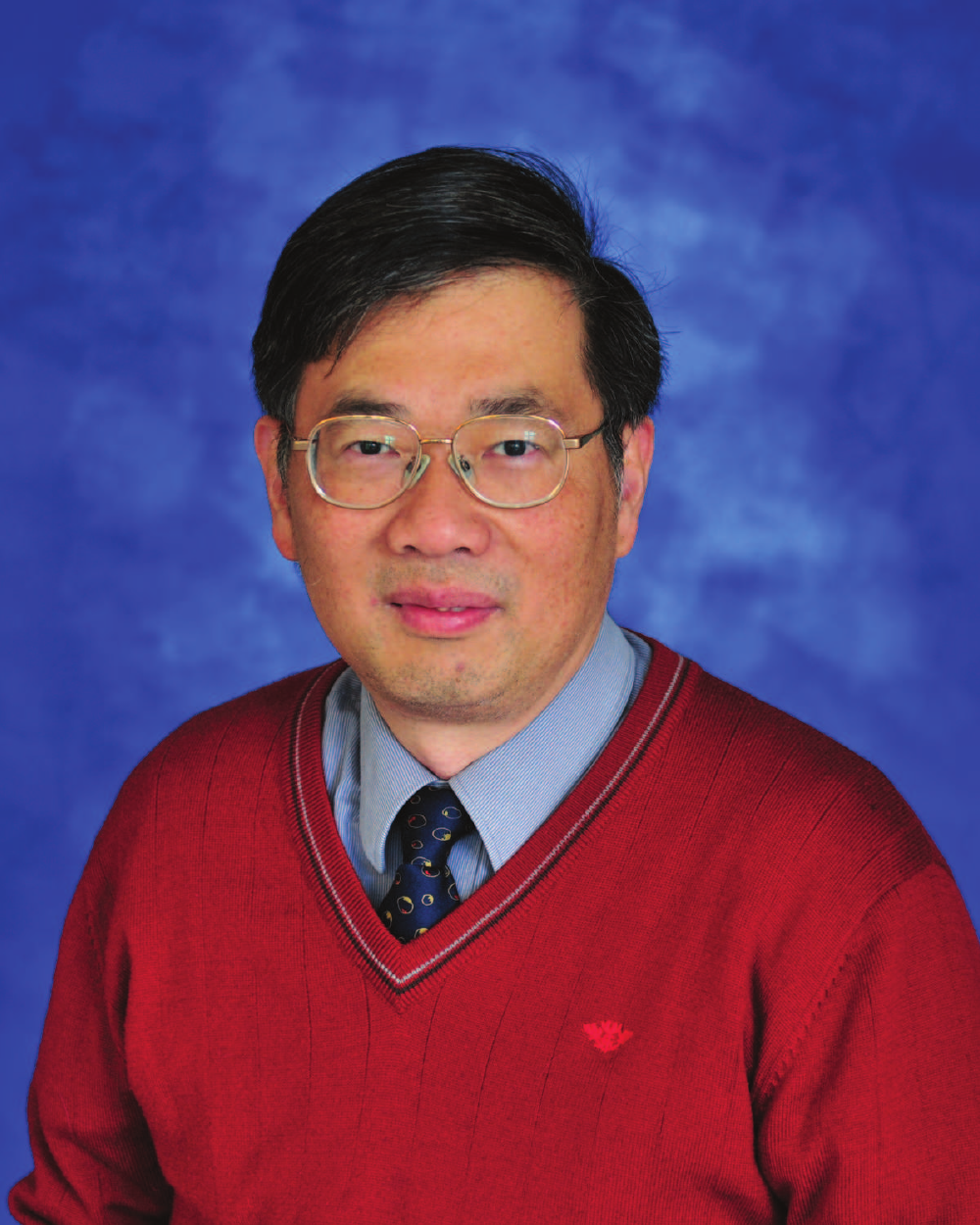}}]
	{Cheng-Shang Chang}
	(S'85-M'86-M'89-SM'93-F'04)
	received the B.S. degree from National Taiwan
	University, Taipei, Taiwan, in 1983, and the M.S.
	and Ph.D. degrees from Columbia University, New
	York, NY, USA, in 1986 and 1989, respectively, all
	in electrical engineering.
	
	From 1989 to 1993, he was employed as a
	Research Staff Member with the IBM Thomas J.
	Watson Research Center, Yorktown Heights, NY,
	USA. Since 1993, he has been with the Department
	of Electrical Engineering, National Tsing Hua
	University, Taiwan, where he is a Tsing Hua Distinguished Chair Professor. He is the author
	of the book Performance Guarantees in Communication Networks (Springer,
	2000) and the coauthor of the book Principles, Architectures and Mathematical
	Theory of High Performance Packet Switches (Ministry of Education, R.O.C.,
	2006). His current research interests are concerned with network science, big data analytics,
	mathematical modeling of the Internet, and high-speed switching.
	
	Dr. Chang served as an Editor for Operations Research from 1992 to 1999,
	an Editor for the {\em IEEE/ACM TRANSACTIONS ON NETWORKING} from 2007
	to 2009, and an Editor for the {\em IEEE TRANSACTIONS
		ON NETWORK SCIENCE AND ENGINEERING} from 2014 to 2017. He is currently serving as an Editor-at-Large for the {\em IEEE/ACM
		TRANSACTIONS ON NETWORKING}. He is a member of IFIP Working
	Group 7.3. He received an IBM Outstanding Innovation Award in 1992, an
	IBM Faculty Partnership Award in 2001, and Outstanding Research Awards
	from the National Science Council, Taiwan, in 1998, 2000, and 2002, respectively.
	He also received Outstanding Teaching Awards from both the College
	of EECS and the university itself in 2003. He was appointed as the first Y. Z.
	Hsu Scientific Chair Professor in 2002. He received the Merit NSC Research Fellow Award from the
	National Science Council, R.O.C. in 2011. He also received the Academic Award in 2011 and the National Chair Professorship in 2017 from
	the Ministry of Education, R.O.C. He is the recipient of the 2017 IEEE INFOCOM Achievement Award.
\end{IEEEbiography}

\begin{IEEEbiography}
	[{\includegraphics[width=1in,height=1.25in,clip,keepaspectratio]{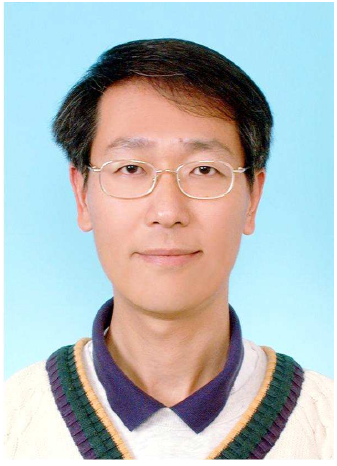}}]
	{Duan-Shin Lee}(S'89-M'90-SM'98) received the B.S. degree from National Tsing Hua
	University, Taiwan, in 1983, and the MS and Ph.D. degrees from
	Columbia University, New York, in 1987 and 1990, all in electrical
	engineering.  He worked as a research staff member at the C\&C Research Laboratory
	of NEC USA, Inc. in Princeton, New Jersey from 1990 to 1998.  He joined the
	Department of Computer Science of National Tsing Hua University in Hsinchu,
	Taiwan, in 1998.  Since August 2003, he has been a professor.  He received
	a best paper award from the Y.Z. Hsu Foundation in 2006.  He served as
	an editor for the Journal of Information Science and Engineering between
	2013 and 2015.  He is currently an editor for Performance Evaluation.
	Dr. Lee's current research interests are network science, game theory,
	machine learning and high-speed networks.  He is a senior IEEE member.
\end{IEEEbiography}



\end{document}